\documentclass[final,12pt]{colt2023} 

\title[A new ranking scheme for modern  data]{A new ranking scheme for modern  data and its application to two-sample hypothesis testing}
\usepackage{times}
\usepackage{bbm}
\usepackage{enumerate}
\usepackage{multirow}

\def\RR{\mathbb{R}}

\def \diff {\mathrm{diff}}
\def \mB {\mathrm{B}}

\def \Pp {{\mathbbm P}}
\def \Ep {{\mathbbm E}}
\def \Vp { {\rm Var}}
\def \Covp {{\rm C o v}}

\def \Eb {{\mathbbm E}_\mB}
\def \Vb {{\rm V a r}_\mB}
\def \Covb {{\rm C o v}_\mB}

\def\indi{\mathbbm{1}}

\def\cX{\mathcal{X}}
\def\cS{\mathcal{S}}
\def\cG{\mathcal{G}}

\def \A {\mathbf{A}}

\def \I {\mathbf{I}}

\def \U {\mathbf{U}}

\def \R {\mathbf{R}}
\def \S {\mathbf{S}}

\def \bSigma {\mathbf{\Sigma}}

\def\cN{\mathcal{N}}

\def \node{{\rm node}}
\def\trans{^{\scriptscriptstyle \sf T}}

\def\bzero{\mathbf{0}}
\def \bone{\mathbf{1}}

\def \bmu{\boldsymbol{\mu}}
\def \bSigma{\boldsymbol{\Sigma}}
\def \Rg{\rm R_g}
\def \Ro{\rm R_o}

\coltauthor{%
 \Name{Doudou Zhou} \Email{douzh@ucdavis.edu}\\
 \addr University of California, Davis
 \AND
 \Name{Hao Chen} \Email{hxchen@ucdavis.edu}\\
 \addr University of California, Davis%
}

\graphicspath{{figures/}}
\begin{document}

\maketitle

\begin{abstract}%
Rank-based approaches are among the most popular nonparametric methods for univariate data in tackling statistical problems such as hypothesis testing due to their robustness and effectiveness. However, they are unsatisfactory for more complex data. 
In the era of big data, high-dimensional and non-Euclidean data, such as networks and images, are ubiquitous and pose challenges for statistical analysis. Existing multivariate ranks such as component-wise, spatial, and depth-based ranks do not apply to non-Euclidean data and have limited performance for high-dimensional data. Instead of dealing with the ranks of observations, we  
propose two types of ranks applicable to complex data 
based on a similarity graph constructed on observations: a graph-induced rank defined by the inductive nature of the graph and an overall rank defined by the weight of edges in the graph. To illustrate their utilization, both the new ranks are used to construct test statistics for the two-sample hypothesis testing, which converge to the $\chi_2^2$ distribution under the permutation null distribution and some mild conditions of the ranks, enabling an easy type-I error control. Simulation studies show that the new method exhibits good power under a wide range of alternatives compared to existing methods. The new test is illustrated on the New York City taxi data for comparing travel patterns in consecutive months and a brain network dataset comparing male and female subjects.
\end{abstract}

\begin{keywords}%
  Rank-based method; high-dimensional/nonparametric statistics;    similarity graph; non-Euclidean data
\end{keywords}

\section{Introduction}

\subsection{Multivariate ranks}
High-dimensional and non-Euclidean data have become ubiquitous in the era of big data, such as networks and images, which poses challenges for statistical analysis \citep{bullmore2012economy, tian2016analysis,menafoglio2017statistical}. Parametric approaches are limited when many nuisance parameters need to be estimated. Among the nonparametric methods, rank-based methods are attractive due to their robustness and effectiveness and have been extensively studied for univariate data. However, univariate ranks can not be easily extended to multivariate data due to the lack of natural ordering of the values. 
 The existing extensions of ranks to multivariate data include the component-wise rank \citep{bickel1965some,hallin1995multivariate,puri2013class}, the spatial rank \citep{chaudhuri1996geometric,oja2010multivariate}, the depth-based rank \citep{Liu1993rank,zuo2000general}, the Mahalanobis rank \citep{hallin2002optimal,hallin2004rank,hallin2006parametric}, the metric rank \citep{pan2018ball} and the measure transportation-based rank \citep{deb2021multivariate}. Specifically, given $N$ observations $Z_1,\ldots, Z_N \in \RR^d$:
\begin{itemize}
    \item The component-wise rank $R_i \in \RR^d $ is the rank vector for each dimension of $Z_i$, e.g., $R_{ij}$ is the rank of $Z_{ij}$ among $Z_{1j},\ldots,Z_{Nj}$ for $j = 1, \ldots, d$.  
    Since it is defined for each dimension, this rank suffers from correlated covariates and is not invariant to affine transformations. 
    \item The spatial rank function is defined as $R(Z) = \sum_{i=1}^N U(Z-Z_i)/N$ where $U(Z) = Z/\|Z\|$ for $Z \neq \bzero_d$ and $U(\bzero_d) = \bzero_d$. 
The rank is powerful for detecting location differences, but not for distinguishing scale parameters due to the normalizing procedure involved in $U(\cdot)$. 

\item The depth-based rank measures the centrality of the observations. It depends on the choice of depth function. For example, the Mahalanobis's depth is defined as 
${\rm M}_h{\rm D} (Z) = \big\{ 1 + (Z-\bar Z)\trans \S^{-1}  (Z-\bar Z) \big\}^{-1}\,,$
where $\bar Z = \sum_{i=1}^N Z_i/N$ is the sample mean and $\S$ is the sample covariance matrix, and the Tukey's depth is defined as ${\rm TD} (Z) = \inf_{\cX}\{ F_N (\cX): \cX \text{ is a closed half-space containing } Z\}$, 
where $F_N$ is the empirical cumulative distribution function. Given a depth function, the depth-based ranks are the ranks of the depth values.  The depth ${\rm M}_h {\rm D}$ does not work when the dimension is larger than the number of observations.  Other depth functions are computationally extensive for high-dimensional data, for example, ${\rm TD}$ has the computational complexity $O(N^{d-1} \log N)$ \citep{liu2017fast} and the simplicial depth \citep{liu1988notion} has the computational complexity $O(N^{d} \log N)$ \citep{afshani2016approximating}. 

\item The Mahalanobis rank is designed for multivariate one-sample testing, which is defined as the rank of the pseudo-Mahalanobis distance $d(Z,\theta_0) = (Z-\theta_0)\trans \hat \bSigma^{-1}(Z-\theta_0)$, where $\theta_0$ is the location parameter of interest and specified under $H_0$, and $\hat \bSigma$ is an M-estimator of the covariance matrix due to \cite{tyler1987distribution}. It is powerful for elliptical symmetric distribution but is not robust to heavy-tailed distributions.  

\item The metric rank measures the difference between two probability distributions. Assume $Z_1,\dots, Z_m \overset{iid}{\sim} F_X, Z_{m+1},\dots Z_N\overset{iid}{\sim} F_Y$, and define 
$n A_{ij}^X, i,j\in\{1,\ldots,m\}$ be the rank of $d(Z_i,Z_j)$ among $\{ d(Z_i,Z_u), u=1,\ldots,m\}$ where $d(Z_i,Z_j)$ is the distance between $Z_i$ and $Z_j$, $m A_{ij}^Y, i,j\in\{1,\ldots,m\}$ be the rank of $d(Z_i,Z_j)$ among $\{ d(Z_i,Z_u), u=j,m+1,\ldots,N\}$,  $n C_{ij}^X, i,j \in \{m+1,\ldots,N\}$ be the rank of $d(Z_i,Z_j)$ among $\{ d(Z_i,Z_u), u=1,\ldots,m,j\}$, and $m C_{ij}^Y, i,j \in \{m+1,\ldots,N\}$ be the rank of $d(Z_i,Z_j)$ among $\{ d(Z_i,Z_u), u=m+1,\ldots,N\}$. Then the differences $A_{ij}^X - A_{ij}^Y$ and $C_{ij}^X - C_{ij}^Y$ are used to compare the two distributions. However, the limiting distribution of the test statistic is not easy to approximate, so a resampling procedure is usually used to obtain the $p$-value.
\item The measure transportation-based ranks are defined by the optimization problem 
\[\hat \sigma  = \arg \min_{\sigma \in \cS_N } \sum_{i=1}^N \|Z_i - c_{\sigma(i)}\|^2\,,\]
where $\sigma = ( \sigma(1), \ldots, \sigma(N) ) $
and $\cS_N$ is the set of all permutations of $\{1,\ldots,N\}$,  the multivariate rank vectors  $\{c_1,\ldots,c_N\}$ are a sequence of `uniform-like' points in $[0,1]^d$ generated from Halton sequences \citep{hofer2009distribution,hofer2010existence}. As a result, the rank vector of $Z_i$ will be $c_{\hat \sigma(i) }$. These ranks are also useful in detecting location differences. However, when the dimension is high, it is difficult to generate `uniformly' distributed rank vectors, which suffers from the curse of dimensionality. 
\end{itemize}

Noticing the limitations of the existing multivariate ranks, we propose ranks that rely on a similarity graph (Section \ref{sec:new rank}). We then build test statistics based on the new ranks for two-sample hypothesis testing (Section \ref{sec: new stat}). The asymptotic properties of the new test statistics are studied (Section \ref{sec: asymptotic}) and the performance of the new tests is explored through extensive simulation studies (Section \ref{sec: simulation}) and two real data applications (Section \ref{sec: real data}). The paper concludes with discussions in Section \ref{sec:discussion}.

\section{Graph-based ranks}
\label{sec:new rank}

One way of dealing with high-dimensional data is using inter-point distances, which has been shown to capture much information from data \citep{hall2005geometric,biswas2014nonparametric,Angiulli2018}. However, the distance-based methods suffer from outlier and heavy-tailed distributions. Specifically, many distance-based methods require the existence of some moments for their key theoretical properties to hold (e.g., \cite{li2018asymptotic,guo2020nonparametric,chakraborty2021new,zhu2021interpoint}). On the other hand, the graph-based methods are robust to outlier and heavy-tailed distributions. These methods construct unweighted similarity graphs using the pairwise similarities/distances of the observations,  then conduct statistical analysis based on the graphs (e.g., \cite{friedman1979multivariate,schilling1986multivariate,henze1988multivariate,rosenbaum2005exact, chen2017new}). 
We thus want to combine the advantage of both approaches by using more information compared to the graph-based methods while still keeping their robustness and propose the following graph-based ranks.

\begin{figure}[t]
 \centering
 \begin{tabular}{ccc}          \includegraphics[width=0.3\textwidth]{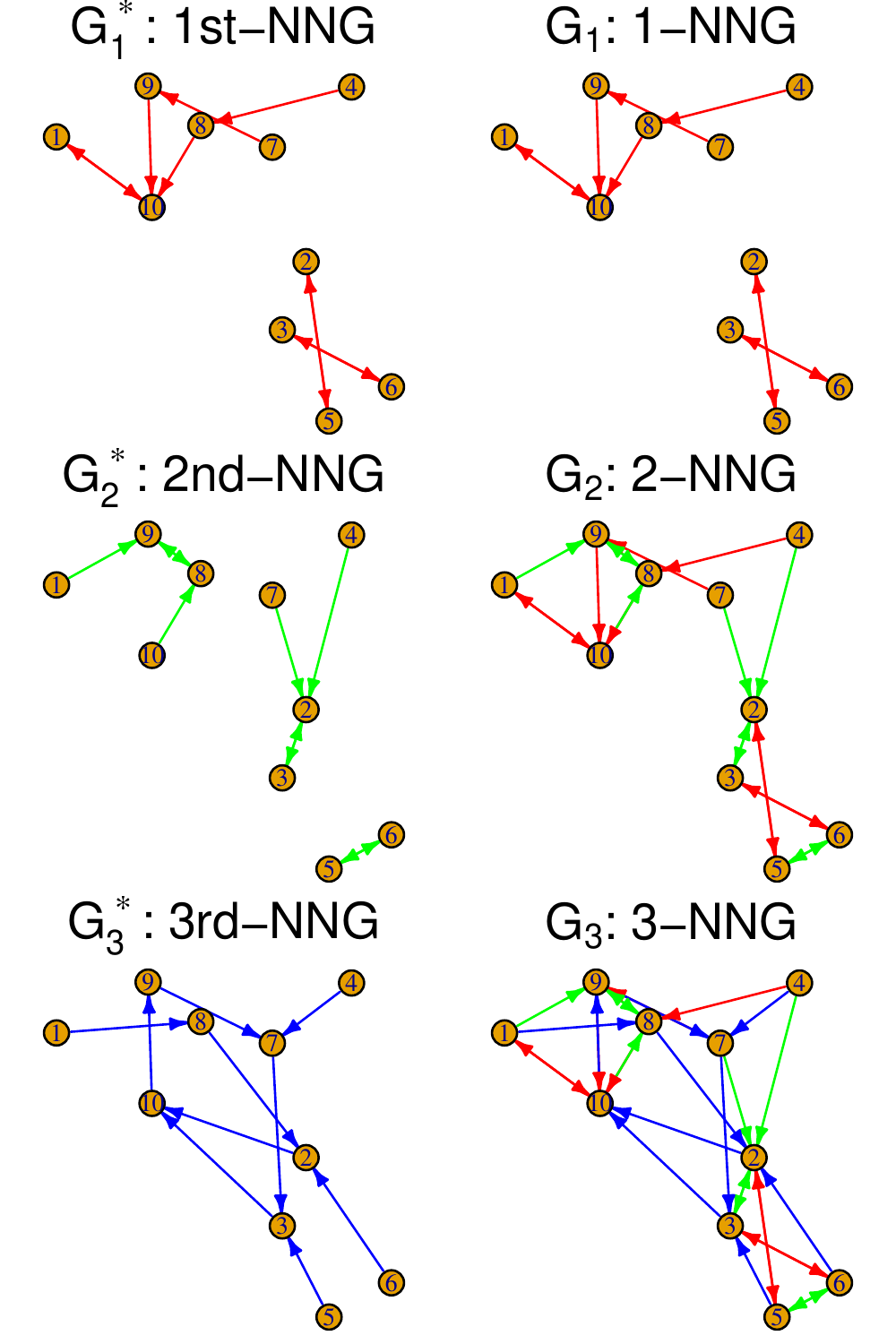} & 
           \includegraphics[width=0.3\textwidth]{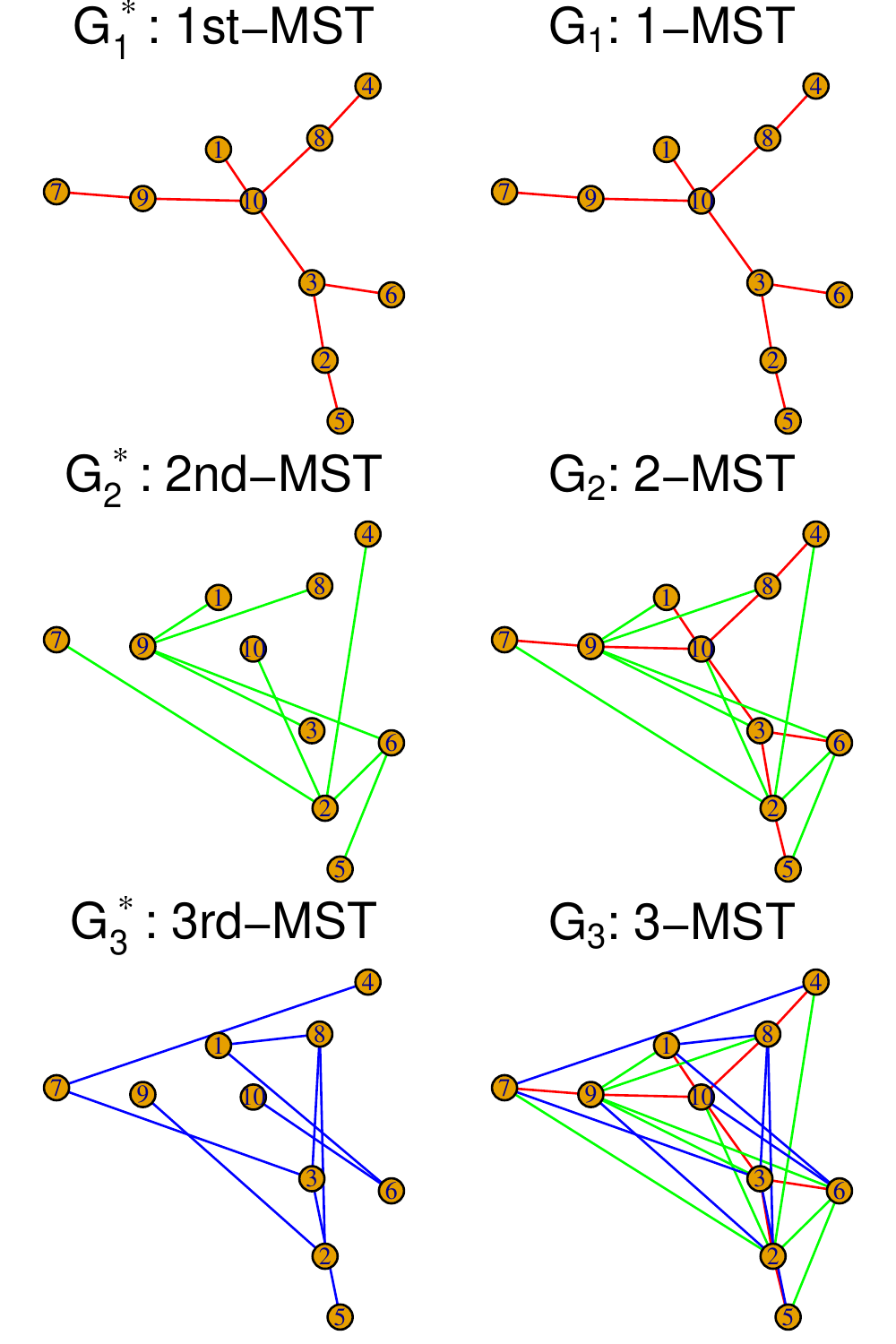} &           \includegraphics[width=0.3\textwidth]{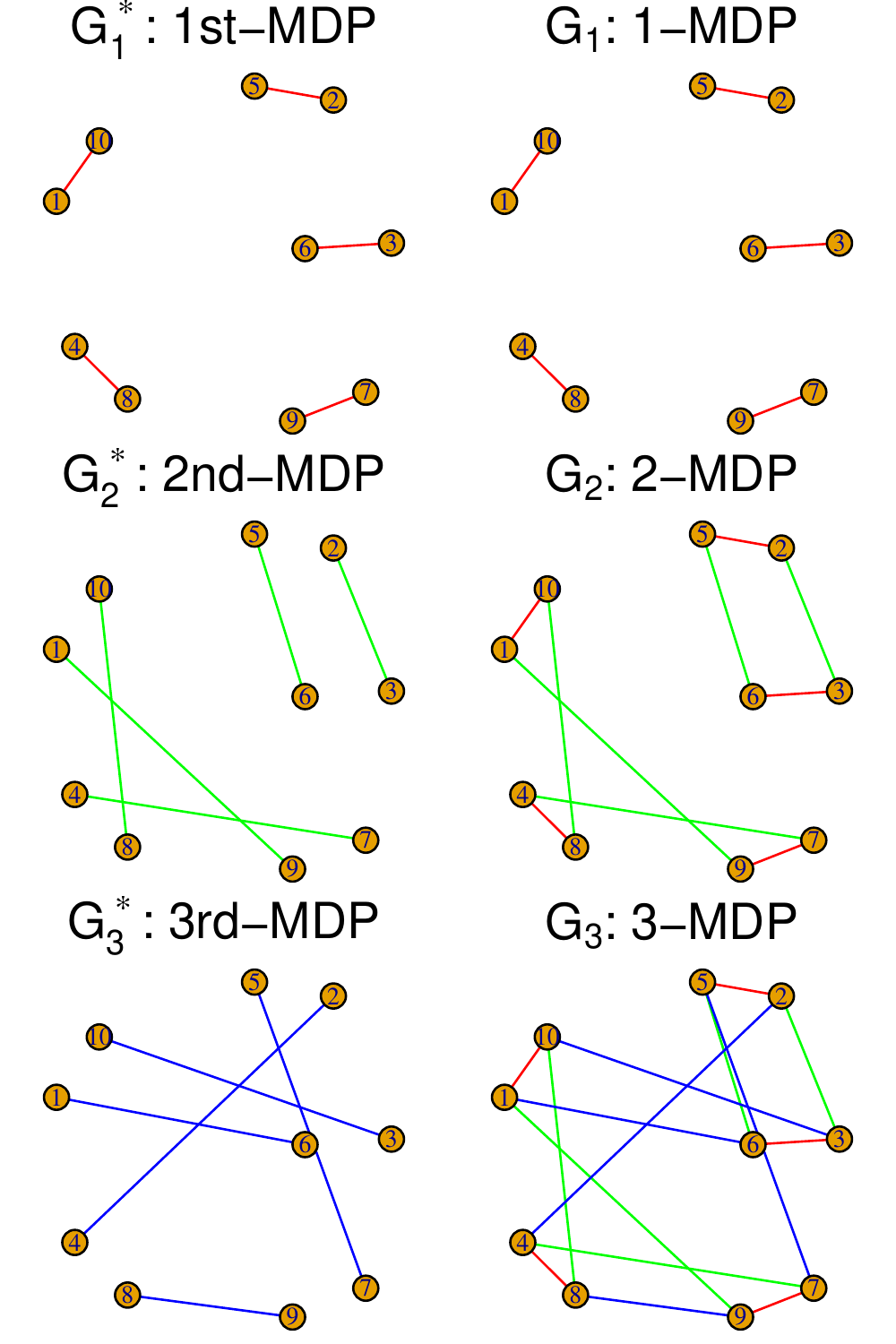} \\
(i) $k$-NNG. & (ii) $k$-MST. & (iii) $k$-MDP. \\
\end{tabular}
     \caption{Examples of different similarity graphs.}
     \label{fig:graph}
 \end{figure}

For two graphs $G_1$ and $G_2$  with identical vertices, define $G_1 \cap G_2 = \emptyset$ if they have no overlapping edges and  $G_1 \cup G_2$ as the graph with the same vertex set as them and the edge set their union. Given $N$ independent observations $\{ Z_i \}_{i=1}^N$, and a pre-specified integer $k$, we can construct a sequence of simple similarity graphs\footnote{A simple graph is a graph without self-loops and multiple edges between any two 
vertices.} $\{ G_l \}_{l=0}^k$ in an inductive way such that $G_0$ has no edges and 
$$G_{l+1} = G_{l} \cup G_{l+1}^{*}\,  \text{  with } G_{l+1}^{*} = \arg \max_{G' \in \cG_{l+1} } \sum_{(i,j) \in G'} S(Z_i,Z_j) \,,$$
where $\cG_{l+1} = \{G' \in \cG: G'\cap G_{l} =  \emptyset\}$ and $\cG$ is a graph set whose elements satisfy specific user-defined constraints. Here $S(\cdot,\cdot)$ is a similarity measure, for example,  $S(Z_i,Z_j) =  - \| Z_i - Z_j\|$ for Euclidean data.  For other choices of the similarity measures, see \cite{chen2013graph,sarkar2018some,sarkar2020some}. 
Many widely used similarity graphs can be constructed in this way with different constraints, for example, 
\begin{itemize}
    \item $k$-nearest neighbor graph  ($k$-NNG): $\cG=\{ G':$ each vertex $i$ connects to another vertex $j$ $\}$;
    \item  $k$-minimum spanning tree ($k$-MST)\footnote{The MST is a spanning tree connecting all observations while minimizing the sum of distances of edges in the tree. The $k$-MST is the union of the $1$st, \ldots, $k$th MSTs, where the $k$th MST is a spanning tree that connects all observations while minimizing the sum of distances across edges excluding edges in the $(k-1)$-MST.}\citep{friedman1979multivariate}: $\cG=\{ G':$  $G'$ is a tree that connects all vertices$\}$; 
    \item $k$-minimum distance non-bipartite pairing ($k$-MDP)\footnote{A non-bipartite pairing divides the $N$ observations into $N/2$ (assuming $N$ is even) non-overlapping pairs while edges exist within pairs. The MDP is constructed by minimizing the $N/2$ distances within pairs. The $k$-MDP is the union of the $1$st, \ldots, $k$th MDPs, where the $k$th MDP is a minimum distance non-bipartite pairing while minimizing the sum of distances within pairs excluding the pairs in the $(k-1)$-MDP.}\citep{rosenbaum2005exact}: $\cG=\{ G':$ $G'$ is a non-bipartite pairing$\}$; 
    \item $k$-shortest Hamiltonian path ($k$-SHP) \citep{biswas2014distribution}: $\cG=\{ G':$ $G'$ is a Hamiltonian path\footnote{A Hamiltonian path with $N$  vertices is a connected and acyclic graph with $N-1$ edges, where each node has degree at most two.}$\}$.
\end{itemize}
Take the $k$-NNG as an example. By definition, $G_1$ is the $1$-NNG as the summation of the edges' similarities is maximized if and only if each vertex connects to its nearest neighbor. With similar arguments, $G_{l+1}^{*}$ is the $(l+1)$th NNG for any $l \geq 1$. Thus, $G_{l+1}$ is the $(l+1)$-NNG. Similarly, for MSTs, $G_{1}$ is the $1$-MST, $G_{l+1}^{*}$ is the $(l+1)$th MST for any $l \geq 1$,  and $G_{l+1}$ is the $(l+1)$-MST. An illustration of these graphs is presented in Figure~\ref{fig:graph}.
 
 With $\{ G_l \}_{l=1}^k$, we define two types of graph-based rank matrices $\R = (R_{ij})_{i,j=1}^{N}  \in \RR^{N \times N}$ as follows. For an event $A$, $\indi(A)$ is an indicator function that equals to one if event $A$ occurs, and equals to zero otherwise.
\begin{itemize}
    \item Graph-induced rank 
    \begin{equation}
        R_{i j } = \sum_{l=1}^k \indi\big( (i,j) \in G_l \big) \,.
    \end{equation}
    \item Overall rank
    \begin{equation}
        R_{i j } =  {\rm rank}( S(Z_i,Z_j), G_k ) \,,
    \end{equation}
where 
${\rm rank}(S(Z_i,Z_j),G_k)$ is the rank of $S(Z_i,Z_j)$ among $\{ S(Z_u,Z_v) \}_{(u,v) \in G_k}$ if $(i,j) \in G_k$ and is zero if $(i,j) \notin G_k$.
\end{itemize}
{Both ranks  depend implicitly on $k$, whose choice is discussed in Sections \ref{sec: simulation} and \ref{sec: con}. The graph-induced rank $R_{ij}$ is the number of graphs that the edge $(i,j)$ appears in the sequence of graphs $\{ G_1, \ldots, G_k\}$. For instance, the graph-induced rank of edges in the $l$th NNG or the $l$th MST will be $k-l+1$ for $k$-NNG and $k$-MST, respectively.} The overall rank is the rank of the similarity of edges in the graph $G_k$. These graph-based ranks impose more weights on edges with higher similarity, thus incorporating more similarity information than the unweighted graph. In the meantime, the robustness property of the ranks makes the weights less sensitive to outliers compared to the direct utilization of similarity. With these ranks, we are ready to build different test statistics for different problems.

\section{A new  two-sample test statistic  for high-dimensional data and non-Euclidean data}
\label{sec: new stat}

\subsection{Two-sample test problem and background}
For two independent random samples $X_1$, \ldots, $X_m \stackrel{i.i.d}{\sim} F_X$ and $Y_1$, \ldots, $Y_n \stackrel{i.i.d}{\sim} F_Y$, we consider the test
$$ H_0: F_X = F_Y \quad \text{ against } \quad H_1: F_X \neq F_Y \, .$$ 
For many high-dimensional or non-Euclidean data problems, 
one has little information on $F_X$ and $F_Y$, which makes parametric approaches not applicable. 
A number of nonparametric tests have been proposed for high-dimensional data such as the graph-based tests \citep{friedman1979multivariate, schilling1986multivariate, henze1988multivariate, rosenbaum2005exact, chen2013graph, chen2017new, chen2018weighted, zhang2020graph}, the classification-based tests \citep{hediger2019use,lopez2016revisiting,kim2021classification}, the interpoint distances-based tests \citep{szekely2013energy, biswas2014nonparametric, li2018asymptotic},  and the kernel-based tests \citep{gretton2008kernel, harchaoui2007testing, gretton2009fast, gretton2012optimal,song2020generalized}.

Recently, \cite{pan2018ball} introduced Ball Divergence (BD) to measure the
difference between the two distributions and proposed a metric rank test procedure.  \cite{deb2021multivariate} proposed to define the multivariate ranks through the theory of measure transportation \citep{hallin2021distribution}, based on which they built the multivariate rank-based distribution-free nonparametric testing.  Both tests can be applied to high-dimensional data and achieve good performance for some useful settings.  However, they also lose power under some common alternatives, 
which will be detailed in Section \ref{sec: simulation}. Besides, even though their asymptotic properties were studied, they were not useful to obtain analytic $p$-value approximations. The random permutation procedure was recommended by the authors to obtain their $p$-values. 

\allowdisplaybreaks

\subsection{Test statistics on graph-based ranks}
Let $Z_i = X_i, i =1,\dots m; Z_{m+j} = Y_j, j =1,\dots n$ be the pooled samples and $N = m+n$. Let $\R \in \RR^{N \times N}$ be the graph-based rank matrix constructed on $\{Z_i\}_{i=1}^N$ (details see Section \ref{sec:new rank}). We first define two basic quantities based on $\R$: 
\begin{equation*}
  U_x = \sum_{i=1}^m \sum_{j=1}^m R_{ij} \quad {\rm and}\quad U_y = \sum_{i=m+1}^N \sum_{j=m+1}^N R_{ij} \,,
\end{equation*}
which are the within-sample rank sums of sample $X$ and sample $Y$, respectively. We can symmetrize $\R$ by using $\frac{1}{2} (\R + \R\trans)$. This does not change the values of $U_x$ and $U_y$ by their definitions; while the derivation for their expectations and variances would be much simpler. With a slight notation abuse, in the following, $\R$ is used for the symmetric version. Before we propose the test statistic, we illustrate the behaviors of $U_x$ and $U_y$ under different scenarios through toy examples. Here we set $n=m=50$ and consider multivariate Gaussian distribution with dimension $d = 100$:  (a) null: $F_X = F_Y = N(\bzero_d,\I_d)$; (b) location alternative: $F_Y = N(\bone_d,\I_d)$; (c) scale alternative: $F_Y = N(\bzero_d,4\I_d)$;  (d) mixed alternative: $F_Y = N(0.5\bone_d,2\I_d)$.

\begin{figure}[ht]
\begin{tabular}{c}
     \includegraphics[width=0.9\textwidth]{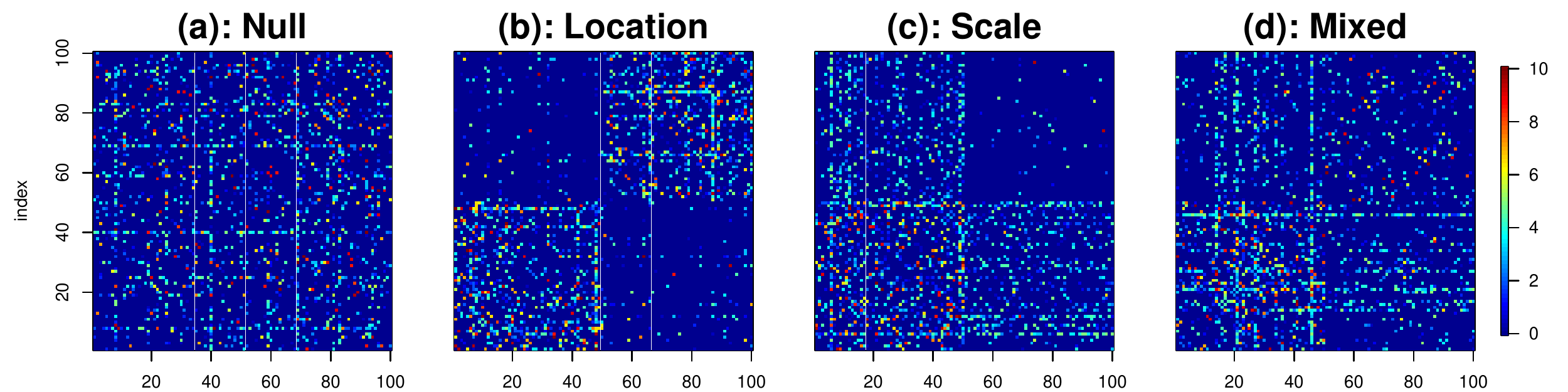}  \\
 \includegraphics[width=0.9\textwidth]{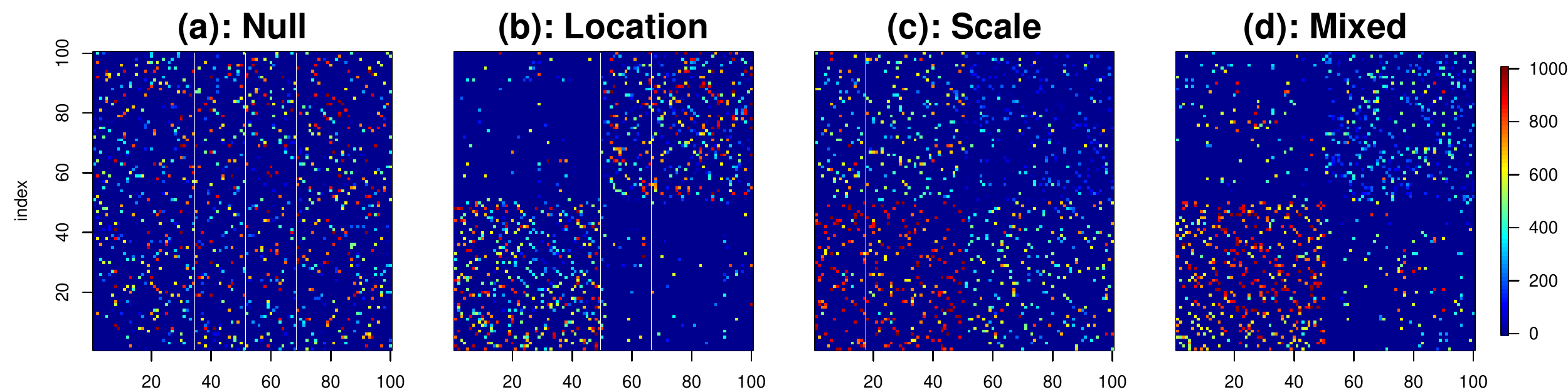}  
\end{tabular}
\caption{Heatmap of the graph-based rank matrix. Top: graph-induced ranks in $10$-NNG. Bottom: overall ranks in $10$-MDP.
}
\label{fig:rank}
\end{figure}
Figure~\ref{fig:rank} shows the heatmaps of the graph-induced rank matrix in the $10$-NNG and the overall rank matrix in the  $10$-MDP. When the two distributions are different in the location parameter, both $U_x$ and $U_y$ tend to be larger than their corresponding values under the null; while for scale alternative, one of $U_x$ and $U_y$ tends to be larger while the other one tends to be smaller than their corresponding values under the null. For both location and scale differences, $U_x$ and $U_y$ will also be different from their values under the null. Thus, $U_x$ and $U_y$ can capture different scenarios. 
The proposed  {\bf R}ank {\bf I}n {\bf S}imilarity graph {\bf E}dge-count two-sample test (RISE) statistic is defined as 
\begin{equation}
  T_R = (U_x - \mu_x, U_y - \mu_y) \bSigma^{-1} (U_x - \mu_x, U_y - \mu_y) \trans 
\,,
\end{equation}
where $\mu_x =  \Ep( U_x )$, $\mu_y = \Ep (U_y)$ and $\bSigma = \Covp\big( (U_x,U_y)\trans \big)$. Under the null hypothesis, the group labels of $X$ and $Y$ are exchangeable. Thus, we can work under the permutation null distribution, which places $1/\tbinom{N}{m}$ probability on each of the $\tbinom{N}{m}$ permutations of the group labels where the first group has $m$ observations and the second group has $n$ observations. We use $\Pp$, $\Ep$, $\Vp$, and $\Covp$ to denote the probability, expectation, variance, and covariance under the permutation null distribution, respectively. 
\begin{theorem}
Under the permutation null distribution, we have that
\begin{align*}
 & \mu_x = \Ep(U_x)  = m(m-1) r_0 \, , \quad  \mu_y = \Ep(U_y) = n(n-1)r_0  \\
  & \Vp ( U_x)  = \frac{ 2 m n (m-1)}{(N-2)(N-3)} \big(  (n-1)V_d + 2 (m-2)(N-1)V_r\big) \, , \\
& \Vp( U_y)   = \frac{2 m n (n-1)}{(N-2)(N-3)} \big( (m-1)V_d + 2(n-2)(N-1)V_r\big) \, , \\
& \Covp(U_x, U_y)  = \frac{2 m(m-1)n(n-1)}{(N-2)(N-3)} \big(  V_d -  2 (N-1) V_r \big) \, ,
\end{align*}
where $V_r =  r_1^2 - r_0^2$ and $V_d = r_d^2 - r_0^2$ with $\bar R_{i \cdot} = \frac{1}{N-1} \sum_{j \neq i}^N R_{ij}$, $r_0 = \frac{1}{N} \sum_{i=1}^N  \bar R_{i \cdot}$, $r_1^2 = \frac{1}{N} \sum_{i=1}^N \bar R_{i \cdot}^2$ and $r_d^2 = \frac{1}{N(N-1)} \sum_{i=1}^N \sum_{j \neq i}^N R_{ij}^2$.
\label{theorem: stat}
\end{theorem}
The proof of Theorem \ref{theorem: stat} is provided in Appendix \ref{proof:stat}. To assure that $T_R$ is well-defined, the covariance matrix $\bSigma$ should be invertible. Here we present the sufficient and necessary conditions. 
\begin{theorem}
Given $m,n \geq 2$, the covariance matrix $\bSigma$ is positive-definite unless (C1) $V_r = 0$ or (C2) $(N-2)V_d = 2(N-1) V_r$. 
\label{thm:well-defined}
\end{theorem}
The proof of Theorem \ref{thm:well-defined} is provided in Appendix \ref{proof:well}. Except for some special graphs, it is rare to have graphs that satisfy (C1) or (C2). For example,  the graph-induced rank in the $k$-NNG and the overall rank in the $k$-MDP would hardly ever run into either (C1) or (C2) (detailed in  Appendix \ref{proof:well}). 
\begin{theorem}
When $T_R$ is well-defined, we have
\begin{equation}
      T_R  =   Z_{w}^2 +  Z_{\diff}^2 \; \text{ and } \; \Covp(Z_{w},Z_{\diff}) = 0 \,,
 \label{decom:rS} 
\end{equation}
where $Z_{w} = \frac{U_w - \Ep(U_w)}{\sqrt{ \Vp(U_w)}}, Z_{\diff} =  \frac{U_{\diff} - \Ep(U_{\diff}) }{\sqrt{ \Vp(U_{\diff})}}$ with $U_w = \frac{n-1}{N-2} U_x + \frac{m-1}{N-2} U_y$ and 
$U_{\diff} = U_x - U_y$. 
\label{thm: decompostion}
\end{theorem}
The proof of Theorem \ref{thm: decompostion} is provided in Appendix \ref{proof:thm: decompostion}. Under the alternative hypothesis, it is possible that (i) both  $U_x$ and $U_y$ are larger than their null expectations (a typical scenario under location alternatives) and (ii)  one of them is larger than while the other one is smaller than its corresponding null expectation (a typical scenario under scale alternatives). See \cite{chen2017new} for more discussions on these scenarios.  For (i), $Z_{w}$ will be large and for (ii), $|Z_{\diff}|$ will be large. 
Some test statistics other than $T_R$ can also be considered. For instance, the weighted rank sum statistic $Z_w$ corresponding to the weighted edge-count test 
\citep{chen2018weighted} that should work well for the location alternative and unbalanced sample sizes, and the max-rank test statistics $R_{\max} \equiv \max\{Z_{w}, |Z_{\diff}|\}$ that corresponds to the  max-type edge-count test statistic 
\citep{chu2019asymptotic}, which is preferred under the change-point setting. 

\section{Asymptotic properties}
\label{sec: asymptotic}
Obtaining the exact $p$-value of $T_R$ by examining all permutations could be feasible for small sample sizes, but is time-prohibitive when the sample size is large. We thus work on the asymptotic distribution of $T_R$. 
Let $a_{n}\prec b_{n}$ be that $a_{n}$ is dominated by $b_{n}$ asymptotically,  $a_{n} \asymp b_{n}$ be that $a_{n}$ is bounded both above and below by $b_{n}$ asymptotically, $a_{n} \precsim b_{n}$ be that  $a_{n}$ is bounded above by $b_{n}$ asymptotically, and `the usual limit regime' be that $m, n \rightarrow \infty$ and $m/(m+n) \rightarrow p \in (0,1)$. 
\begin{theorem}[Limiting distribution under the null hypothesis]
Let $\R = (R_{ij})_{i\in[N]}^{j\in[N]} \in \RR^{N \times N}$ be the graph-induced rank or the overall rank matrix defined in Section \ref{sec: new stat} in the sequence of graphs $\{G_l\}_{l=0}^k$. In the usual limit regime, under Conditions (1) $r_1 \prec r_d$; (2) $\sum_{i=1}^N \big(\sum_{j=1}^N R_{ij}^2 \big)^2 \precsim N^3 r_d^4$; (3) $\sum_{i=1}^{N} \big|\widetilde R_{i \cdot} \big|^3  \prec ( NV_r)^{1.5}$;  (4) $\sum_{i=1}^N  \widetilde R_{i \cdot}^3 \prec N r_d V_r$;    (5) $\big| \sum_{i=1}^N  \sum_{j=1}^N \sum_{s=1, s \neq j}^N R_{ij} R_{is}   \widetilde R_{j \cdot} \widetilde R_{s \cdot} \big| \prec N^3 r_d^2 V_r$;  (6) $\sum_{i=1}^N \sum_{j=1}^N  \sum_{
s \neq i,j}^N  \sum_{
l \neq i,j}^N  R_{ij} R_{js} R_{sl}R_{li}\prec N^4 r_d^4$, where $\widetilde R_{i \cdot} = \bar R_{i \cdot} - r_0$, 
we have that 
$\big(Z_{w}, Z_{\diff}\big)\trans \stackrel{\mathcal{D}}{\rightarrow} N_2(\bzero_2,\mathbf{I}_2)  \text{ and } T_R \stackrel{\mathcal{D}}{\rightarrow} \chi_{2}^{2}$
under the permutation null distribution, where $\stackrel{\mathcal{D}}{\rightarrow}$ is convergence in distribution.
\label{thm: bi normal}
\end{theorem}
The proof of Theorem \ref{thm: bi normal} is provided in Appendix \ref{proof:bi}. Theorem \ref{thm: bi normal} holds for a general matrix $\R$ with some additional conditions (discussed in Section \ref{sec:discussion}). As a result, we can use different ways to weigh the similarity graph such as kernel values. These conditions also assure the invertibility of $\bSigma$. Specifically, Condition (3) requires that $V_r > 0$. By Cauchy–Schwarz inequality $r_0 \leq r_1 \leq r_d$. Then Condition (1) implies that $V_r \leq r_1^2 \prec r_d^2 \asymp V_d$. Thus (C1) and (C2) in Theorem \ref{thm:well-defined} are prohibited. We discuss these conditions more in Appendix \ref{app:dis}. For $k$-MDP, all vertices have the same degree $k$, we thus have the following lemma.  
\begin{lemma}
The overall rank in $k$-MDP satisfies Conditions (1), (2), (4), and (6) when $k = o(N)$. 
\label{lemma:MDP}
\end{lemma}
The proof of Lemma \ref{lemma:MDP} is provided in Appendix \ref{proof:MDP}. When $k = 1$, the other Conditions (3) and (5) will also be satisfied. Specifically, $T_R$ constructed on the overall rank in $1$-MDP is exactly distribution-free, while its distribution can be approximated by $\chi_2^2$ when $N$ is large enough. 


\begin{remark}
The above theoretical results allow the similarity graph to be very dense such as $k \asymp N^{\beta} $ for some $\beta \in (0,1)$. Besides, the conditions in Theorem \ref{thm: bi normal} are only sufficient conditions. As we observed in numeric experiments, even if some conditions are violated, the tail probability of $T_R$ can usually be well controlled by the tail probability of $\chi_2^2$. 
\end{remark}

\begin{theorem}[Consistency]
For two continuous multivariate distributions $F_X$ and $F_Y$, if the graph-induced rank is used with the $k$-MST or $k$-NNG based on the Euclidean distance, where $k = O(1)$, then the power of RISE of level $\alpha \in (0,1)$ goes to one in the usual limiting regime. 
\label{thm: consistency}
\end{theorem}
The proof of Theorem \ref{thm: consistency} is provided in Appendix \ref{proof: thm: consistency}. It follows straightforwardly from \cite{schilling1986multivariate} and \cite{henze1999multivariate}, which involves the (stochastic) limit of the statistic $T_R$.  

\begin{theorem}\label{thm:HDLSS}
Assume that $F_X$ and $F_Y$ satisfy Assumptions 1-2 in \cite{biswas2014distribution}, and there exist $\sigma_1^2, \sigma_2^2 > 0$ and $\upsilon^2$ such that for $X \sim F_X$ and $Y \sim F_Y$ independently, 
$\lim_{d \rightarrow \infty}E\|X-E X\|_2^2/d  = \sigma_1^2$, $\lim_{d \rightarrow \infty}E\|Y-E Y\|_2^2/d  = \sigma_2^2$, and $\lim_{d \rightarrow \infty}\|E X- E Y\|_2^2/d  = \upsilon^2$, where $d$ is the dimension of the data.  Without loss of generality, assume that $\sigma_1^2 \geq \sigma_2^2$. When $m, n \geq 2$, for a fixed $\alpha \in (0,1)$, we have $\lim_{d\rightarrow\infty} P(T_R > \chi_2^2(1-\alpha)) = 1$ for
\begin{enumerate}[ (1)]
    \item $\Rg$-NN with $k < \min\{n,m\}$ when either of the following conditions hold:
    \begin{enumerate}
        \item $|\sigma_1^2 - \sigma_2^2| < \upsilon^2$,  $N \geq C_\alpha$ for a constant $C_\alpha>0$ depending only on $\alpha$,
        \item $\sigma_1^2 - \sigma_2^2 > \upsilon^2$, the degrees of the $k$-NNG are bounded by $c m/n N^{1/2-\beta}$ for constants $c, \beta >0$, and $N \geq C_{\alpha,c,\beta}$ for a constant $C_{\alpha,c,\beta} > 0$  depending only on $\alpha$ and $c$ and $\beta$,  
    \end{enumerate}
    \item $\Ro$-MDP with $k \leq \min\{n,m\}/2$, $\sigma_1^2 > \sigma_2^2$, $\upsilon^2 > 0$, $m/N=p$, $N \geq {C}_{\alpha,p}$ for a constant ${C}_{\alpha,p} > 0$ depending only on $\alpha$ and $p$.
\end{enumerate}
\end{theorem}
Theorem \ref{thm:HDLSS} studies the consistency of the test in the HDLSS (high-dimension low-sample size) regime.  The proof the theorem is provided in Appendix \ref{HDLSS}.

\section{Simulation studies}
\label{sec: simulation}

In this section, we conduct simulations to examine the performance of t RISE. We mainly focus on the graph-induced rank in the $k$-NNG and the overall rank in the $k$-MDP as the representation of the two types of ranks. Supplement S8.3 provides results on other combinations as well.
Specifically, we consider a wide range of null and alternative distributions in moderate/high dimensions, including multivariate Gaussian distribution,  Gaussian mixture distribution, multivariate log-normal distribution, and multivariate $t_5$ distribution. These different distributions range from light-tails to heavy-tails, and the alternatives range from location difference, and scale difference to mixed alternatives, with the hope that these simulation settings can cover real-world scenarios. The details of these settings are in Appendix \ref{setting}. \cite{chen2017new} suggested using $k = 5$ for GET based on $k$-MST to achieve moderate power. For the $k$-NNG and $k$-MDP, the largest value of $k$ can be $N-1$, while for the $k$-MST, the largest value of $k$ can only be $N/2$. So it is reasonable to choose $k$ for the $k$-NNG and $k$-MDP as twice $k$ for the $k$-MST. Hence, we use $k=10$ for simplicity in both simulation and real data analysis. We denote our methods as $\Rg$-NN and $\Ro$-MDP for RISE on the $10$-NNG with the graph-induced rank and on the $10$-MDP with the overall rank, respectively. Besides, a detailed comparison between RISE and GET including the results of RISE on the $k$-MST with the graph-induced rank and the overall rank is provided in Appendix  \ref{sec:rise and get}. 

We compare the type-I error and statistical power with seven state-of-art methods, including two graph-based methods: GET on $5$-MST using the R package \textit{gTests} \citep{chen2017new}, Rosenbaum’s cross-matching test (CM) using the R package \textit{crossmatch} \citep{rosenbaum2005exact}; 
two rank-based methods: a multivariate rank-based test using measure transportation (MT) \citep{deb2021multivariate} and a non-parametric two-sample test based on ball divergence (BD) using the R package \textit{Ball} \citep{pan2018ball}; and three other tests: an LP-nonparametric test statistic (GLP) using the R package \textit{LPKsample} \citep{mukhopadhyay2020nonparametric},  a high-dimensional low sample size $k$-sample tests (HD) using the R package \textit{HDLSSkST}  \citep{HDLSSkST} and a kernel-based two-sample test (MMD) using the R package \textit{kerTests} \citep{gretton2012kernel}. The tuning parameters of these comparable methods are set as their default values.

Here we present the results for $m = n = 50$ and $d \in \{200, 500, 1000\}$. The results for $m = 50, n = 100$ show  similar patterns and are deferred to Tables~\ref{tabm1}-\ref{tabm4} in  Appendix \ref{app: add}. The empirical sizes are presented in Table~\ref{tab1} of Appendix \ref{app: add}. RISE can control the type-I error well for different significant levels and settings, which validates the effectiveness of the asymptotic approximation even for relatively small sample sizes ($m=n = 50$). For other tests, MMD seems a little conservative and GLP has a somewhat inflated type-I error for some settings, while all of the other tests can control the type-I error well.

\begin{table}[h]
    \centering
 \caption{Estimated power (in percent) ($\alpha = 0.05$) under multivariate Gaussian I: (a) simple location,  (b) directed location, (c) simple scale, (d) correlated scale, and (e) location and scale mixed and the Gaussian mixture II: (a) location, (b) scale, and (c) location and scale mixed.}
    \label{tab2} 
       \begin{tabular}{|c|ccc|ccc|ccc|ccc|}
\hline
$d$ & 200 & 500 & 1000 & 200 & 500 & 1000 & 200 & 500 & 1000& 200 & 500 & 1000\\
\hline
 $m = n = 50$ &  \multicolumn{3}{c|}{Setting I (a)} & \multicolumn{3}{c|}{Setting I (b)} & \multicolumn{3}{c|}{Setting I (c)} & \multicolumn{3}{c|}{Setting I (d)} \\
\hline 
$\Rg$-NN & 68 & 64 & 60 & 89 & \textbf{78} & \textbf{67} & 64 & 78 & 84 & \textbf{94} & \textbf{92} & \textbf{91}\\
$\Ro$-MDP & 66 & 58 & 53 & 84 & 71 & 57 & 75 & 87 & 91 & \textbf{92} & \textbf{93} & \textbf{91}\\
GET & 62 & 56 & 50 & 81 & 68 & 56 & 59 & 71 & 80 & 81 & 78 & 75\\
CM & 30 & 27 & 22 & 38 & 29 & 24 & 4 & 4 & 4 & 63 & 63 & 63\\
MT & \textbf{98} & \textbf{96} & \textbf{93}& 7 & 6 & 7 & 5 & 5 & 4 & 13 & 14 & 14\\
BD & 79 & 61 & 41 & 52 & 37 & 23 & \textbf{82} & \textbf{94} & \textbf{97} & 15 & 16 & 14\\
GLP & 55 & 49 & 22 & 15 & 15 & 8 & 6 & 5 & 5 & 7 & 6 & 6\\
HD & 4 & 4 & 3 & 3 & 3 & 4 & 55 & 71 & 84 & 8 & 9 & 7\\
MMD & 90 & 54 & 6 & \textbf{98} & 54 & 3 & 0 & 0 & 0 & 0 & 0 & 0\\
\hline
& \multicolumn{3}{c|}{Setting I (e)}& \multicolumn{3}{c|}{Setting II (a)} & \multicolumn{3}{c|}{Setting II (b)} & \multicolumn{3}{c|}{Setting II (c)} \\
\hline 
$\Rg$-NN & \textbf{98} & \textbf{96} & \textbf{96} & \textbf{53} & \textbf{69} & \textbf{85} & \textbf{62} & \textbf{63} & \textbf{64} & \textbf{68} & \textbf{57} & \textbf{54}\\
$\Ro$-MDP & \textbf{97} & \textbf{95} & \textbf{96} & 41 & 50 & 58 & 23 & 25 & 26 & 48 & 47 & 50\\
GET & 91 & 87 & 86 & 44 & 59 & 75 & \textbf{63} & \textbf{65} & \textbf{66} & 51 & 40 & 38\\
CM & 71 & 69 & 71 & 14 & 20 & 23 & 4 & 4 & 4 & 53 & \textbf{55} & \textbf{57}\\
MT & 16 & 14 & 11 & 49 & 54 & 56 & 4 & 5 & 5 & 7 & 11 & 12\\
BD & 20 & 19 & 18 & 37 & 47 & 63 & 39 & 29 & 30 & 6 & 9 & 11\\
GLP & 9 & 9 & 5 & 8 & 8 & 8 & 8 & 8 & 8 & 8 & 8 & 8\\
HD & 8 & 8 & 7 & 2 & 4 & 2 & 3 & 4 & 3 & 2 & 4 & 2\\
MMD & 1 & 0 & 0 & 1 & 2 & 1 & 0 & 1 & 0 & 1 & 1 & 0\\
\hline
\end{tabular}
\end{table}

The estimated power of these tests  (in percent)  is presented in Tables~\ref{tab2}-\ref{tab4}. The highest power for each setting and those with power higher than $95\%$ of the highest one are highlighted in bold type. Table \ref{tab2} shows the results for the multivariate Gaussian distribution and the Gaussian mixture distribution settings. From Table \ref{tab2}, we see that for the multivariate Gaussian distribution, under the simple location alternative (a), MT performs the best, followed immediately by BD, $\Rg$-NN and $\Ro$-MDP. MMD is also good for $d=200$ and $500$. 
Under the directed location alternative (b), $\Rg$-NN outperforms all of the other tests, followed immediately by $\Ro$-MDP, then by GET. MMD is also good for $d = 200$, while all of other tests have low power. Under the simple sale alternative (c), BD performs the best and $\Ro$-MDP performs the second best.  $\Rg$-NN, GET and HD also have satisfactory performance, while all of other tests have much lower power. Under the correlated scale alternative (d), $\Rg$-NN and $\Ro$-MDP exhibit the highest power and GET is also good enough. 
Under the location and scale mixed alternative (e), $\Rg$-NN and $\Ro$-MDP perform the best again, CM and GET have moderate power, and all other tests have low power. In these settings, $\Rg$-NN, $\Ro$-MDP, and GET perform well in the multivariate Gaussian distribution setting, across a wide range of alternatives, while other tests can perform well in some alternatives, but have low power in other alternatives. For the Gaussian mixture distribution setting II, we see that under the location alternative (a), $\Rg$-NN performs the best. $\Ro$-MDP, GET, MT, and BD have moderate power while all of the other tests have low power. Under the scale alternative (b), GET and $\Rg$-NN outperform all other tests. Under the location and scale mixed alternative (c), $\Rg$-NN and CM perform the best. So the overall performance of $\Rg$-NN is the best in the Gaussian mixture setting.

\begin{table}[t]
    \centering
 \caption{Estimated power  (in percent)  ($\alpha = 0.05$) under the  multivariate log-normal distribution III: (a) simple location, (b) sparse location, (c) scale, and (d) location and scale mixed. }
    \label{tab3} 
       \begin{tabular}{|c|ccc|ccc|ccc|ccc|}
\hline
$d$ & 200 & 500 & 1000 & 200 & 500 & 1000 & 200 & 500 & 1000& 200 & 500 & 1000\\
\hline
 $m = n = 50$
& \multicolumn{3}{c|}{Setting III (a)} & \multicolumn{3}{c|}{Setting III (b)} & \multicolumn{3}{c|}{Setting III (c)} & \multicolumn{3}{c|}{Setting III (d)} \\
\hline 
$\Rg$-NN & 75 & 71 & 68 & \textbf{94} & \textbf{86} & \textbf{71} & 26 & 30 & 32 & 53 & 59 & 58\\
$\Ro$-MDP & \textbf{94} & \textbf{95} & \textbf{95} & 85 & 80 & \textbf{68} & 46 & 58 & 63 &\textbf{ 80} & \textbf{88} & \textbf{93}\\
GET & 68 & 61 & 56 & 85 & 69 & 49 & 24 & 26 & 27 & 49 & 51 & 50\\
CM & 18 & 17 & 15 & 32 & 30 & 25 & 6 & 6 & 6 & 9 & 10 & 12\\
MT & \textbf{97} & \textbf{94} & 88 & 11 & 25 & 43 & 17 & 19 & 13 & 68 & 65 & 60\\
BD & 91 & \textbf{93} & \textbf{94} & 17 & 14 & 10 &  \textbf{56} & \textbf{68} & \textbf{72} & \textbf{82} & \textbf{91} & \textbf{94}\\
GLP & 70 & 65 & 30 & 23 & 36 & 15 & 12 & 9 & 10 & 22 & 18 & 11\\
HD & 29 & 36 & 43 & 4 & 4 & 4 & 16 & 19 & 23 & 24 & 34 & 44\\
MMD & 83 & 57 & 20 & \textbf{98} & 79 & 8 & 19 & 7 & 0 & 54 & 32 & 10\\ 
\hline
\end{tabular}
\end{table}
  \begin{table}[th]
    \centering
 \caption{Estimated power  (in percent)  ($\alpha = 0.05$) under the multivariate $t_5$ distribution IV: (a) simple location, (b) sparse location,  (c) scale and (d) location and scale mixed.}
    \label{tab4}  
       \begin{tabular}{|c|ccc|ccc|ccc|ccc|}
\hline
$d$ & 200 & 500 & 1000 & 200 & 500 & 1000 & 200 & 500 & 1000& 200 & 500 & 1000\\
\hline
 $m = n = 50$ &  \multicolumn{3}{c|}{Setting IV (a)} & \multicolumn{3}{c|}{Setting IV (b)}& \multicolumn{3}{c|}{Setting IV (c)}& \multicolumn{3}{c|}{Setting IV (d)} \\
\hline 
$\Rg$-NN & 82 & 66 & 57 & \textbf{81} & \textbf{62} & \textbf{49} & 81 & 65 & 58 & \textbf{88} & 73 & 63\\
$\Ro$-MDP & 70 & 63 & 53 & 68 & 55 & 44 & \textbf{95} & \textbf{93} & \textbf{93} & 82 & \textbf{78} & \textbf{74}\\
GET & 66 & 44 & 33 & 58 & 36 & 24 & 70 & 46 & 39 & 76 & 56 & 43\\
CM & 24 & 21 & 18 & 24 & 20 & 17 & 72 & 68 & 67 & 45 & 41 & 42\\
MT & \textbf{95} & \textbf{92} & \textbf{88} & 10 & 9 & 6 & 17 & 19 & 19 & 75 & 72 & 67\\
BD & 6 & 6 & 5 & 5 & 5 & 5 & 66 & 66 & 69 & 7 & 6 & 5\\
GLP & 52 & 40 & 18 & 8 & 10 & 6 & 39 & 39 & 39 & 51 & 39 & 30\\
HD & 2 & 2 & 2 & 3 & 2 & 2 & 13 & 11 & 11 & 2 & 3 & 1\\
MMD & 62 & 17 & 4 & 42 & 8 & 3 & 30 & 29 & 35 & 60 & 20 & 5\\
\hline
\end{tabular}
\end{table}

Table \ref{tab3} shows the result of the multivariate log-normal distribution. Under the simple location alternative (a), MT performs the best when $d$ is $200$, and $\Ro$-MDP performs the best when $d$ is $500$ and $1000$. $\Rg$-NN, GET, GLP, and BD also perform well. Under the sparse location alternative (b), $\Rg$-NN outperforms all of the other tests, followed by $\Ro$-MDP. MMD also performs well for $d=200$. Under the scale alternative (c), BD performs the best and $\Ro$-MDP performs the second best. 
Under the mixed alternative (d),  $\Ro$-MDP and BD perform the best, followed immediately by MT, $\Rg$-NN, and GET. So the overall performance of $\Ro$-MDP is the best under Setting III. 

Finally, Table \ref{tab4} shows the result of the multivariate $t_5$ distribution. MT performs the best under the simple location alternative (a), while $\Rg$-NN and $\Ro$-MDP are also good and outperform other tests. Under the sparse location alternative (b), $\Rg$-NN performs the best. $\Ro$-MDP performs the best in the scale alternative (c) and both $\Rg$-NN and  $\Ro$-MDP perform the best in the mixed alternative (d). In these settings, $\Rg$-NN and $\Ro$-MDP are doing well consistently. 

To summarize, we observe that RISE performs well in a wide range of alternatives under different distributions. Besides, MT performs well in the simple location alternative, e.g., Setting I (a), III (a), IV (a), but lacks power in directed or sparse location alternative and scale alternatives, while
BD performs well in the simple scale alternative but lacks power in the location alternatives. GET is doing a good job overall, but it is outperformed by RISE in most of the settings. 


\section{Real data analysis}
\label{sec: real data}

\subsection{New York City taxi data} 
To illustrate the proposed tests, we here conduct an analysis of whether
the travel patterns are different in consecutive months in New York City. We use New York City taxi data from the NYC Taxi Limousine Commission (TLC) website\footnote{ \href{https://www1.nyc.gov/site/
tlc/about/tlc-trip-record-data.page}{https://www1.nyc.gov/site/
tlc/about/tlc-trip-record-data.page}}. The data contains rich information such as the taxi pickup and drop-off
date/times, longitude, and latitude coordinates of pickup and drop-off locations. Specifically, we are interested in the travel pattern from the John F. Kennedy International Airport of the year $2015$. Similarly to \cite{chu2019asymptotic}, we set the boundary of JFK airport from $40.63$ to $40.66$ latitude and from $-73.80$ to $-73.77$ longitude. Additionally, we set the boundary of New York City from $40.577$ to $41.5$ latitude and from $-74.2$ to $-73.6$ longitude. We only consider those trips that began with a pickup at JFK and ended with a drop-off in New York City. The New York City is then split into a $30 \times 30$ grid with equal
size and the number of taxi drop-offs that fall within each cell is counted for each day. Thus each day is represented by a $30 \times 30$ matrix 
and we use the negative Frobenius norm as the similarity measure. 

\begin{table}[h]
    \centering
    \caption{The $p$-values of the tests for the NYC taxi data.}
    \label{tab:real}   
\begin{tabular}{|c|ccccc|}
\hline \hline 
Method & $\Rg$-NN  & $\Ro$-MDP& GET & MT & BD \\
\hline \hline
Jan/Feb & \textbf{0.007} & \textbf{0.002}  & 0.090  &0.528 & 0.340 \\
Feb/Mar & \textbf{0.005}  & \textbf{0.000}  & \textbf{0.013}  & 0.053 & \textbf{0.050}  \\
Mar/Apr & \textbf{0.000} & \textbf{0.008}  & \textbf{0.000}  & \textbf{0.030} & \textbf{0.020}\\
\hline \hline
\end{tabular}
\end{table}

We conduct three comparisons over consecutive months: January vs February, February vs March, and March vs April. 
With the aim of illustration,
 we treat them as three separate tests rather than multiple testing problems. For simplicity, we only compare our method with GET and two rank-based methods MT and BD that show merits in simulation studies. The $p$-values of the five tests are presented in Table \ref{tab:real}, where those smaller than $0.05$ are highlighted by bold type. For February vs March, all methods other than MT can reject the null hypothesis at the significance level of $0.05$, while RISE is the only method that can reject the null hypothesis at the significance level of $0.01$. A similar pattern can be observed in the comparison of March and April (MT rejects this comparison at the $0.05$ level as well). 
It indicates that RISE may be more powerful than other methods in both comparisons. For the comparison of  January and February, RISE is the only test that can reject at the $0.05$ level. We then take a closer look at GET to understand this better in Appendix \ref{detail of New York}.

\subsection{Brain network data}

We here evaluate the performance of RISE in distinguishing differences in brain connectivity between male and female subjects using brain networks constructed from diffusion magnetic resonance imaging (dMRI). The data from the HNU1 study \citep{zuo2014open} consists of dMRI records of fifteen male and fifteen female healthy subjects that were scanned ten times each over a period of one month. Processing the data in the same way as \cite{arroyo2021inference}, we constructed $300$ weighted networks (one per subject and scan) with $200$ nodes registered to the CC200 atlas using the NeuroData’s MRI to Graphs pipeline \citep{kiar2018high}. 
The non-Ecludiean network data are then represented by $200 \times 200$ weighted adjacency matrices. For each subject, we use the average of their ten networks from different scans as their brain network representation, then we obtain fifteen networks for the male and female groups, respectively. Here, we also use the negative Frobenius norm as the similarity measure. 

The results are presented in Table~\ref{tab:real3}. Since the sample size is small ($N=30$), to check the validity of the asymptotic $p$-value approximation, we also show the $p$-values of GET and RISE from $1000$ permutations, which are shown in the brackets. We notice that for RISE, the approximate $p$-values are very close to the $p$-values from permutations even in such a small sample size. All of these tests have small $p$-values. BD shows some evidence of difference with a $p$-value slightly larger than $0.05$ while MT shows less evidence of difference, 
but RISE can provide a more confident conclusion with smaller $p$-values.

Besides, a heat map of the distance matrix of the $30$ subjects is presented in Figure~\ref{fig:brain_heat} in Appendix \ref{detail of New York}    where the first $15$ subjects are male and the following $15$ subjects are female. We see an obvious difference between male and female subjects from the heat map, where the male subjects have larger within-sample distances, but the female subjects have smaller within-sample distances. This is evidence of scale difference. 

\begin{table}[t]
    \centering
    \caption{The $p$-values of the tests for the brain network data.}
    \label{tab:real3}   
\begin{tabular}{|c|ccccc|}
\hline \hline 
Method & $\Rg$-NN  & $\Ro$-MDP& GET & MT & BD \\
\hline \hline
$p$-values & \textbf{0.003 (0.007)} & \textbf{0.019 (0.019)} & \textbf{0.005 (0.011)} & 0.095 & 0.057\\
\hline \hline
\end{tabular}
\end{table}

\section{Discussion and conclusion}
\label{sec:discussion}


\subsection{Potential applications of graph-based ranks}

Besides the two-sample hypothesis testing detailed in this paper, the new ranking scheme can also be applied to other statistical problems, such as the multi-sample tests \citep{song2022new} and independence tests \citep{friedman1983graph,heller2016multivariate,shi2022distribution}.
 For example, we can propose test statistics based on the within-sample and between-sample ranks to test the equality of the multi-samples similarly to \cite{song2022new}. We can also define a rank-based association measure for multivariate data by constructing rank matrices for two sets of multivariate variables following the procedure of \cite{friedman1983graph}.





\subsection{Kernel and Distance IN Graph}
\label{sec:final discussion}
The approach proposed in this paper can be extended to weights other than ranks in weighting the edges in the similarity graph. By incorporating different weights, the performance of the test can be different.  For example, kernel-based methods are popular since they can be applied to any data and distance-based methods are intuitive. Here we discuss extending our framework to these methods for the two-sample testing problem. Specifically, we can define $R_{i j } = K(y_i,y_j) \indi\big( (i,j) \in G_k \big)$,
where $K$ is a kernel function or a negative distance function, for example, the Gaussian kernel $K(y_i,y_j) = \exp \big(-\|y_i - y_j\|^2/(2 \sigma^2) \big)$ with the kernel bandwidth $\sigma$. 
We then define statistics based on Kernel IN Graph (KING) or Distance IN Graph (DING). By Theorem \ref{thm:king}, the asymptotic property of the two-sample test statistic $T_R$ holds. 
\begin{theorem}
Let $\R = (R_{ij})_{i\in[N]}^{j\in[N]} \in \RR^{N \times N}$ be a symmetric matrix with non-negative entries and zero diagonal elements. Suppose further $R_{ij}\geq 1$ if $R_{ij} >0$ and $\max_{i,j} R_{ij} = o\big(N^2r_d^2\big)$. In the usual limit regime, under the permutation null distribution and Conditions (1)-(6), we have that 
$\big(Z_{w}, Z_{\diff}\big)\trans \stackrel{\mathcal{D}}{\rightarrow} N_2(\bzero_2,\mathbf{I}_2) \text{ and } T_R \stackrel{\mathcal{D}}{\rightarrow} \chi_{2}^{2}.$
\label{thm:king}
\end{theorem}

\subsection{Other graph-based ranks}
Besides the two graph-based ranks proposed in the paper, we can also define other types of graph-based ranks. For example, we can define the graph-depth rank which lies between the graph-induced rank and the overall rank. For all $(i,j) \in G_k$, by definition, there exists $1 \leq l \leq k$ such that $(i,j) \in G_l/G_{l-1}$. Let $r_{ij}$ be the normalized rank (e.g., the largest one ranks $1$ and the smallest one ranks $1/M$, where $M$ is the number of edges to be ranked) 
of $S(Z_i,Z_j)$ among $S(Z_l,Z_s), (l,s) \in G_l/G_{l-1}$. We then define the graph-depth rank as $R_{i j } = \sum_{l=1}^k \indi\big( (i,j) \in G_l \big) - 1 + r_{ij}$. 
This graph-depth rank utilizes more information from the graphs than the overall rank by keeping the order of the graph sequence. Specifically, an edge from $G_l/G_{l-1}$ will rank higher than an edge from $G_{l+1}/G_{l}$ since the former one is added to the graph earlier, while the overall rank will lose the information. On the other hand, the graph-depth rank exploits more similarity information by imposing more weights on the edges with higher similarity within a graph. We explored the performance of the graph-depth rank and it shows similar results to the other two ranks. 

\subsection{Conclusion}
\label{sec: con}
We propose a new framework of an asymptotically distribution-free rank-based test, which shows superior performance under a wide range of alternatives. {The computational times for $k$-NN, $k$-MST, and $k$-MDP are $O(N^2d)$, $O\big(N^2 (d + \log N) \big)$ \citep{friedman1979multivariate} and $O\big(N^2(d + kN)\big)$ \citep{rosenbaum2005exact}  respectively, while computing shortest Hamiltonian path (SHP) \citep{biswas2014distribution} is NP-hard. If we use the $kd$-tree algorithm to search for the approximate nearest neighbors, it takes $O(d N (\log N + k \log d)$ time \citep{beygelzimer2013fnn}. }Specifically, we suggest using $\Rg$-NN because of its robust performance and lower computational complexity. In most settings of the paper, we fix $k=10$ for $\Rg$-NN, which is already good enough in terms of power.  For tests based on similarity graphs, the choice of the graph is still an open question. 
Some previous works \citep{friedman1979multivariate,zhang2020graph, chen2017new,chen2018weighted} suggested to use the $k$-MST and set $k$ as a small constant number, e.g., $k=3$ or $k=5$. 
Recently, \cite{yejiong} observed that a denser graph can improve the power of the tests such that $k = O(N^{\lambda})$ for some $0< \lambda <1$ where $N$ is the total number of observations. Following this, \cite{zhang2021graph} compared the power for different $\lambda$’s under various simulation settings and suggested using $\lambda = 0.5$ for GET, where it showed adequate power across different simulation settings. Here we adopt a similar procedure to explore $k$ for RISE with details in Appendix \ref{discussion k}. Based on these numerical results, 
we found that if the sample size is large enough, it can be sufficient to use $k=10$, otherwise, using $k = [N^{0.65}]$ for $k$-NNG or $k$-MDP could be a good choice when computation is not an issue. Another  plausible way could be to select a few representative values of $k$’s to run the test and then combine the results.

\acks{
The authors were partly supported by NSF DMS-1848579.}

\bibliography{yourbibfile}

\appendix





\renewcommand{\theequation}{A.\arabic{equation}}
\renewcommand{\thetheorem}{A.\arabic{theorem}}
\renewcommand{\thelemma}{A.\arabic{lemma}}
\renewcommand{\thetable}{A.\arabic{table}}  
\renewcommand{\thefigure}{A.\arabic{figure}}

\section{Proof of Theorem \ref{theorem: stat} }
\label{proof:stat}

Let $g_i = 1$ if the $i$th sample is from $F_X$ and $g_i = 0$ if from $F_Y$. Then $U_x$ and $U_y$ can be rewritten as
\begin{equation*}
U_x = \sum_{i=1}^N \sum_{j=1}^N  g_i g_j  R_{ij} \quad {\rm and}\quad U_y = \sum_{i=1}^N \sum_{j=1}^N (1-g_i)(1-g_j)  R_{ij} \, . 
\end{equation*} 
Under the permutation null distribution, for $i,j,s,k$ all different, we have
$$
\begin{aligned}
\Ep ( g_i ) & = \frac{m}{N} \,, & \Ep ( g_i g_j)  & = \frac{m(m-1)}{N(N-1)}\,, \\
 \Ep(g_i g_j g_k) & = \frac{m(m-1)(m-2)}{N(N-1)(N-2)}\,, \quad & \Ep (g_i g_j g_k g_s) & = \frac{m(m-1)(m-2)(m-3)}{N(N-1)(N-2)(N-3)}\,.
\end{aligned}
$$
Recall that $\R$ is symmetric with zero diagonal elements, then
$$
\Ep (U_x ) = \sum_{i=1}^N \sum_{j\neq i}^N R_{i j} \Ep(g_i g_j) = \frac{m(m-1)}{N(N-1)} \sum_{i=1}^N \sum_{j\neq i}^N R_{i j}  = m(m-1) r_0 \,,
$$
and similarly $\Ep ( U_y ) = n(n-1) r_0$.
Then we have 
\begin{equation*}
\begin{aligned}
\Ep ( U_x^2 )  =& \sum_{i=1}^N \sum_{j=1}^N \sum_{s=1}^N \sum_{l=1}^N R_{i j} R_{s l} \Ep(g_i g_j g_s g_l)  \\
 = & 2  \sum_{i=1}^N \sum_{j=1}^N R_{i j}^2 \Ep(g_i g_j)  + 4 \sum_{i=1}^N \sum_{j=1}^N \sum_{s \neq i,j}^N R_{i j} R_{i s} \Ep(g_i g_j g_s ) \\
& + \sum_{i=1}^N \sum_{j=1}^N \sum_{s \neq i,j}^N \sum_{l \neq i,j,s}^N R_{i j} R_{s l} \Ep(g_i g_j g_s g_l)  \\
 =  &\frac{m(m-1)n \Big ( 2 (n-1) r_d^2 + 4 (m-2)(N-1) r_1^2 + \frac{N(N-1)(m-2)(m-3)}{n} r_0^2 \Big)  }{(N-2)(N-3)} \,. 
\end{aligned}    
\end{equation*}
Combing with $\Vp(U_x) = \Ep( U_x^2)  - \Ep( U_x )^2$, we can obtain the variance of $U_x$ under the permutation null distribution. A similar result can be obtained  for $\Vp(U_y)$. Finally, we have $\Covp (U_x, U_y)  =  \Ep ( U_x U_y ) - \Ep( U_x) \Ep( U_y)$, where
\begin{align*}
\Ep \big( U_x U_y \big)
= &  \sum_{i=1}^N \sum_{j=1}^N \sum_{s=1}^N \sum_{l=1}^N R_{i j} R_{s l} \Ep\big(g_i g_j (1-g_s) (1-g_l) \big)  \\
= &   \sum_{i=1}^N \sum_{j=1}^N \sum_{s=1}^N \sum_{l=1}^N R_{i j} R_{s l} \big( \Ep(g_i g_j) - \Ep(g_i g_j g_s) -  \Ep(g_i g_j g_l) + \Ep(g_i g_j g_s g_l) \big) \\
= &  m(m-1) N(N-1) r_0^2 - 2 \frac{m(m-1)}{N}  \sum_{i=1}^N \sum_{j=1}^N R_{i j} (  \bar R_{i \cdot} + \bar R_{j \cdot} ) \\
&  - 2 \frac{m(m-1)}{N}  \sum_{i=1}^N \sum_{j=1}^N R_{i j} (  \bar R_{i \cdot} + \bar R_{j \cdot} ) + \Vp(U_x) \\
 =& m(m-1) N(N-1) r_0^2 - 4 m(m-1)(N-1) r_1^2 \\
&  - 2 \frac{m(m-1)(m-2)}{N(N-1)(N-2)} \big( N^2(N-1)^2 r_0^2 - 2 N(N-1)^2 r_1^2 \big )  + \Vp(U_x)\,.
\end{align*}    
We then finish the proof by plugging in the expression of $\Vp(U_x)$. 

\section{Proof of Theorem \ref{thm:well-defined}}
\label{proof:well}

We have 
$$
\begin{aligned}
{\rm det}(\bSigma) & = \Vp ( U_x) \Vp ( U_y) - \Covp(U_x, U_y)^2 \\
& = \frac{32m^2 n^2 (m-1)^2 (n-1)^2 (N-1)V_r\big( (N-2)V_d - 2(N-1)V_r \big)}{(N-2)^2(N-3)} \\
& \neq 0 \text{ if } V_r \neq 0 \text{ and } (N-2)V_d - 2(N-1) V_r \neq 0. 
\end{aligned}
$$
In the following, we briefly discuss the two cases. It is obvious that (C1) happens when $\bar R_{i \cdot} = r_0$. For instance, the graph-induced rank in the $k$-MDP satisfies (C1) as all vertices are required to have the exact same degree $k$ for the $k$-MDP and thus $ \bar R_{i \cdot} = r_0$ for all $i$. We can also show that (C2) happens only for some special graphs.
For example, when $|G_k| \leq N-1$ where $|\cdot|$ 
denotes the cardinality of a set and the number of edges for a graph,  we have 
$$N (N-1)^2 r_1^2 \leq \frac{N^2(N-1)^2}{4} r_0^2 + \frac{N(N-1)}{2}r_d^2$$
and
\begin{align*}
(N-2)V_d - 2(N-1) V_r & = (N-2) r_d^2 - 2(N-1)r_1^2 + N r_0^2 \\
& \geq (N-2)(N-1) r_1^2 - 2(N-1)r_1^2 + N r_0^2 \\
& = N ( (N-1) r_0^2 - r_1^2) 
\\ & \geq (N-3) r_d^2 - \frac{N(N-3)}{2} r_0^2  \\
& = \frac{N-3}{N(N-1)} \Big( \sum_{i=1}^N\sum_{j=1}^N R_{ij}^2 - \frac{(\sum_{i=1}^N\sum_{j=1}^N R_{ij} )^2}{2(N-1)} \Big) \geq 0
\end{align*}
by  Cauchy–Schwarz inequality and $|G_k| \leq N-1$. The equalities hold if and only if for some $i$, we have $R_{i j} = R_{j i} = c$ for some constant $c$ and all $j \neq i$, and $R_{j l} = 0$ for all $j,l\neq i$. As a result, $G_k$ is perfectly star-shaped with the hub vertex $i$, and all other vertices have the same rank $c$ related to the vertex $i$.

\begin{figure}[ht]
\centering
\begin{tabular}{cc}
 \includegraphics[width=195pt]{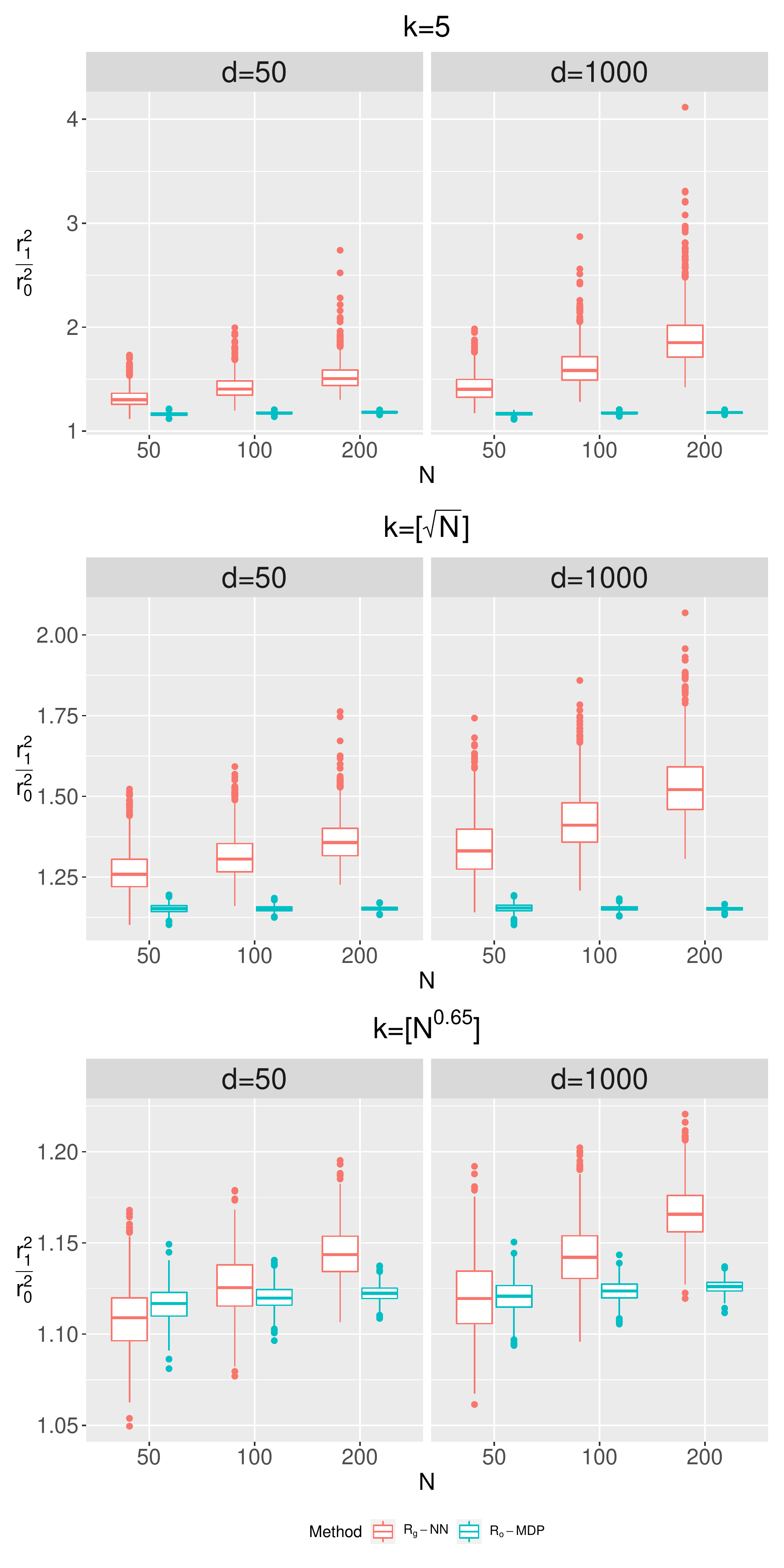} 
    &  \includegraphics[width=195pt]{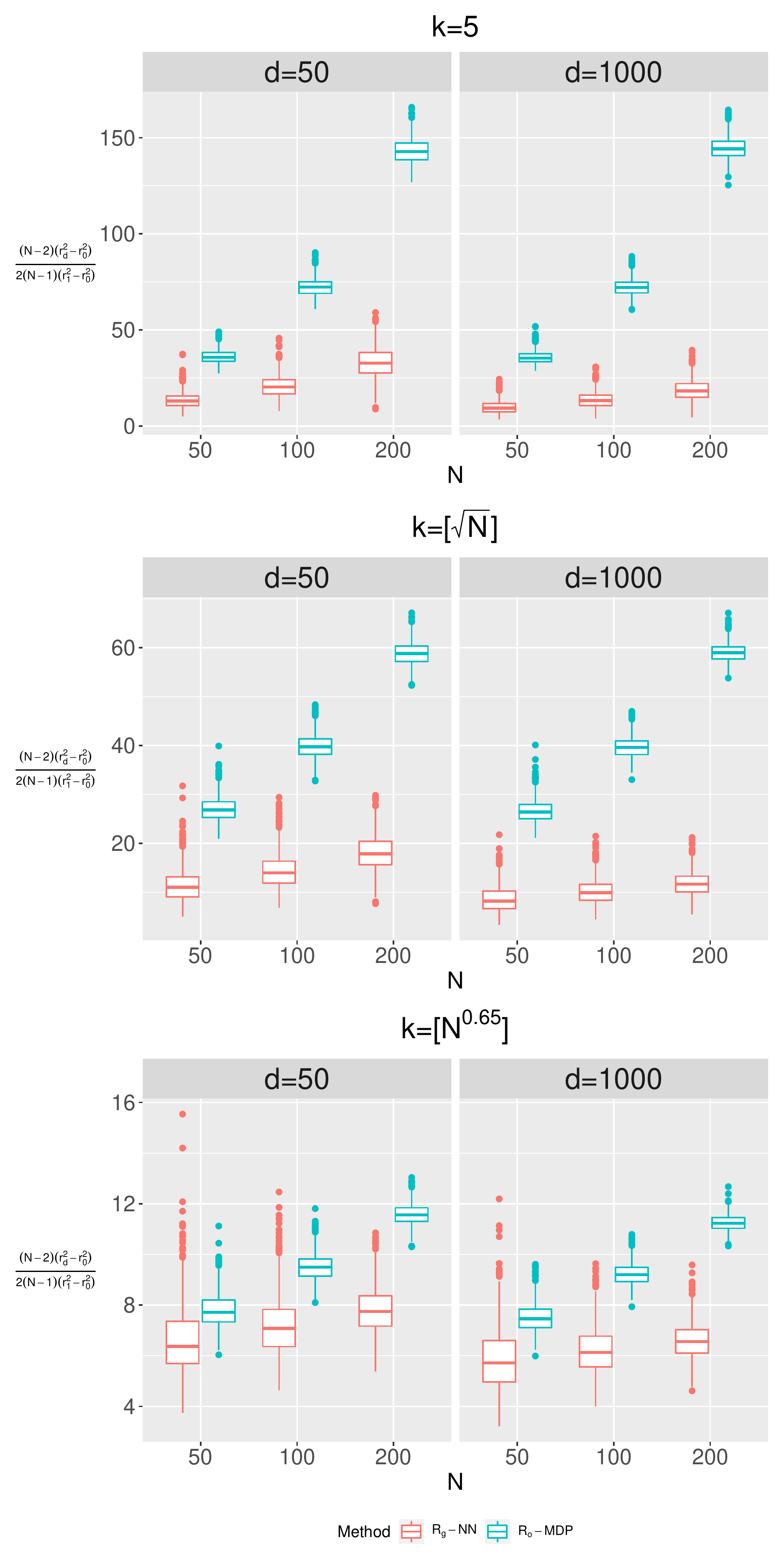}
 \\
 Corner case (C1)    & Corner case (C2)
\end{tabular}
\caption{Boxplots of the two corner conditions.}
\label{fig: corner}
\end{figure}

Except for such special graphs, it is rare to have graphs that satisfy (C1) or (C2). For example, the graph-induced rank in the $k$-NNG and the overall rank in the $k$-MDP would hardly ever run into either (C1) or (C2). We check it through Monte Carlo simulations by generating datasets from the standard multivariate Gaussian distribution with different sample sizes $N$'s and dimension $d$'s. For each dataset, we calculate the two ratios $r_1^2/r_0^2$ and $(N-2)V_d/\big( 2(N-1) V_r \big)$. The procedure is repeated $1,000$ times for each combination of $N \in \{\,50,100,200\,\}$ and $d \in \{\,50,1000\,\}$ using $\R$ constructed by the graph-induced rank in the $k$-NNG and the overall rank in the $k$-MDP, respectively, where $k$ is set as $5$, $[N^{0.5}]$ and $[N^{0.8}]$, respectively. Among these $18,000$ simulation runs, the smallest $r_1^2/r_0^2$ value is $1.049$ and the smallest  $(N-2)V_d/\big( 2(N-1) V_r\big)$ value is $3.219$.  They are all larger than $1$.
The boxplots of the two corner conditions under each combination of $k$, $d$, and $N$ are shown in Figure~\ref{fig: corner}. We find that neither (C1) nor (C2) happens in any of these simulation runs. In practice, when we apply the method, we can easily check whether the two cases happen. If it unfortunately happens, we could always use a different type of similarity graph to avoid the problem.

\section{Proof of Theorem \ref{thm: decompostion}}
\label{proof:thm: decompostion}

Denote $\overline \U = (U_x - \mu_x, U_y - \mu_y)\trans$ and $\A = \begin{pmatrix} 1& -1 \\
\frac{n-1}{N-2} & \frac{m-1}{N-2}
\end{pmatrix}$. Since $\A$ is invertible, we have
$$T_R =\overline \U \trans \bSigma^{-1} \overline \U = \overline \U \trans \A\trans (\A \bSigma \A\trans)^{-1} \A \overline \U \,.$$
It is easy to see that
$$\A \bSigma \A\trans = \begin{pmatrix} \sigma_{\diff}^2 & 0 \\
0 & \sigma_{w}^2\end{pmatrix}$$
and $\A \overline \U = \big(U_{\diff} - \Ep(U_{\diff}), U_{w} - \Ep(U_{w})\big)\trans$, thus finishing the proof.



\section{Proof of Theorems \ref{thm: bi normal}}
\label{proof:bi}

At first, we consider the bootstrap null distribution, which places probability $1/2^{N}$ on each of the $2^{N}$ assignments of $N$ observations to either of the two samples, i.e., each observation is assigned to sample $X$ with probability $m/N$ and to sample $Y$ with probability $n/N$, independently from any other observations. Let $\Eb$, $\Vb$, $\Covb$ be expectation, variance, and covariance under the bootstrap null distribution. It is not hard to see that the number of observations assigned to sample $X$ may not be $m$. Let $n_{X}$ be this number and $Z_{X}=(n_{X}-m)/\sigma^{B}$ where $\sigma^{\text {B }}$ is the standard deviation of $n_{X}$ under the bootstrap null distribution. Notice that the bootstrap null distribution becomes the permutation null distribution conditioning on $n_{X}=m$. 

By applying Theorem \ref{theorem: stat} and making simplifications, we have that
$$\mu_{w} = \Ep(U_w) = \frac{N(n-1)(m-1)}{N-2}r_0 \, ; \quad  \mu_{\diff} = \Ep(U_{\diff}) = (N-1)(m-n)r_0 \, ; $$
$$\sigma_{w}^2 = \Vp(U_w) = \frac{ 2  m(m-1) n(n-1)  }{(N-2)^2(N-3)}\{ (N-2) (r_d^2 - r_0^2) - 2(N-1)(r_1^2 - r_0^2)\}$$
and 
$$\sigma_{\diff}^2 = \Vp(U_{\diff}) = 4(N-1) m n (r_1^2 - r_0^2)\, .$$
Since $g_i$'s are independent under the bootstrap null distribution, it’s not hard to derive that
\begin{equation*}
\begin{aligned}
\Eb ( U_x) &= \frac{m^2 (N-1)}{N} r_0\, ; \quad
\Eb (U_y) = \frac{n^2 (N-1)}{N} r_0\, , \\
\Vb (U_x) &= \frac{2 m^{2} n^{2} (N-1) }{N^3} r_d^2 +\frac{4 n m^{3} (N-1)^2}{N^3} r_1^2 \,,
\\
\Vb (U_y) &= \frac{ 2 m^{2} n^{2} (N-1) }{N^3} r_d^2 +\frac{4 n^{3} m (N-1)^2}{N^3} r_1^2 \,,
\\
\Covb(U_x, U_y) &=\frac{2 m^{2} n^{2} (N-1) }{N^3} r_d^2 - \frac{4 n^2 m^2 (N-1)^2}{N^3} r_1^2\, ,
\end{aligned}
\label{Eq: Eb}
\end{equation*}
which implies that
$$\mu_{w}^{\mB} = \Eb(U_w)
= \frac{N-1}{N(N-2)}(Nmn - m^2 - n^2) r_0 \, ,$$
$$\mu_{\diff}^{\mB} = \Eb(U_{\diff}) 
= (N-1)(m-n) r_0\, ,$$
and
$$(\sigma_{w}^{\mB})^2 = \Vb(U_w) = \frac{ 2 (N-1) m^{2} n^{2}  }{N^3} r_d^2 + \frac{4(N-1)^2 n m (m-n)^2}{(N-2)^2 N^3} r_1^2\, ,$$
$$(\sigma_{\diff}^{\mB})^2 = \Vb(U_{\diff}) = \frac{4(N-1)^2 n m }{N} r_1^2\, , \; \text{ and } (\sigma^{\mB})^{2}= \Vb (n_{X}) = \frac{m n}{N}\, .$$
By defining $Z_{w}^{\mB}= (U_{w}-\mu_{w}^{\mB})/\sigma_{w}^{\mB},  Z_{\diff}^{\mB} = (U_{\diff}-\mu_{\diff}^{\mB})/\sigma_{\diff}^{\mB}$, we express $(Z_{w}, Z_{\diff})$ in the following way:
\begin{equation}
\begin{aligned}
\begin{pmatrix}
Z_{w} \\ Z_{\diff}
\end{pmatrix}
& = \begin{pmatrix}
\sigma_{w}^{\mB}/\sigma_{w} & 0 \\
0 & \sigma_{\diff}^{\mB}/{\sigma_{\diff}}
\end{pmatrix} \begin{pmatrix}
Z_{w}^{\mB} \\
Z_{\diff}^{\mB}
\end{pmatrix} +
\begin{pmatrix}
(\mu_{w}^{\mB}-\mu_{w})/\sigma_{w} \\
(\mu_{\diff}^{\mB}-\mu_{\diff})/{\sigma_{\diff}}
\end{pmatrix} \\
& =\begin{pmatrix}
\sigma_{w}^{\mB}/{\sigma_{w}} & 0 \\
0 & \sqrt{(N-1)/N}
\end{pmatrix} \begin{pmatrix}
Z_{w}^{\mB} \\
\sqrt{ T} Z_{\diff}^{\mB}
\end{pmatrix}+ \begin{pmatrix}
(\mu_{w}^{\mB}-\mu_{w})/\sigma_{w} \\
(\mu_{\diff}^{\mB}-\mu_{\diff})/{\sigma_{\diff}}
\end{pmatrix}\,,
\end{aligned}
\label{eq: dec}
\end{equation}
where $T =  r_1^2/(r_1^2 - r_0^2)$. 
Since the distribution of $(Z_{w}, Z_{\diff})$
under the permutation null distribution is equivalent to the distribution of $(Z_{w}^{\mB}, Z_{\diff}^{\mB}) \mid Z_{X}=0$ under the bootstrap null distribution, we only need show following two statements for proving Theorem \ref{thm: bi normal}:
\begin{enumerate}
     \item[(i)] $\big(Z_{w}^{\mB},\sqrt{T} ( Z_{\diff}^{\mB} - \sqrt{1-1/T}Z_{X}), Z_{X}\big)$ is asymptotically multivariate Gaussian distributed under the bootstrap null distribution and the covariance matrix of the limiting distribution is of full rank. \label{statement i}
    \item[(ii)] ${\sigma_{w}^{\mB}}/{\sigma_{w}} \rightarrow c_{w} ; 
    ({\mu_{w}^{\mB}-\mu_{w}})/{\sigma_{w}} \rightarrow 0 ; ({\mu_{\diff}^{\mB}-\mu_{\diff}})/{\sigma_{\diff}} \rightarrow 0 
$ where $c_{w}$  is a  positive constant. 
\end{enumerate}

From Statement (i), the asymptotic distribution of $\big(Z_{w}^{\mB},\sqrt{T} ( Z_{\diff}^{\mB} - \sqrt{1-1/T}Z_{X}) \big)$ conditioning on $Z_{X}=0$ is a bivariate Gaussian distribution under the bootstrap null distribution when the joint distribution of $\big(Z_{w}^{\mB},\sqrt{T} ( Z_{\diff}^{\mB} - \sqrt{1-1/T}Z_{X}), Z_{X}\big)$ is smooth at $Z_X = 0$, which further implies that the asymptotic distribution of $(Z_{w}^{\mB}, \sqrt{T} Z_{\diff}^{\mB})$ under the permutation null distribution is a bivariate Gaussian distribution. Then, with Statement (ii) and equation \eqref{eq: dec}, we have $(Z_{w}, Z_{\diff})$ is asymptotically bivariate Gaussian distributed under the permutation null distribution. Finally, with the fact that $\Vp(Z_{w})=\Vp(Z_{\diff})=1$ and $\Covp(Z_{w}, Z_{\diff})=0,$ we have that $T_R \stackrel{\mathcal{D}}{\rightarrow} \chi_{2}^{2}$. 

The proof of Statement (i) is deferred to Appendix \ref{proof:s1}.
{Here, we show the joint distribution of $\big(Z_{w}^{\mB},\sqrt{T} ( Z_{\diff}^{\mB} - \sqrt{1-1/T}Z_{X}), Z_{X}\big)$ is smooth at $Z_X = 0$. It can be noticed that $Z_X = 0$ is not a singular point and the behavior of the three random variables has nothing special at $Z_X = 0$. This can be roughly shown as follows. Let $(\bar U_x, \bar U_y)$ be the statistics from the bootstrap data which only has one different label with $(U_x,U_y)$. Without loss of generality, assume that $(\bar U_x, \bar U_y)$ have  $\bar m = m + 1 > 1$ observations with label $X$ and $\bar n = n - 1 > 0$ observations with label $Y$. Let $\bar U_w = \frac{\bar n - 1}{N-2} \bar U_x + \frac{\bar m - 1}{N-2} \bar U_y$ and $\bar U_{\rm diff} =  \bar U_x -  \bar U_y$. 
Then
$$\max \big \{ |U_w - \bar U_w|,  |(U_{\rm diff} -\bar U_{\rm diff}|  \big\} \leq 2 \max_{i=1,\ldots,N} R_{i \cdot}\,.$$
We have 
$$
\begin{aligned}
|Z_w^{\mB} - \bar Z_w^{\mB}| &= \frac{|U_w^{\mB} - \bar U_w^{\mB}|}{\sigma_{w}^{\mB} }  \leq C \frac{  \max_{i=1,\ldots,N}  R_{i \cdot}}{ \sqrt{N^2 r_d^2} } \precsim \frac{\sqrt{N^2 r_1^2}}{ \sqrt{N^2 r_d^2}} \rightarrow 0
\end{aligned}
$$
by Condition (1) and $(\sigma_{w}^{\mB})^2 \asymp N^2 r_d^2$. 
We also have 
$$
\begin{aligned}
|Z_{\diff}^{\mB} - \bar Z_{\diff}^{\mB}| & \leq \frac{|U_{\diff}^{\mB} - \bar U_{\diff}^{\mB} |}{\sigma_{{\diff}}^{\mB} }  \leq C \frac{  \max_{i=1,\ldots,N} R_{i \cdot}}{ \sqrt{N^3 r_1^2} }  \precsim \frac{1}{ \sqrt{N}}  \rightarrow 0
\end{aligned}
$$
since $(\sigma_{\diff}^{\mB})^2 \asymp N^3 r_1^2$. 
As a result, the joint distribution of $\big(Z_{w}^{\mB},\sqrt{T} ( Z_{\diff}^{\mB} - \sqrt{1-1/T}Z_{X}), Z_{X}\big)$ is  smooth at $Z_X = 0$. }

For Statement (ii), by Condition (1) that  $r_1 \prec r_d$ and Cauchy–Schwarz inequality that $r_d^2 \geq r_1^2 \geq r_0^2$, we have
$$\sigma_{w}^2  \asymp N^2 (r_d^2 - 2 r_1^2 + r_0^2) \asymp N^2 r_d^2;\; (\sigma_{w}^{\mB})^2  \asymp N^2 r_d^2; \; \sigma_{\diff}^2 \asymp N^3 (r_1^2 - r_0^2); \; (\sigma_{\diff}^{\mB})^2 \asymp N^3 r_1^2 \,.$$
Since $\mu_{\diff}^{\mB} - \mu_{\diff} = 0$ and 
$$\mu_{w}^{\mB} - \mu_{w} = \frac{m n}{N} r_0 \asymp N r_0 ,$$
by Condition (1), we have 
 $$ \frac{\mu_{w}^{\mB} - \mu_{w}}{\sigma_{w}} \asymp r_0/r_d \precsim r_1/r_d \rightarrow 0 \, .$$
We then finish the proof of Statement (ii).  


\section{Discussion on Conditions of the Asymptotic Null Distribution}
\label{app:dis}
Denote $K = \max R_{ij}$ (for example, $K = k$ for the graph-induced rank in $k$-NNG or $k$-MST and $K = Nk/2$ for the overall rank in $k$-MDP). Usually we have $r_0 \asymp K|G_k|/N^2$ and $r_d^2 \asymp K^2|G_k|/N^2$ where $|G_k| \asymp Nk$, which hold for the three types of graphs in Section \ref{sec:new rank}. Conditions (1)-(4) essentially require the absence of hubs that nodes with a large degree or a cluster of small hubs. For instance, assuming the largest degree of $G_k$ is bounded by $C k$ for some constant $C$, we have Conditions (1), (2), (4), and (6) always hold such as 
\begin{align*}
& r_1^2 = \frac{1}{N(N-1)^2} \sum_{i=1}^N  (\sum_{j \neq i}^N R_{ij} )^2 \lesssim  \frac{ K^2 k^2}{N^2} \prec r_d^2 \,, \\
& \sum_{i=1}^N \big(\sum_{j=1}^N R_{ij}^2 \big)^2 \precsim  N (k K^2)^2  \asymp N^3 r_d^4 \asymp K^4|G_k|^2/N \,,\\
& \sum_{i=1}^N  \widetilde R_{i \cdot}^3 \leq \max_{i} |\widetilde R_{i \cdot}|  N V_r \lesssim N V_r kK/N \prec N r_d V_r\,,\\
& \sum_{i=1}^N \sum_{j=1}^N  \sum_{s \neq i,j}^N \sum_{l \neq i,j}^N R_{ij} R_{js} R_{sl}R_{li} \precsim  K N^3 \sum_{i=1}^N \bar R_{i \cdot}^3 \precsim  K^4 N k^3  \prec K^4 N^2 k^2  \asymp N^4 r_d^4\,,
\end{align*}
when $k = o(N)$. Particularly, Condition (6) can be viewed the constraint on the number of squares in $G_k$, denoted as $N_{\rm sq}$. We then have 
$$\sum_{i=1}^N \sum_{j=1}^N  \sum_{l\neq i,j}^N \sum_{s \neq i,j}^N R_{ij} R_{jl} R_{ls}R_{si} \lesssim K^4 N_{\rm sq} \text{ and } N^4 r_d^4 \asymp K^4|G_k|^2\,.$$
Thus, if $N_{\rm sq} \prec |G_k|^2$, Condition (6)  will hold even if the degrees are not asymptotically bounded by $k$. For Condition (3), by $\sum_{i=1}^{N} \big|\widetilde R_{i \cdot} \big|^3  \lesssim \max_{i} |\widetilde R_{i \cdot}|  N V_r$, it holds if 
\begin{equation}
    \max_{i} |\widetilde R_{i \cdot}| \prec \sqrt{NV_r} = \big( \sum_{i=1}^N \widetilde R_{i \cdot}^2 \big)^{0.5} \,, 
\label{eq:vr6}
\end{equation}
which may be satisfied unless the variation of the average row-wise ranks $V_r$ is dominated by some vertices such that $\sum_{i=1}^N \widetilde R_{i \cdot}^2 \approx \widetilde R_{j \cdot}^2$ for some vertex $j$. Finally, for Condition (5), by Cauchy–Schwarz inequality, 
$$
\begin{aligned}
\big| \sum_{i=1}^N  \sum_{j=1}^N \sum_{s=1, s \neq j}^N R_{ij} R_{is}   \widetilde R_{j \cdot} \widetilde R_{s \cdot} \big| & = \big| \sum_{i=1}^N  \big( \sum_{j=1}^N R_{ij} \widetilde R_{j \cdot} \big)^2  - \sum_{i=1}^N  \sum_{j=1}^N R_{ij}^2   \widetilde R_{j \cdot}^2 \big| \\ 
& \leq \sum_{i=1}^N  \big( \sum_{j=1}^N R_{ij} \widetilde R_{j \cdot} \big)^2    \leq \sum_{i=1}^N Ck K^2 \max_j \widetilde R_{j \cdot}^2  \\
&  = C Nk K^2 \max_j \widetilde R_{j \cdot}^2\asymp N^2 r_d^2 \max_j \widetilde R_{j \cdot}^2 .
\end{aligned}
$$
As a result, Condition (5) holds if $\max_j \widetilde R_{j \cdot}^2 \prec  N V_r$, which is equivalent to \eqref{eq:vr6}.

\begin{figure}[h!]
  \centering
 \includegraphics[width=360pt]{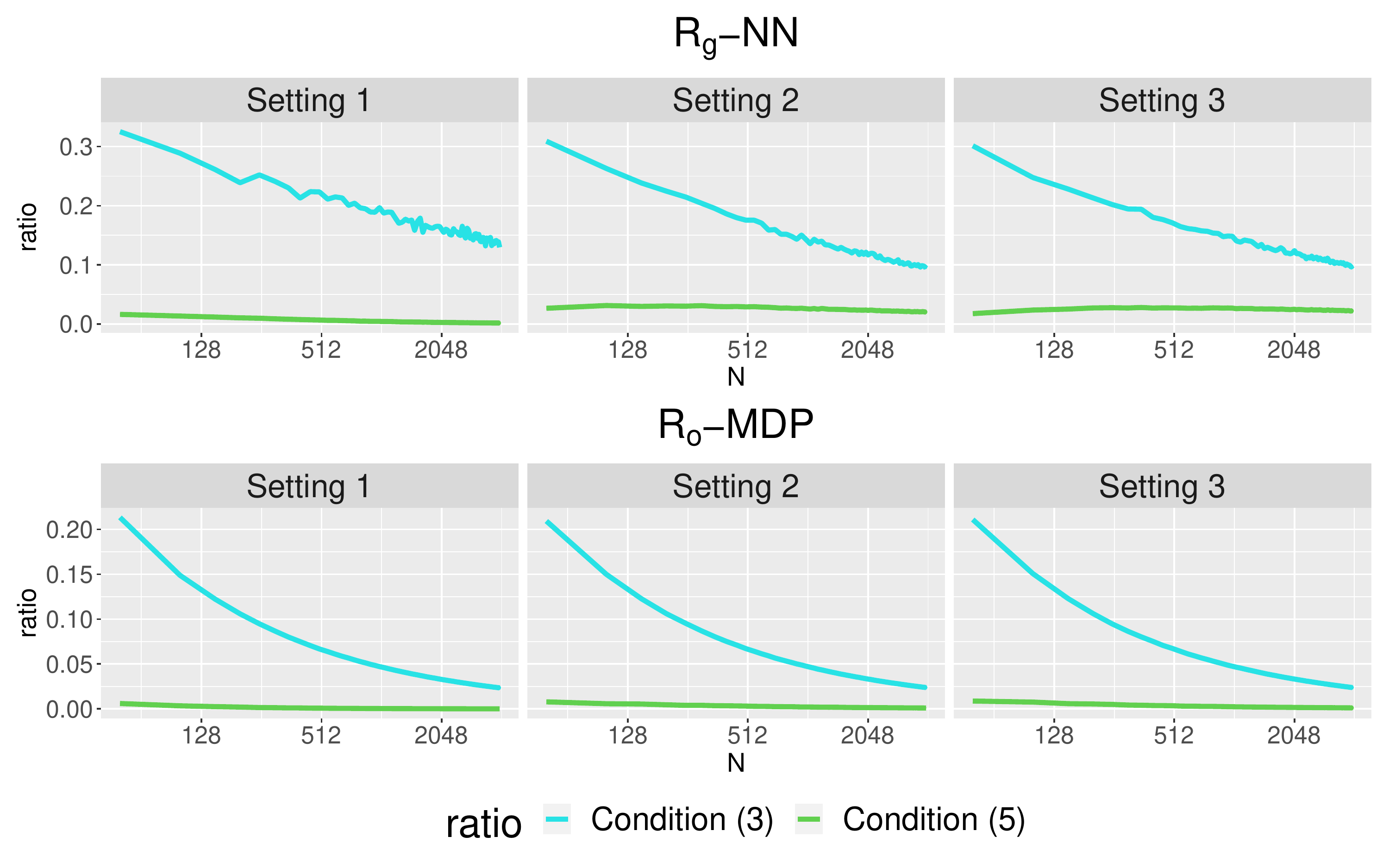}
 \caption{The average ratios of Conditions (3) and (5) based on $100$ simulations for $\Rg$-NN and $\Ro$-MDP. 
 }
\label{fig:ratio} 
\end{figure}

We verify conditions (3) and (5) on simulation as follows. We set $m = n = N/2$ and increase $N$ from $50$ to $4000$ and generate the observations from  $F_x = F_y =  N_{d}(\bzero_d,\I_d)$. We consider three combinations of the data dimension $d$ and the $k$ for $k$-NNG and $k$-MDP: (1) $(d,k) = (40,5)$; (2) $(d,k) = (1000,[\sqrt{N}])$; and (3) $(d,k) = (N,[\sqrt{N}])$. We calculate the two ratios $A_3 = \sum_{i=1}^{N} \big|\widetilde R_{i \cdot} \big|^3  / ( NV_r)^{1.5}$ and $A_5 = \big| \sum_{i=1}^N  \sum_{j=1}^N \sum_{s=1, s \neq j}^N R_{ij} R_{is}   \widetilde R_{j \cdot} \widetilde R_{s \cdot} \big| / N^3 r_d^2 V_r$ for Conditions (3) and (5), respectively and show the average values based on $100$ simulations in Figure~\ref{fig:ratio}. We can see that for both $\Rg$-NN and $\Ro$-MDP, the two ratios converge to zero or are very close to zero, which verifies that the two conditions are satisfied.

Upon examining the New York City taxi data, it is observed that the highest values for $A_3$ and $A_5$ in the eleven comparisons are $0.249$ and $0.025$ for $\Rg$-NN, and $0.169$ and $0.107$ for $\Ro$-MDP, respectively. In the brain network data, these two ratios are recorded as $0.317$ and $0.032$ for $\Rg$-NN, and $0.243$ and $0.063$ for $\Ro$-MDP, respectively. These values are in alignment with the simulation results in Figure~\ref{fig:ratio}.

Even though the value of $A_3$ is not sufficiently low, the corresponding asymptotic $p$-values maintain a high level of accuracy, as can be seen in Table~\ref{tab:real3}. This suggests that the present sufficient conditions leave room for potential enhancements, an aspect that warrants future exploration.


\section{Proof of Lemma \ref{lemma:MDP}}
\label{proof:MDP}
\begin{proof}
A $k$-MDP is an undirected graph where each vertex has degree $k$, thus it has $Nk/2$ edges in total (assuming that $N$ is even for simplicity). We then have 
$$
\begin{aligned}
& r_0 = \frac{2}{N(N-1)}  \sum_{l=1}^{Nk/2} l =  \frac{ k (1+Nk/2)}{2 (N-1)} \asymp k^2 \,, \\
& r_d^2 = \frac{2}{N(N-1)}  \sum_{l=1}^{Nk/2} l^2  =  \frac{k(1+Nk/2)(1+Nk)}{6 (N-1)} \asymp Nk^3 \,,\\
& r_1^2 = \frac{1}{N} \sum_{i=1}^N \bar R_{i \cdot}^2 \in [ r_0^2, \frac{1}{N (N-1)^2 }\sum_{i=1}^N (2 k i+1)^2 k^2] \asymp k^4 \,,
\end{aligned}
$$
which implies Condition (1) since $k \prec N$. 
For Condition (2), we have 
$$\sum_{i=1}^N \big(\sum_{j=1}^N R_{ij}^2 \big)^2 \leq N \big(k (Nk/2)^2 \big)^2 \asymp N^5 k^6 \asymp N^3 r_d^4\;.$$
For Condition (4), by
$$
 \bar R_{i \cdot} \in [\frac{1}{N-1} \sum_{l=1}^k l, \frac{1}{N-1} \sum_{l=1}^k (Nk/2-l+1) ] = [O(k^2/N), O(k^2)] \,,
$$
we have 
 $$\sum_{i=1}^{N} \big|\widetilde R_{i \cdot} \big|^3  \leq \max_{i} |\widetilde R_{i \cdot}|  \sum_{i=1}^{N} \widetilde R_{i \cdot} ^2 \leq k^2 N V_r \leq N^{0.5} k^{1.5} N V_R \prec N r_d V_r \,.$$
 Finally, for Condition (6), we have 
\begin{align*}
& \sum_{i=1}^N \sum_{j=1}^N  \sum_{l \neq i,j}^N \sum_{s \neq i,j}^N R_{i j} R_{j l} R_{ls} R_{s i} \precsim k N^2 \sum_{i=1}^N \sum_{j=1}^N   \sum_{s \neq i,j}^N R_{i j}  R_{s i} \min \{ \bar R_{j \cdot} \, \bar R_{s \cdot} \}  \\
& \leq k N^2   \sum_{i=1}^N \sum_{j=1}^N   \sum_{s \neq i,j}^N R_{i j}  R_{s i}  \bar R_{j \cdot}  \leq k N^3 \sum_{i=1}^N \sum_{j=1}^N  R_{i j} \bar R_{i \cdot}  \bar R_{j \cdot} \\
& \leq  k N^3  \sqrt{ \Big( \sum_{i=1}^N \sum_{j=1}^N   R_{i j} \bar R_{i \cdot}^2 \Big) \Big( \sum_{i=1}^N \sum_{j=1}^N   R_{i j} \bar R_{j \cdot}^2 \Big)} = k N^4 \sum_{i=1}^N \bar R_{i \cdot}^3 \\
&\precsim  N^5 k^7  \prec  N^6 k^6 \asymp N^4 r_d^4\,.
\end{align*}
\end{proof}

\section{Proof of Theorem \ref{thm: consistency}}
\label{proof: thm: consistency}
\begin{proof}
Let $f_x$ and $f_y$ be the density function of $F_X$ and $F_Y$, respectively. When $k = O(1)$, if the similarity graph is the $k$-MST or the $k$-NNG, following the approach of \cite{henze1999multivariate} or \cite{schilling1986multivariate}, we have 
$$
\frac{U_j}{N} \rightarrow \frac{k(k+1)}{2} \int \frac{p_j^2 f_j^2(z)}{\sum_{i=x,y} p_i f_i(z)} {\rm d}z \quad \text{ almost surely, }
$$
where $j = x,y$, $p_x = \lim_{m,n \rightarrow \infty} {m}/(m+n)$ and $p_y = 1 - p_y$. Let $\delta_j = \lim_{N \rightarrow \infty} (U_j - \mu_j)/{N}$ for $j = x,y$. We then have 
$$
\begin{aligned}
 \lim_{N \rightarrow \infty} \frac{T_R}{N} = \lim_{N \rightarrow \infty} (\delta_x, \delta_y) \Big( \frac{\bSigma}{N}\Big) ^{-1}  (\delta_x, \delta_y) \trans =  a(\delta_x - \delta_y)^2 + b (p_y \delta_x + p_x \delta_y)^2 
\,,
\end{aligned}
$$
where 
$a = \lim_{N \rightarrow \infty} {N}/{\sigma_{\diff}^2}$ and $b = \lim_{N \rightarrow \infty} {N}/{\sigma_w^2}$. By Theorem \ref{theorem: stat}, $\Vp(U_w) = O(N)$, so $b>0$. It can be shown that $p_y \delta_x + p_x \delta_y > 0$ when $f_1$ and $f_2$ differ on a set of positive measure: 
$$
\begin{aligned}
p_y \delta_x + p_x \delta_y & = \frac{k(k+1)p_x p_y}{2}  \Big(  \int \frac{\sum_{i=x,y} p_i f_i(z)^2 }{\sum_{i=x,y} p_i f_i(z)} {\rm d}z - 1 \Big) \\
& = \frac{k(k+1)p_x^2 p_y^2}{2}  \int \frac{\big( f_x(z) - f_y(z) \big)^2 }{\sum_{i=x,y} p_i f_i(z)} {\rm d}z > 0 \,.
\end{aligned}
$$
Thus, RISE is consistent. 
\end{proof}

\section{Proof of Theorem \ref{thm:HDLSS}} \label{HDLSS}

\begin{proof}
    We first show (1). For $k$-NNG, Let $\{N_{ij}\}_{i,j \in \{X, Y\}}$ be the number of edges pointing from sample $i$ to sample $j$.  Then, it is easy to see that $N_{XX} + N_{XY} = km$ and $N_{YX} + N_{YY} = kn$.  As shown in Section 4 of \cite{biswas2014distribution}, when $d \rightarrow \infty$, $\|X_1  - X_2\|_2^2/d$, $\|Y_1  - Y_2\|_2^2/d$, and $\|X_1  - Y_1\|_2^2/d$ converge to $2 \sigma_1^2$, 
 $2 \sigma_2^2$, and $\sigma_1^2 + \sigma_2^2 + \upsilon^2$ in probability, respectively. Then the sum of distances of the edges in the $k$-NNG divided by $d$ converges in probability to $2 N_{XX} \sigma_1^2 + 2 N_{YY} \sigma_2^2 + (N_{XY} + N_{YX})(\sigma_1^2 + \sigma_2^2 + \upsilon^2) = 2k m \sigma_1^2 +2k n \sigma_2^2 + N_{XY} (\upsilon^2 - (\sigma_1^2 - \sigma_2^2)) + N_{YX}(\upsilon^2 + (\sigma_1^2 - \sigma_2^2))$.  

For (a), when $|\sigma_1^2 - \sigma_2^2| < \upsilon^2$, the above sum is minimized when $N_{XY} = N_{YX}=0$, so all edges in the $k$-NNG are within samples. Then for $\Rg$-NN, we have $U_x = m \sum_{i=1}^k i = \frac{k(k+1)m}{2}$ and  $U_y = n \sum_{i=1}^k i = \frac{k(k+1)n}{2}$. Besides, we have $r_0 = \frac{1}{N(N-1)} N \sum_{i=1}^k i  = \frac{k(k+1)}{2(N-1)}$ and $r_d^2 \leq \frac{2}{N(N-1)} N \sum_{i=1}^{k}  i^2  = \frac{k(k+1)(2k+1)}{3(N-1)}$. Then 
$$\sigma_w^2 \leq \frac{ 2  m(m-1) n(n-1)  }{(N-2)^2(N-3)} (N-2) (r_d^2 - r_0^2) \leq  \frac{8 n^2m^2 r_d^2}{N^2}   \leq \frac{32 k^3 m^2 n^2 }{N^3}.$$
In addition, 
$$
\begin{aligned}
    Z_w &  = \frac{(n-1)m k (k+1) + (m-1) n k (k+1) } { 2(N-2) \sigma_w } - \frac{N(n-1)(m-1)}{(N-2)\sigma_w} \frac{k(k+1)}{2(N-1) } \\
    & =  \frac{ mn (N-2) k (k+1) } { 2(N-1)\sigma_w  } \geq \frac{m n k^2}{4 \sigma_w}  
\end{aligned}
$$
We then get 
$$T_R \geq Z_w^2 \geq \frac{k N^3 }{512}  > \chi_2^2(1-\alpha)$$
when $N \geq C_{\alpha}$ for a constant $C_{\alpha} >0$ depending only on $\alpha$. 

For (b), when $\sigma_1^2 - \sigma_2^2 > \upsilon^2$, the sum is minimized when $N_{XY}=km$, $N_{YX}=0$.  Then for $\Rg$-NN, we have $U_x = 0$ and $U_y = n \sum_{i=1}^k i = \frac{k(k+1)n}{2}$. 
By the condition in (b) that the degrees of the $k$-NNG are bounded by $d_k = c \sqrt{m/n} N^{1/2-\beta}$, we have 
$$r_1^2 \leq \frac{1}{N(N-1)^2} \sum_{i=1}^N (k d_k)^2 \leq \frac{ 4 k^2 d_k^2 }{N^2}$$
and 
$$\sigma_\diff^2 = 4(N-1) m n (r_1^2 - r_0^2) \leq 4(N-1) m n r_1^2 \leq \frac{16 mn k^2 d_k^2}{N}.$$
We then get 
$$
\begin{aligned}
    Z_\diff &  = \frac{  -n k (k+1) } { 2 \sigma_\diff } - \frac{(N-1)(m-n) k(k+1)}{2(N-1) \sigma_\diff} = - \frac{ m k (k+1) } { 2\sigma_\diff  } \leq - \frac{k}{8 \sqrt{ c} } N^{\beta} ,
\end{aligned}
$$
and as a result, 
$$T_R \geq Z_\diff^2 \geq \frac{k^2}{4 c} N^{2\beta} > \chi_2^2(1-\alpha)$$
when $N \geq C_{\alpha,c,\beta}$ for a constant $C_{\alpha,c,\beta} > 0$  depending only on $\alpha$, $c$ and $\beta$.

We next show (2). For simplicity, assume that $m$ and $n$ are even. When $m$ or $n$ is not even, a similar proof can be applied with a more tedious procedure, thus leaving it out here. For $k$-MDP, let $A$, $B$ and $C$ be the number of edges connecting within sample $X$, within sample $Y$, and between sample $X$ and sample $Y$, respectively.  With a similar argument as in proving (1), when $d \rightarrow \infty$, the sum of distances of the edges in $k$-MDP  divided by $d$ converges in probability to $2 k A \sigma_1^2 + 2k  B \sigma_2^2 + k C(\sigma_1^2 + \sigma_2^2 + \upsilon^2) = m k( \sigma_1^2 + n \sigma_2^2 + C \upsilon^2)$, 
 which is minimized if and only if $C = 0$ since $\upsilon^2>0$. Thus, the $k$-MDP is constructed with all pairs, with both observations coming from the same distribution. Then by the Proof of Lemma \ref{lemma:MDP}, we obtain $ r_0  =  \frac{ k (Nk+2 )}{4 (N-1)}$ and 
$r_d^2  =  \frac{k(Nk+2)(1+Nk)}{12 (N-1)} $. 
Besides,  $U_x = \sum_{j=1}^{km/2} 2 j = \frac{km(km+2)}{4}$ and $U_y = \sum_{j=km/2+1}^{kN/2} 2 j = \frac{kN(kN+2)}{4} - U_x$. We then get
$$
\begin{aligned}
    Z_w = &  \frac{q U_x + p U_y - \mu_w }{\sigma_w} \\
    = &  \frac{(n-1) k m(km+2) + (m-1) \{ kN(kN+2) - km (km+2)\} } { 4(N-2)\sigma_w } \\
    & - \frac{N(n-1)(m-1)}{(N-2)\sigma_w} \frac{k(kN+2)}{4(N-1) } \\
     = &  \frac{(n-m) k m( k m +2) + (m-1)k N(k N+2) } { 4(N-2)\sigma_w  } - \frac{kN(kN+2)(n-1)(m-1)}{4 (N-2) (N-1) \sigma_w }  \\
    = &  \frac{ k m n } { 4(N-2)  (N-1) \sigma_w  } \{  (kN+2)(N-2) - k (n-m) (N-1)\} 
    \geq \frac{ k^2 m^2 n  } { 4  (N-1) \sigma_w  }
\end{aligned}
$$
and 
$$
\begin{aligned}
\sigma_{w}^2 & = \frac{ 2  m(m-1) n(n-1)  }{(N-2)^2(N-3)}\{ (N-2) (r_d^2 - r_0^2) - 2(N-1)(r_1^2 - r_0^2)\}  \\
& \leq \frac{ 2  m^2 n^2 r_d^2  }{  (N-2)(N-3)} \leq \frac{16 m^2 n^2 k^3}{3 N}. \\
\end{aligned}
$$
Then 
$$T_R \geq Z_w^2 \geq \frac{k m^2}{256 N} = \frac{k p^2 N }{256}  > \chi_2^2(1-\alpha)$$
when $N \geq C_{\alpha,p}$ for some constant $C_{\alpha,p} >0$ depending only on $\alpha$ and $p$. 

\end{proof}

\section{Addition Simulation Details and Results}

\subsection{Detailed Settings}
\label{setting}
The four settings are as follows: 
\begin{itemize}
    \item[I.] $F_X = N_{d}(\bzero_d,\bSigma_X)$ is the multivariate Gaussian distribution, where $\Sigma_{X,ij} = 0.6^{|i-j|}$.
    \begin{itemize}
         \item[(a)] Simple location:  $F_Y = N_{d}( \delta \bone_d, \bSigma_X)$ where $\delta = 0.5 \log d /\sqrt{d}$.
         
        \item[(b)] Directed location:  $F_Y = N_{d}(\bmu, \bSigma_X)$ where $\bmu = 0.5 \log d \bmu'/\|\bmu'\|_2$ and $\bmu' \sim N_{d}(\bzero_d, \I_d)$ is fixed.
          
        \item[(c)] Simple scale: $F_Y = N_{d}(\bzero_d,\sigma^2 \bSigma_X)$ where $\sigma = 1 + 0.12\log d /\sqrt{d}$. 
        
         \item[(d)] Correlated scale:  $F_Y = N_{d}(\bzero_d, \bSigma_Y)$ where $\Sigma_{Y,ij} = 0.15^{|i-j|}$. 
        
         \item[(e)] Location and scale mixed: $F_Y = N_{d}(\bmu, \bSigma_Y)$ where 
        $\bmu = 0.2\log d \bmu'/\|\bmu'\|_2$ and  $\bmu' \sim N_{d}(\bzero_d,\I_d)$ is fixed. 
    \end{itemize}
    
    \item[II.] $F_X = W N_{d}(0.3 \boldsymbol{1}_d,\I_d) + (1-W) N_{d}(-0.3 \boldsymbol{1}_d, 2 \I_d)$ is the Gaussian mixture distribution, where $W \sim {\rm Bernoulli}(0.5)$. 
    \begin{itemize}
        \item[(a)] Location:
        $F_Y = W N_{d}\big( (0.3 + 0.75/\log d ) \bone_d,\I_d \big) + (1-W) N_{d}\big(-(0.3 + 0.75/\log d ) \bone_d, 2 \I_d\big)$. 
        
        \item[(b)] Scale: $F_Y = W N_{d}(0.3 \bone_d, (1+\sigma)^2 \I_d) + (1-W) N_{d}(-0.3 \bone_d, (\sqrt{2}+\sigma)^2 \I_d)$, where $\sigma = 0.12 \sqrt{50/d}$. 
        
        \item[(c)] Location and  scale mixed: $F_Y = W N_{d}(0.35 \bone_d,  \bSigma_Y) + (1-W) N_{d}(-0.35 \bone_d, 2 \bSigma_Y)$, where $\Sigma_{Y,ij} = 0.5^{|i-j|}$. 
    \end{itemize}

     \item[III.] $F_X = \exp\big(N_{d}(\bzero_d,\bSigma_X)\big)$ is the multivariate log-normal distribution, where $\Sigma_{X,ij} = 0.6^{|i-j|}$.
     \begin{itemize}
        \item[(a)]  Simple location: $F_Y = \exp \big( N_{d}(\delta \bone_d,\bSigma_X) \big)$ where $\delta = 0.5 \log d /\sqrt{d}$.
        \item[(b)]  Sparse location: $F_Y = \exp \big( N_{d}(\boldsymbol  \mu, \boldsymbol {\Sigma}_X) \big)$ where $\mu_j = (-1)^{j} 2.8 \log d /\sqrt{d},j=1,\ldots, [ 0.05 d ]$,  $\mu_j = 0, j = [ 0.05 d ] + 1, \ldots, d$.
        \item[(c)]  Scale: $F_Y = \exp \big( N_{d}(\bzero_d, \sigma^2 \boldsymbol {\Sigma}_X ))$, where 
        $\sigma = 1 + 0.15 \log d / \sqrt{d}$. 
        \item[(d)]  Location and scale mixed: $F_Y = \exp \big( N_{d}(\delta \bone_d, \sigma \boldsymbol {\Sigma}_X) \big)$ where $\sigma = 1 + 0.1(50/d)^{0.25}$ and $\delta = 0.25 \log d /\sqrt{d}$. 
    \end{itemize}

    \item[IV.] $F_X = t_{5}\big(\bzero_d, \boldsymbol{\Sigma}_X\big)$ is the multivariate $t_5$ distribution, where $\Sigma_{X,ij} = 0.6^{|i-j|}$.
     \begin{itemize}
         \item[(a)]  Simple location: $F_Y = t_{5}\big(\delta \bone_d, \boldsymbol {\Sigma}_X \big)$ where $\delta = 0.5 \log d /\sqrt{d}$.
         \item[(b)]  Sparse location: $F_Y = t_{5}\big(\boldsymbol  \mu, \boldsymbol {\Sigma}_X \big)$ where $\mu_j = (-1)^{j} 2.1 \log d /\sqrt{d},j=1,\ldots, [ 0.05 d ]$,  $\mu_j = 0, j =  [ 0.05 d ] + 1, \ldots, d$.
        \item[(c)]  Scale: $F_Y =  t_{5}\big( \bzero_d, \boldsymbol {\Sigma}_Y))$, where $\Sigma_{Y,ij} = 0.7 (0.1)^{|i-j|}$. 
        \item[(d)]  Location and scale mixed: $F_Y = t_{5}\big(\delta \bone_d, \boldsymbol {\Sigma}_Y) \big)$ where $\Sigma_{Y,ij} = (0.8)^{|i-j|}$ and $\delta = 0.5 \log d /\sqrt{d}$. 
    \end{itemize}
\end{itemize}

\subsection{Addition Simulation Results}
\label{app: add}

See Tables~\ref{tab1}-\ref{tabm4}.

\begin{table}[h]
    \centering
 \caption{Empirical sizes of the tests under the four settings when the nominal significance level $\alpha =0.01$ and $0.05$, respectively, for  $m = n = 50$ and $d = 200$, $500$, $1000$.}
    \label{tab1} 
       \begin{tabular}{|c|ccc|ccc|ccc|ccc|}
       \hline
       $d$  & 200 & 500 & 1000 & 200 & 500 & 1000 & 200 & 500 & 1000& 200 & 500 & 1000\\
\hline
{$\alpha = 0.01$}
& \multicolumn{3}{c|}{Setting I} & \multicolumn{3}{c|}{Setting II} & \multicolumn{3}{c|}{Setting III} & \multicolumn{3}{c|}{Setting IV} \\ 
\hline 
$\Rg$-NN & 0.01 & 0.00 & 0.01 & 0.01 & 0.01 & 0.01 & 0.01 & 0.01 & 0.01 & 0.01 & 0.00 & 0.01\\
$\Ro$-MDP & 0.01 & 0.01 & 0.01 & 0.01 & 0.02 & 0.01 & 0.01 & 0.02 & 0.01 & 0.01 & 0.00 & 0.01\\
GET & 0.01 & 0.01 & 0.01 & 0.01 & 0.01 & 0.00 & 0.01 & 0.00 & 0.00 & 0.01 & 0.01 & 0.01\\
CM & 0.01 & 0.01 & 0.00 & 0.01 & 0.01 & 0.01 & 0.01 & 0.01 & 0.01 & 0.01 & 0.00 & 0.01\\
MT & 0.01 & 0.01 & 0.02 & 0.01 & 0.01 & 0.00 & 0.01 & 0.01 & 0.01 & 0.01 & 0.01 & 0.01\\
BD & 0.01 & 0.01 & 0.01 & 0.01 & 0.01 & 0.01 & 0.01 & 0.01 & 0.00 & 0.01 & 0.01 & 0.01\\
GLP & 0.01 & 0.01 & 0.01 & 0.02 & 0.03 & 0.03 & 0.06 & 0.07 & 0.06 & 0.01 & 0.01 & 0.01\\
HD & 0.00 & 0.01 & 0.00 & 0.00 & 0.01 & 0.00 & 0.00 & 0.01 & 0.00 & 0.00 & 0.00 & 0.00\\
MMD & 0.00 & 0.00 & 0.00 & 0.00 & 0.00 & 0.00 & 0.00 & 0.00 & 0.00 & 0.00 & 0.00 & 0.00\\
\hline
{$\alpha = 0.05$}
 & \multicolumn{3}{c|}{Setting I} & \multicolumn{3}{c|}{Setting II} & \multicolumn{3}{c|}{Setting III} & \multicolumn{3}{c|}{Setting IV} \\ 
\hline 
$\Rg$-NN & 0.05 & 0.05 & 0.04 & 0.05 & 0.05 & 0.04 & 0.04 & 0.04 & 0.03 & 0.06 & 0.04 & 0.05\\
$\Ro$-MDP & 0.06 & 0.05 & 0.04 & 0.04 & 0.06 & 0.04 & 0.05 & 0.06 & 0.04 & 0.05 & 0.04 & 0.05\\
GET & 0.05 & 0.05 & 0.04 & 0.04 & 0.05 & 0.06 & 0.05 & 0.05 & 0.04 & 0.04 & 0.04 & 0.05\\
CM & 0.04 & 0.04 & 0.03 & 0.04 & 0.03 & 0.04 & 0.03 & 0.03 & 0.04 & 0.04 & 0.03 & 0.03\\
MT & 0.05 & 0.05 & 0.06 & 0.04 & 0.05 & 0.05 & 0.05 & 0.06 & 0.07 & 0.05 & 0.05 & 0.04\\
BD & 0.04 & 0.05 & 0.06 & 0.04 & 0.06 & 0.04 & 0.05 & 0.05 & 0.05 & 0.05 & 0.05 & 0.05\\
GLP & 0.06 & 0.05 & 0.06 & 0.07 & 0.08 & 0.07 & 0.10 & 0.09 & 0.09 & 0.06 & 0.06 & 0.05\\
HD & 0.03 & 0.04 & 0.03 & 0.03 & 0.04 & 0.03 & 0.02 & 0.03 & 0.02 & 0.02 & 0.02 & 0.02\\
MMD & 0.00 & 0.00 & 0.00 & 0.00 & 0.01 & 0.00 & 0.01 & 0.00 & 0.00 & 0.01 & 0.00 & 0.01\\
\hline
\end{tabular}
\end{table}

\begin{table}[!htp]
    \centering
 \caption{Empirical sizes of the tests under the four settings when the nominal significance level $\alpha =0.01$ and $0.05$, respectively, for  $m = 50, n = 100$ and $d = 200$, $500$, $1000$.}
    \label{tabm1} 
       \begin{tabular}{|l|ccc|ccc|ccc|ccc|}
\hline
 & \multicolumn{3}{c|}{Setting I} & \multicolumn{3}{c|}{Setting II} & \multicolumn{3}{c|}{Setting III} & \multicolumn{3}{c|}{Setting IV} \\ 
$\alpha = 0.01$ & 200 & 500 & 1000 & 200 & 500 & 1000 & 200 & 500 & 1000& 200 & 500 & 1000\\
\hline 
$\Rg$-NN & 0.01 & 0.01 & 0.01 & 0.01 & 0.01 & 0.01 & 0.01 & 0.01 & 0.01 & 0.01 & 0.01 & 0.01\\
$\Ro$-MDP & 0.01 & 0.02 & 0.01 & 0.01 & 0.01 & 0.01 & 0.01 & 0.01 & 0.01 & 0.02 & 0.01 & 0.01\\
GET & 0.01 & 0.01 & 0.01 & 0.01 & 0.01 & 0.02 & 0.01 & 0.01 & 0.01 & 0.00 & 0.00 & 0.01 \\
CM & 0.00 & 0.00 & 0.00 & 0.00 & 0.00 & 0.00 & 0.00 & 0.00 & 0.00 & 0.00 & 0.00 & 0.00\\
MT & 0.01 & 0.00 & 0.01 & 0.01 & 0.01 & 0.01 & 0.01 & 0.01 & 0.01 & 0.01 & 0.01 & 0.01\\
BD & 0.01 & 0.01 & 0.01 & 0.01 & 0.01 & 0.01 & 0.01 & 0.01 & 0.01 & 0.01 & 0.01 & 0.01\\
GLP & 0.01 & 0.01 & 0.01 & 0.03 & 0.04 & 0.03 & 0.06 & 0.06 & 0.07 & 0.02 & 0.01 & 0.02\\
HD & 0.01 & 0.01 & 0.01 & 0.01 & 0.01 & 0.01 & 0.00 & 0.00 & 0.01 & 0.01 & 0.01 & 0.00\\
MMD & 0.00 & 0.00 & 0.00 & 0.00 & 0.00 & 0.00 & 0.00 & 0.00 & 0.00 & 0.01 & 0.00 & 0.01\\
\hline

 & \multicolumn{3}{c|}{Setting I} & \multicolumn{3}{c|}{Setting II} & \multicolumn{3}{c|}{Setting III} & \multicolumn{3}{c|}{Setting IV} \\ 
$\alpha = 0.05$ & 200 & 500 & 1000 & 200 & 500 & 1000 & 200 & 500 & 1000& 200 & 500 & 1000\\
\hline 
$\Rg$-NN & 0.04 & 0.04 & 0.05 & 0.05 & 0.06 & 0.05 & 0.05 & 0.06 & 0.06 & 0.04 & 0.04 & 0.03\\
$\Ro$-MDP & 0.04 & 0.06 & 0.05 & 0.05 & 0.06 & 0.06 & 0.05 & 0.06 & 0.05 & 0.06 & 0.05 & 0.05\\
GET & 0.04 & 0.06 & 0.04 & 0.04 & 0.06 & 0.05 & 0.05 & 0.05 & 0.04 & 0.04 & 0.04 & 0.04 \\
CM & 0.05 & 0.05 & 0.04 & 0.04 & 0.05 & 0.05 & 0.06 & 0.04 & 0.05 & 0.06 & 0.04 & 0.05\\
MT & 0.05 & 0.06 & 0.06 & 0.05 & 0.06 & 0.04 & 0.06 & 0.06 & 0.05 & 0.05 & 0.05 & 0.05\\
BD & 0.05 & 0.06 & 0.05 & 0.06 & 0.06 & 0.05 & 0.06 & 0.05 & 0.05 & 0.05 & 0.04 & 0.05\\
GLP & 0.04 & 0.05 & 0.05 & 0.08 & 0.09 & 0.09 & 0.08 & 0.08 & 0.09 & 0.06 & 0.05 & 0.06\\
HD & 0.04 & 0.05 & 0.04 & 0.05 & 0.04 & 0.05 & 0.03 & 0.03 & 0.04 & 0.03 & 0.02 & 0.02\\
MMD & 0.00 & 0.00 & 0.00 & 0.01 & 0.01 & 0.01 & 0.01 & 0.00 & 0.00 & 0.02 & 0.01 & 0.01\\
\hline
\end{tabular}
\end{table}


\begin{table}[htp!]
    \centering
 \caption{Estimated power of the tests with $\alpha = 0.05$ under the multivariate Gaussian distribution (Setting I) and the Gaussian mixture distribution (Setting II) for $m = 50, n =100$ and $d = 200$, $500$, $1000$.}
    \label{tabm2} 
       \begin{tabular}{|l|ccc|ccc|ccc|ccc|}
\hline
&  \multicolumn{3}{c|}{Setting I (a)} & \multicolumn{3}{c|}{Setting I (b)} & \multicolumn{3}{c|}{Setting I (c)} & \multicolumn{3}{c|}{Setting I (d)} \\
Method & 200 & 500 & 1000 & 200 & 500 & 1000 & 200 & 500 & 1000& 200 & 500 & 1000\\
\hline
$\Rg$-NN & 80 & 75 & 70 & 97 & 90 & 81 & 82 & 90 & 95 & 100 & 99 & 100\\
$\Ro$-MDP & 74 & 71 & 66 & 94 & 85 & 73 & 88 & 96 & 98 & 99 & 98 & 99\\
GET & 73 & 67 & 61 & 92 & 82 & 71 & 77 & 87 & 92 & 97 & 96 & 96\\
CM & 36 & 35 & 33 & 51 & 40 & 33 & 4 & 6 & 6 & 83 & 81 & 80\\
MT & 100 & 100 & 99 & 8 & 6 & 7 & 5 & 5 & 5 & 17 & 17 & 18\\
BD & 91 & 76 & 56 & 68 & 48 & 30 & 94 & 99 & 100 & 26 & 28 & 26\\
GLP & 73 & 60 & 45 & 15 & 13 & 14 & 7 & 8 & 4 & 8 & 6 & 5\\
HD & 6 & 6 & 5 & 6 & 7 & 5 & 72 & 88 & 93 & 8 & 9 & 7\\
MMD & 99 & 94 & 58 & 100 & 99 & 60 & 0 & 0 & 0 & 1 & 0 & 0\\
\hline
& \multicolumn{3}{c|}{Setting I (e)}& \multicolumn{3}{c|}{Setting II (a)} & \multicolumn{3}{c|}{Setting II (b)} & \multicolumn{3}{c|}{Setting II (c)} \\
Method & 200 & 500 & 1000 & 200 & 500 & 1000 & 200 & 500 & 1000& 200 & 500 & 1000\\
\hline
$\Rg$-NN & 100 & 100 & 100 & 74 & 92 & 99 & 83 & 83 & 83 & 92 & 87 & 81\\
$\Ro$-MDP & 100 & 100 & 99 & 52 & 68 & 78 & 34 & 36 & 36 & 83 & 86 & 89\\
GET & 99 & 99 & 98 & 65 & 88 & 97 & 84 & 83 & 85 & 80 & 72 & 67\\
CM & 88 & 88 & 86 & 20 & 30 & 33 & 6 & 5 & 5 & 78 & 80 & 80\\
MT & 18 & 18 & 19 & 71 & 82 & 84 & 5 & 6 & 4 & 9 & 12 & 16\\
BD & 37 & 35 & 33 & 56 & 69 & 89 & 52 & 42 & 41 & 9 & 12 & 17\\
GLP & 9 & 10 & 4 & 10 & 8 & 8 & 8 & 9 & 9 & 9 & 10 & 9\\
HD & 8 & 9 & 7 & 5 & 4 & 4 & 4 & 5 & 4 & 5 & 5 & 4\\
MMD & 9 & 0 & 0 & 2 & 1 & 2 & 1 & 1 & 1 & 2 & 1 & 1\\
\hline
\end{tabular}
\end{table}

 \begin{table}[htp!]
    \centering
 \caption{Estimated power of the tests with $\alpha = 0.05$ under the  multivariate log-normal distribution (Setting III) for $m = 50, n = 100$ and $d = 200$, $500$, $1000$. }
    \label{tabm3} 
       \begin{tabular}{|l|ccc|ccc|ccc|ccc|}
\hline
& \multicolumn{3}{c|}{Setting III (a)} & \multicolumn{3}{c|}{Setting III (b)} & \multicolumn{3}{c|}{Setting III (c)} & \multicolumn{3}{c|}{Setting III (d)} \\
Method & 200 & 500 & 1000 & 200 & 500 & 1000 & 200 & 500 & 1000& 200 & 500 & 1000\\
\hline
$\Rg$-NN & 88 & 86 & 85 & 98 & 95 & 83 & 42 & 46 & 48 & 72 & 78 & 78\\
$\Ro$-MDP & 98 & 99 & 98 & 91 & 90 & 78 & 60 & 72 & 77 & 91 & 96 & 97\\
GET & 84 & 82 & 78 & 93 & 83 & 61 & 40 & 42 & 44 & 69 & 73 & 74\\
CM & 24 & 23 & 21 & 44 & 38 & 32 & 6 & 7 & 7 & 13 & 13 & 14\\
MT & 99 & 99 & 98 & 13 & 21 & 39 & 22 & 26 & 22 & 84 & 83 & 79\\
BD & 97 & 99 & 98 & 22 & 19 & 14 & 71 & 82 & 84 & 93 & 98 & 98\\
GLP & 85 & 74 & 62 & 22 & 30 & 36 & 12 & 10 & 10 & 26 & 20 & 18\\
HD & 35 & 46 & 49 & 5 & 5 & 4 & 19 & 28 & 31 & 29 & 44 & 50\\
MMD & 96 & 87 & 62 & 100 & 100 & 77 & 32 & 16 & 3 & 76 & 60 & 35\\
\hline
\end{tabular}
\end{table}

\begin{table}[htp!]
    \centering
 \caption{Estimated power of the tests with $\alpha = 0.05$ under the multivariate $t_5$ distribution (Setting IV) for $m = 50, n = 100$ and $d = 200$, $500$, $1000$.}
    \label{tabm4}  
      \begin{tabular}{|c|ccc|ccc|ccc|ccc|}
\hline
&  \multicolumn{3}{c|}{Setting IV (a)} & \multicolumn{3}{c|}{Setting IV (b)}& \multicolumn{3}{c|}{Setting IV (c)}& \multicolumn{3}{c|}{Setting IV (d)} \\
 Method & 200 & 500 & 1000 & 200 & 500 & 1000 & 200 & 500 & 1000& 200 & 500 & 1000\\
\hline
$\Rg$-NN & 91 & 81 & 72 & 93 & 80 & 66 & 87 & 69 & 56 & 95 & 85 & 75\\
$\Ro$-MDP & 81 & 78 & 69 & 85 & 76 & 62 & 100 & 99 & 99 & 95 & 95 & 94\\
GET & 79 & 58 & 47 & 80 & 54 & 38 & 78 & 44 & 21 & 86 & 69 & 56\\
CM & 33 & 29 & 25 & 36 & 31 & 22 & 89 & 88 & 86 & 62 & 64 & 59\\
MT & 99 & 99 & 99 & 10 & 10 & 7 & 22 & 24 & 28 & 92 & 92 & 86\\
BD & 8 & 5 & 6 & 6 & 4 & 6 & 77 & 76 & 81 & 8 & 5 & 6\\
GLP & 67 & 54 & 44 & 7 & 10 & 9 & 53 & 51 & 50 & 66 & 49 & 39\\
HD & 3 & 2 & 3 & 3 & 2 & 2 & 23 & 24 & 23 & 3 & 2 & 2\\
MMD & 90 & 52 & 14 & 88 & 31 & 8 & 51 & 51 & 53 & 87 & 52 & 16\\
\hline
\end{tabular}
\end{table}

\subsection{A detailed comparison between RISE and GET}
\label{sec:rise and get}

Here, we compare the power of RISE and GET by varying $k$'s. We also explore the graph-induced rank (denoted by $\Rg$-MST) and the overall rank (denoted by $\Ro$-MST) in the $k$-MST. To compare different graphs in a more unified fashion, for the $k$-NNG and $k$-MDP, we set $k = 2 [ N^{\lambda} ]$ while for the $k$-MST, we set $k = [N^{\lambda}]$, for $\lambda \in (0,0.8)$, since for the $k$-NNG and $k$-MDP, the largest value of $k$ can be $N-1$, while for the $k$-MST, the largest value of $k$ can only be $
N/2$. The results for different $n$'s and $d$'s show similar patterns, so we only present the results for $m = n = 50$ and $d=500$ here for Settings I-IV in Section \ref{setting} with $\alpha = 0.05$. 
Each configuration is repeated $1000$ times to estimate the empirical size or power. 

\begin{figure}[!htp]
  \centering
 \includegraphics[width=5.7in]{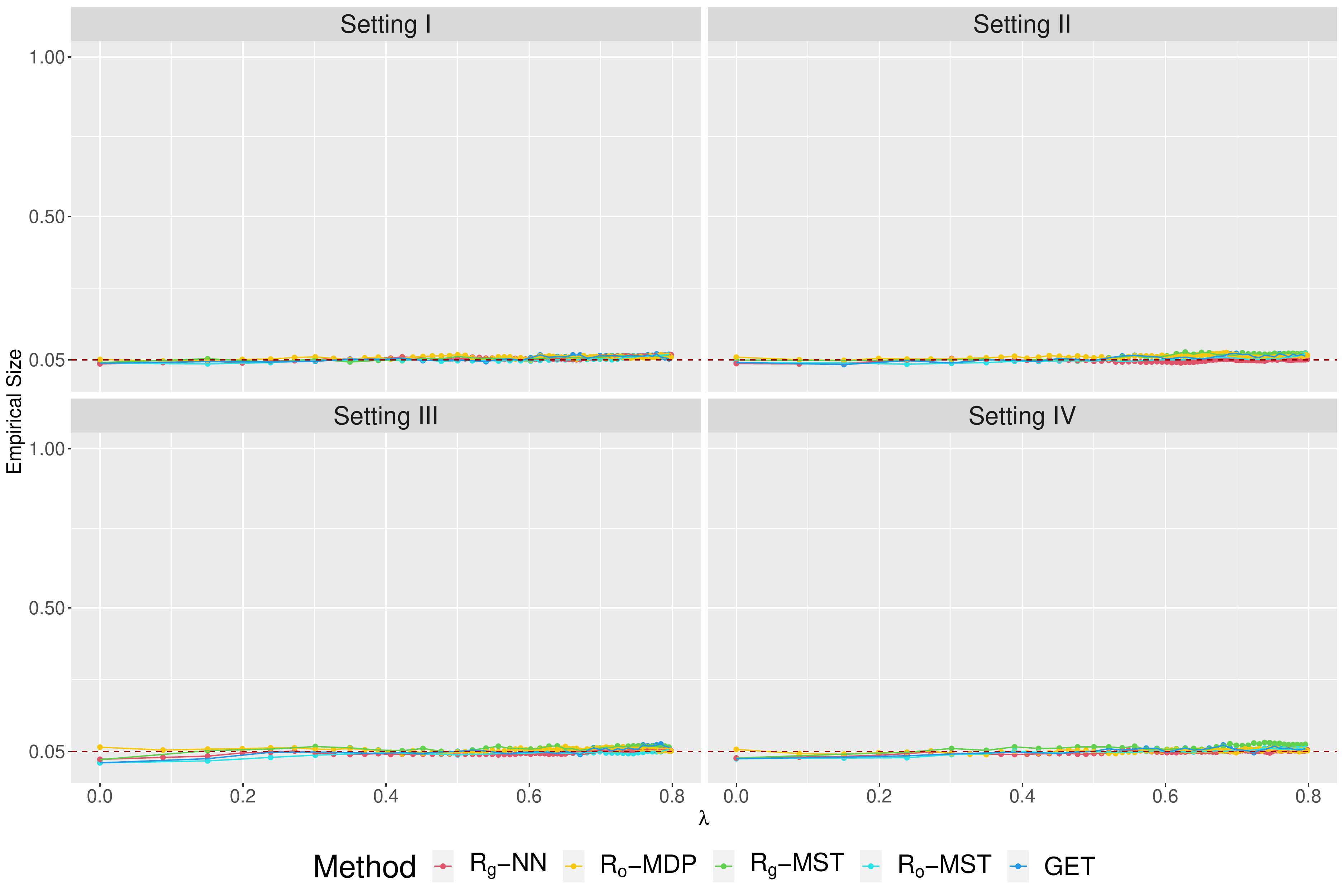}
 \caption{Empirical sizes of RISE and GET for varying $\lambda$. 
 }
 \label{fig:size}
\end{figure}

\begin{figure}[!htp]
  \centering
 \includegraphics[width=5.7in]{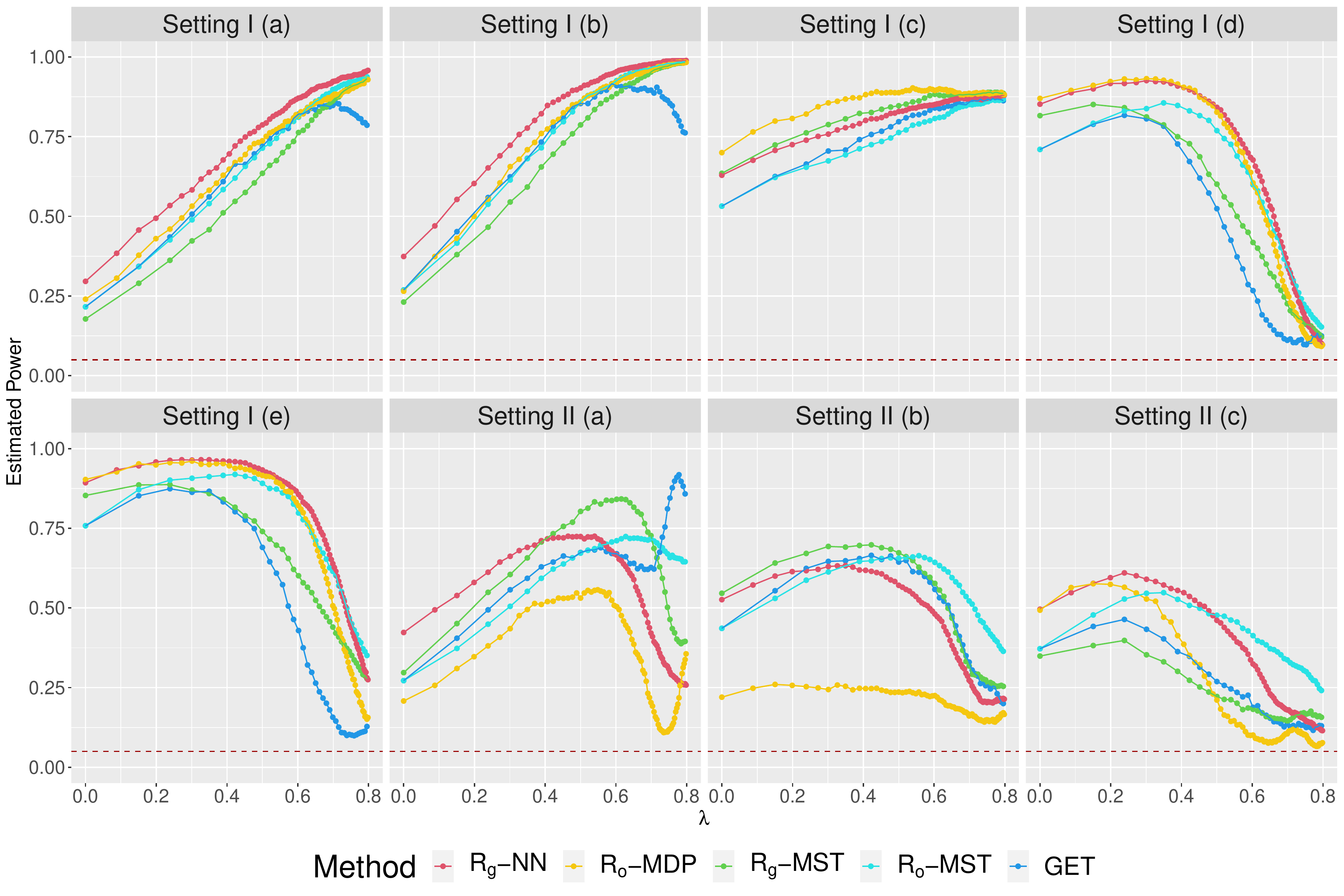}
 \caption{Estimated power of RISE and GET for varying $\lambda$ under Settings I and II.
 }
 \label{fig:setting1}
\end{figure}

\begin{figure}[!htp]
  \centering
 \includegraphics[width=5.7in]{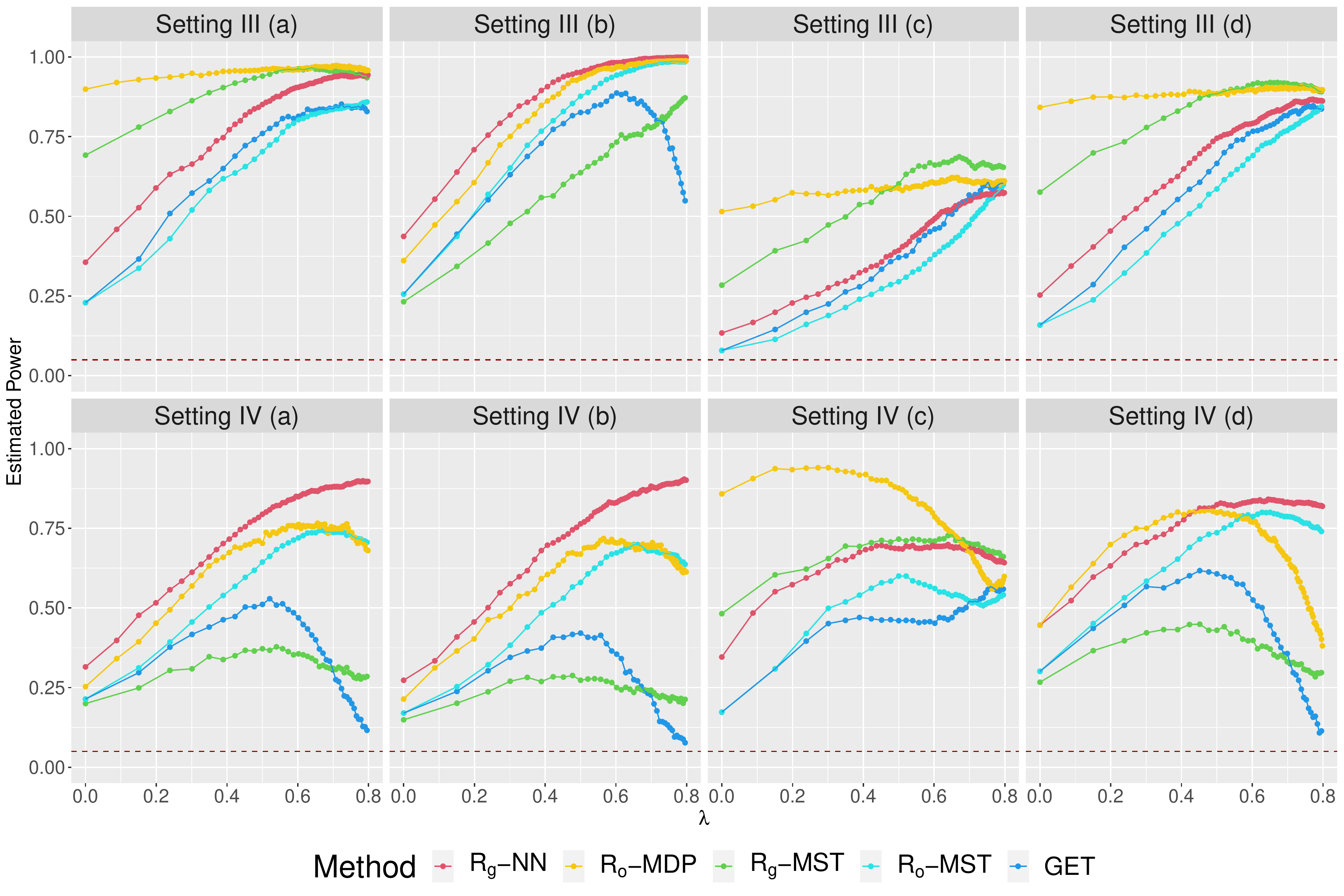}
 \caption{Estimated power of RISE and GET for varying $\lambda$ under Settings III and IV.
 }
 \label{fig:setting2}
\end{figure}

The empirical sizes of the five tests under  Settings I-IV are presented in Figure \ref{fig:size}. We see that all of these tests can control the type-I error well even for large $\lambda$ under all settings. The estimated power for Settings I and II are presented in Figure \ref{fig:setting1} and the estimated power for Settings III and IV are presented in Figure \ref{fig:setting2}. We observe that for some settings, the power of these tests increases first when $\lambda$ increases, then decreases when $\lambda$ is too large. The reason is that a denser graph can contain more similar information among the observations. However, it can also include noisier information when it is too dense. For GET, when $\lambda=1$ which means the graph is a complete graph, its test statistic is not well-defined. Its power may approach zero when $\lambda$ approaches one, while RISE still has power for a complete graph. From these figures, we see that RISE performs better than GET in most of the settings for a wide range of $k$'s. 

We notice that $\Rg$-NN has the best performance in most of the settings for all $k$'s. The improvement of $\Rg$-NN and $\Ro$-MDP over GET is more significant under the heavy-tailed Setting III and IV. However, $\Ro$-MDP is less powerful under the Gaussian mixed Setting II, which may be due to the intrinsic property of MDP. $\Ro$-MST has a moderate performance such that it outperforms GET in most of the settings but is dominated by $\Rg$-NN in most instances. $\Rg$-MST seems not very robust as it can achieve high power in some cases but is outperformed by GET sometimes.

\section{More Discussions on Real Data Analysis}
\label{detail of New York} 
For the comparison of  January and February, RISE is the only test that can reject at the $0.05$ level. We then take a closer look at GET to understand this better.  We first examine each $k$th MST and $k$-MST separately for $k=1,\dots,5$. The test statistic of GET depends on how far the two within-sample edge counts deviate from their expectations under the null distribution, so we check how the two edge-count statistics change when $k$ increases from $1$ to $5$. Table \ref{tab:edge} shows the within-sample edge counts of each sample in each $k$th MST. 
The $p$-values of GET on the $k$th MST and the $k$-MST for different $k$'s are also presented. 
We notice that for most of the $k$th MSTs, at least one of the within-sample edge counts somewhat deviates from their corresponding expectations. 
However, since GET treats all MSTs equally, there are two issues: (i) different MSTs can contain opposite information and (ii) a $k$th MST for a large $k$ can contain noisier information. The first issue is obvious from the edge-count statistics. 
For example, the sample February has the within-sample edge count above its expectation for the first to the fourth MSTs, but below its expectation for the fifth MST. This makes the $p$-value increase from $0.003$ on the $4$-MST to $0.09$ on the $5$-MST. The second issue can be observed from the $p$-values of GET on the $k$th MST. The $p$-value of the comparison on the first MST is small, but it can be very large for other $k$th MSTs. When the $k$th MST does not contain useful information but noise, the consequence for GET is to yield a larger $p$-value. On the other hand, RISE is less affected by the two issues by incorporating weights.

\begin{table}
    \centering
    \caption{The edge-count statistics on the $k$th MST and the $p$-values of GET using the $k$th MST and the $k$-MST, 
    respectively. The expected edges for each MST are $15.76$ and $12.81$ for Samples Jan and Feb, respectively. 
    }
     \label{tab:edge}   
\begin{tabular}{||c||c|ccccc|}
\hline \hline 
    & $k$  &   1 & 2 & 3 & 4 & 5 \\ \hline
\multirow{2}{*}{ Edge-count } &  Jan & 15 & 15 & 14 & 14 & 13 \\
   & Feb & 20 & 18 & 19 & 16 & 8 \\ \hline
\multirow{2}{*}{ $p$-values} &$k$th MST & 0.034 & 0.112 & 0.105 & 0.540 & 0.109\\
  &$k$-MST  & 0.034 &  0.007 & 0.002 & 0.003  & 0.090 \\
\hline \hline
\end{tabular}
\end{table}

\begin{figure}[t]
  \centering
 \includegraphics[width=200pt]{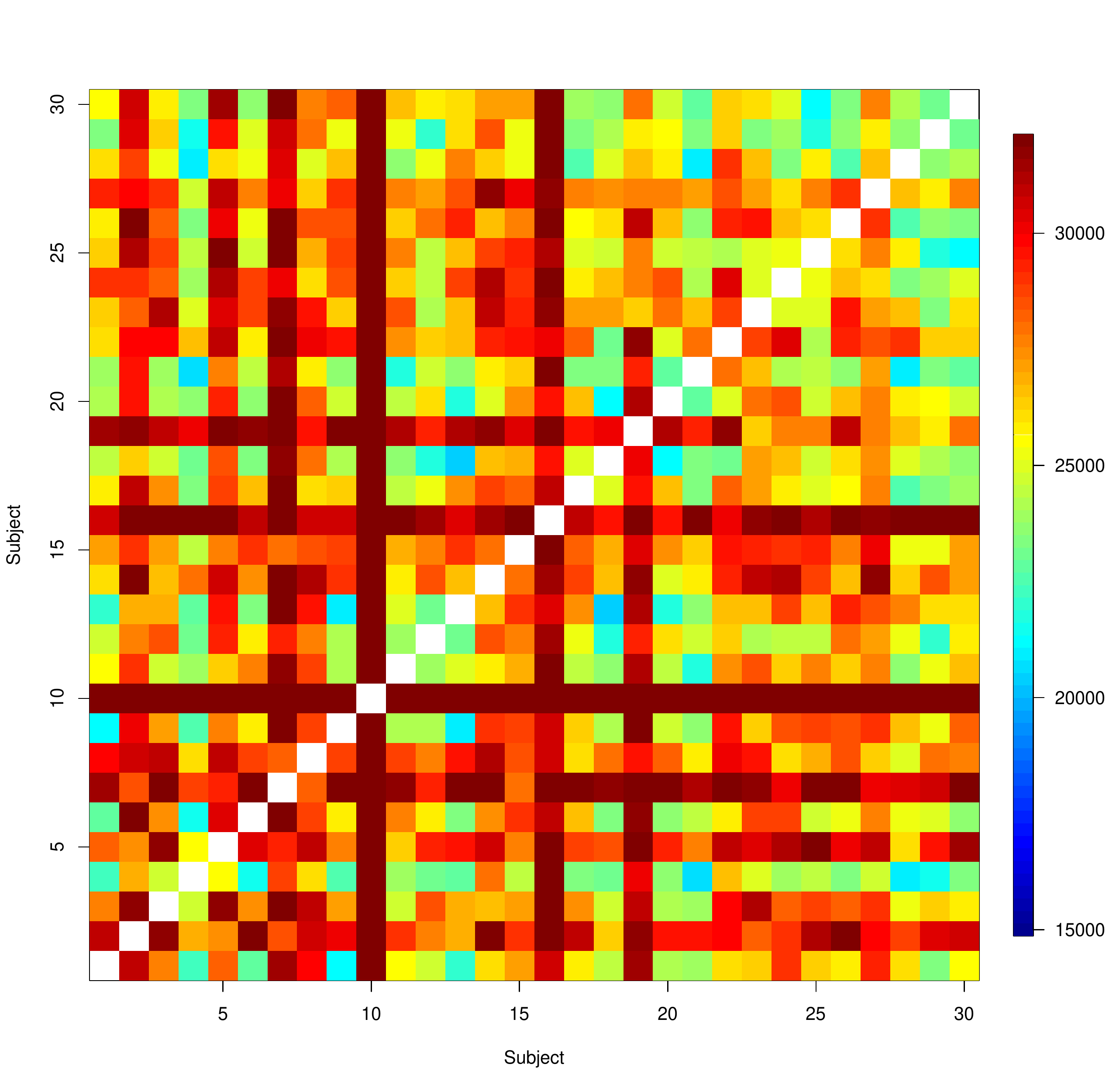}
 \caption{The heatmap of the distance matrix of the $30$ subjects, where the first $15$ subjects are male and the others female. }
\label{fig:brain_heat} 
\end{figure}

\section{Exploration on graphs}
\label{discussion k}

\begin{figure}[htpb!]
  \centering
 \includegraphics[width=370pt]{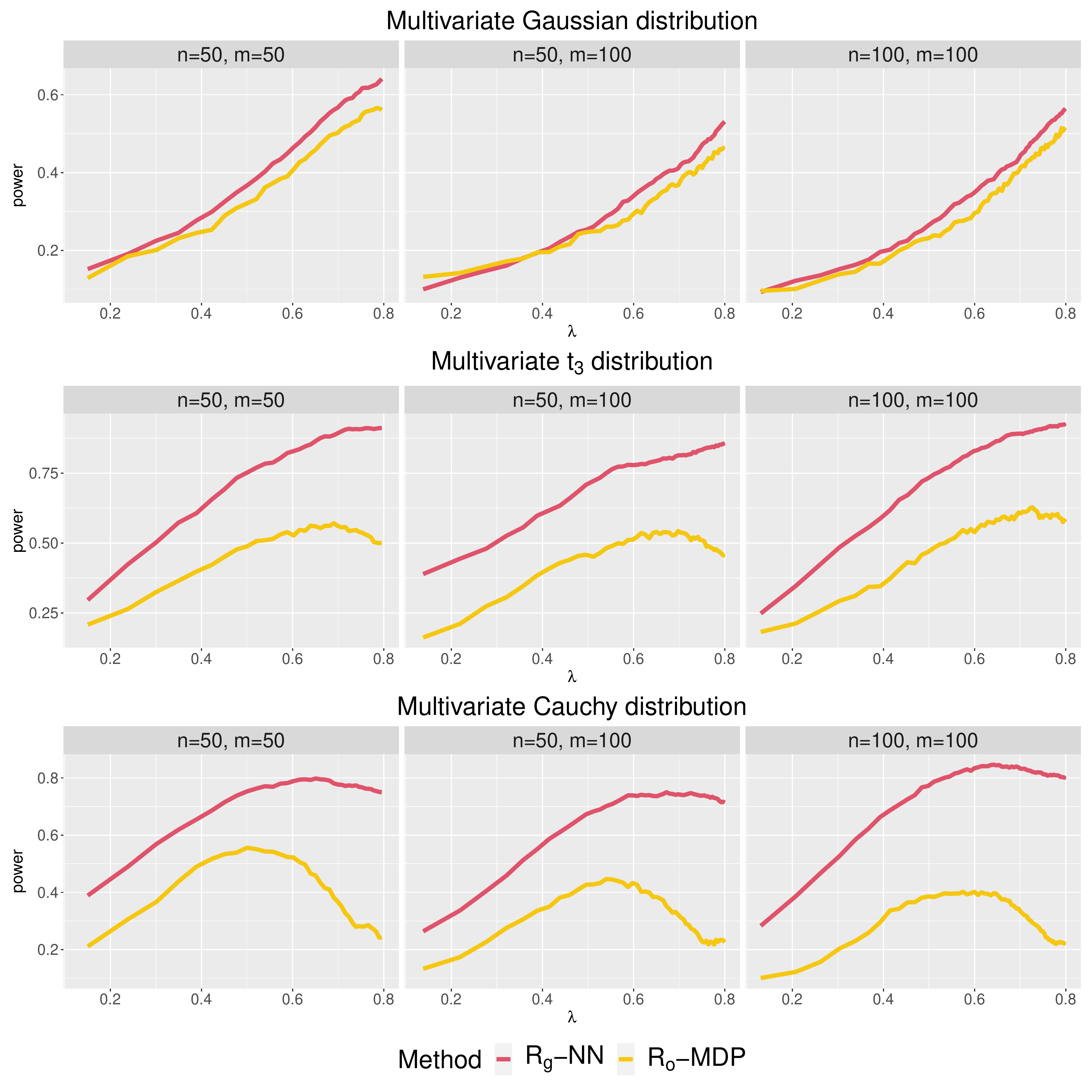}
 \caption{Estimated power of $\Rg$-NN and $\Ro$-MDP with $k = [N^{\lambda}]$ over $1000$ repetitions under each setting. The three settings are: $\big( N_{d}( \bzero_d, \I_d), N_{d}( \delta_1 \bone_d, \I_d) \big)$, $\big( t_{3}( \bzero_d, \I_d), t_{3}( \delta_2 \bone_d, \delta_3 \I_d) \big)$ and  $\big( {\rm Cauchy}_d( \bzero_d, \I_d), {\rm Cauchy}_d( \delta_4 \bone_d, \I_d) \big)$ where $\delta_1 = \frac{20}{\sqrt{Nd}}$, $\delta_2 = \frac{28}{\sqrt{Nd}}$, $\delta_3 = (1 + \frac{25}{\sqrt{Nd}})^2$ and $\delta_4  = \frac{1.44}{\sqrt{Nd}}$. Here $\delta_i$'s are set to make these tests have moderate power. 
 }
\label{fig:powerill} 
\end{figure}

We generate i.i.d. samples of $X_i \sim F_X$ and $Y_i \sim F_Y$, and set $d=500$ and vary the sample sizes $(m,n)$. Three combinations of $(F_X,F_Y)$ are considered. Figure~\ref{fig:powerill} shows how the power varies with $\lambda$ such that $k = [ N^{\lambda} ]$ and the nominal significance level is set as $0.05$. We see that the optimal $k$ varies for different settings and it is reasonable to choose $\lambda = 0.65$ for both the $k$-NNG and the $k$-MDP to achieve adequate power. Besides, $\Rg$-NN performs better than $\Ro$-MDP.

\section{Proof of Statement (i) }
\label{proof:s1}
Let 
$$
\begin{aligned}
W & = a_{1} Z_{w}^{\mB} + a_{2} \sqrt{T} ( Z_{\diff}^{\mB} - \sqrt{1-1/T}Z_{X}) + a_{3} Z_{X} \\
& = a_{1} Z_{w}^{\mB} + a_{2} \sqrt{T}  Z_{\diff}^{\mB} + (a_{3} - a_{2}  \sqrt{T-1}) Z_{X}.
\end{aligned}
$$ 
We firstly show that, in the usual limit regime, 
$$
\lim _{N \rightarrow \infty} \Vb(W)=0 \text { iff } a_{1}=a_{2}=a_{3}=0.
$$
By the independence of $g_{i}$'s under the bootstrap null distribution, it is easy to see that
$$\Covb(Z_{w}^{\mB}, Z_{\diff}^{\mB}) = \frac{4 m n(n-m)}{(N-2) N^{2}} \frac{(N-1)^2 r_1^2}{\sigma_{w}^{\mB} \sigma_{\diff}^{\mB}}\, , $$
$$\Covb(Z_{w}^{\mB}, Z_{X})= \frac{2 (N-1) m n(n-m)}{(N-2) N^{2}} \frac{r_0}{ \sigma_{w}^{\mB} \sigma^{\mB}} \, ,$$
$$\text{ and } \Covb(Z_{\diff}^{\mB}, Z_{X})=\frac{2 (N-1) m n r_0}{N \sigma_{\diff}^{\mB} \sigma^{\mB}} = \frac{r_0}{r_1}\, .$$
As a result, we have $\sqrt{T}  \Covb(Z_{\diff}^{\mB}, Z_{X}) = \sqrt{T - 1}$ and 
\begin{align*}
\Vb(W)=& a_{1}^{2}+a_{2}^{2} (2T-1)+a_{3}^{2} - 2 a_2 a_3 \sqrt{T - 1} 
+2 a_{1} a_{2} \sqrt{T} \Covb(Z_{w}^{\mB}, Z_{\diff}^{\mB})
\\& + 2 a_{1} ( a_{3} - a_2 \sqrt{T-1} ) \Covb(Z_{w}^{\mB}, Z_{X}) \\
&+2 a_{2} ( a_{3} - a_2 \sqrt{T-1} ) \sqrt{T}  \Covb(Z_{\diff}^{\mB}, Z_{X}) \\
= &   a_{1}^{2}+a_{2}^{2}+a_{3}^{2} +  2 a_{1} a_{3}  \Covb(Z_{w}^{\mB}, Z_{X})
\\ &  + 2 a_{1} a_{2} \big( \sqrt{T} \Covb(Z_{w}^{\mB}, Z_{\diff}^{\mB})  - \sqrt{T-1} \Covb(Z_{w}^{\mB}, Z_{X}) \big) \, .
\end{align*}
Besides, we have 
$$\Covb\big(Z_{w}^{\mB}, Z_{X}\big) \asymp \frac{r_0}{\sqrt{N} r_d}  \rightarrow 0\, ,$$
$$
\begin{aligned}
 & \sqrt{T} \Covb\big(Z_{w}^{\mB}, Z_{\diff}^{\mB}\big)  - \sqrt{T-1} \Covb\big(Z_{w}^{\mB}, Z_{X}\big) \\
 & =  \frac{2 (N-1) m n(n-m)}{(N-2) N^{2} \sigma_{w}^{\mB} \sqrt{r_1^2 - r_0^2} } \Big(\frac{ 2 (N-1) r_1^3}{ \sigma_{\diff}^{\mB}} -   \frac{r_0^2}{  \sigma^{\mB}} \Big) \\
& =  \frac{2 (N-1) m n(n-m)}{(N-2) N^{2} \sigma_{w}^{\mB} \sqrt{r_1^2 - r_0^2} } \sqrt{\frac{N}{m n}}  \big(r_1^2 - r_0^2 \big) \\
& \precsim \frac{\sqrt{r_1^2 - r_0^2}}{\sqrt{N^3} r_d} \rightarrow 0\, ,
\end{aligned}
$$
by Cauchy–Schwarz inequality $r_d^2 \geq r_1^2 \geq r_0^2$. Thus, we have $\lim_{N \rightarrow \infty} \Vb(W) = a_1^2 + a_{2}^{2}+a_{3}^{2} >0$ in the usual limit regime.  This implies that the covariance matrix of the joint limiting distribution is of full rank. Then by Cram\'er-Wold device, Statement (i) 
holds if $W$ is 
is asymptotically Gaussian distributed under the bootstrap null distribution when at least one of constants $a_{1}, a_{2}, a_{3}$ is nonzero. We use the Stein’s method \citep{chen2010normal}, in particular, the following theorem. 
\begin{theorem}[Stein's Method, \cite{chen2010normal}, Theorem 4.13]
\label{theorem:stein}
Let $\big\{\xi_{i}, i \in \mathcal{J}\big\}$ be a random field with mean zero, $W=\sum_{i \in \mathcal{J}} \xi_{i}$ and $\Vp(W)=1$, for each $i \in \mathcal{J}$ there exits $K_{i} \subset \mathcal{J}$ such that $\xi_{i}$ and $\xi_{K_{i}^{c}}$ are independent, then
\begin{equation}
\sup _{h \in \operatorname{Lip}(1)}\big|\Ep h(W)- \Ep h(Z) \big| \leq \sqrt{\frac{2}{\pi}} \Ep\Big|\sum_{i \in \mathcal{J}}\big\{\xi_{i} \eta_{i}-\Ep(\xi_{i} \eta_{i})\big\}\Big|+\sum_{i \in \mathcal{J}} \Ep\big|\xi_{i} \eta_{i}^{2}\big|
\label{eq:stein}
\end{equation}
where $\eta_{i}=\sum_{j \in K_{i}} \xi_{j}, Z$ is the standard normal random variable.
\end{theorem}
As long as we show that the right-hand side of \eqref{eq:stein} goes to zero when $N \rightarrow \infty$, $W$ converges to the standard normal distribution by Stein's Theorem. We can represent the graph by 
\begin{equation*}
    G_k \equiv \big( V = \cN,E = \{(i,j): R_{ij}>0, i,j \in\cN\} \big)\, ,
\end{equation*}
where $\cN = \{1,\ldots,N\}$. 
To simplify notations, we let $p={m}/{N}, q={n}/{N}$, 
and for each edge $e = (e^+,e^-) \in G_k$, let
$$
J_{e}=\left \{\begin{array}{ll}
0 & \text { if } g_{e^{+}} \neq g_{e^{-}}  \,,\\
1 & \text { if } g_{e^{+}}=g_{e^{-}}=1  \,,\\
2 & \text { if } g_{e^{+}}=g_{e^{-}}=0 \,.
\end{array}
\right.
$$
We can reorganize $W$ in the following way: 
\begin{align*}
W=& \frac{a_{1} \Big( \frac{n-1}{N-2}\big(U_x- p^{2}N(N-1)r_0 \big) +\frac{m-1}{N-2}\big(U_y- q^{2} N(N-1)r_0 \big) \Big)}{\sigma_{w}^{\mB}} \\
& +\frac{a_{2} \sqrt{T}\big(U_x - U_y - (p^{2} -q^2) N(N-1)r_0 \big)}{\sigma_{\diff}^{\mB}}  +\frac{   (a_{3} - a_{2}  \sqrt{T-1}) \big(n_{X}-m\big)}{\sigma^{\mB}} \\
=& \sum_{e \in G}  \frac{2 R_e a_1}{N-2} \Big(\frac{N}{\sigma_{w}^{\mB}} \big(\indi( g_{e^+}=1 )-p\big) \big( \indi( g_{e^-}=1 ) -p) - \frac{ \indi(J_e = 1) + \indi(J_e = 2) -p^{2}-q^{2}}{\sigma_{w}^{\mB}}\Big) \\
&+\sum_{e \in G} 2 R_e \frac{a_{2} \sqrt{T} }{\sigma_{\diff}^{\mB}}\big( \indi(g_{e^+} = 1) + \indi(g_{e^-} = 1) -2 p \big) \\
& + \sum_{i =1}^N \frac{ (a_{3} - a_{2}  \sqrt{T-1}) \big( \indi(g_i = 1) -p\big)}{\sigma^{\mB}} \, .
\end{align*}
Define the function $h: \cN \rightarrow \mathbb{R}$ such that $h(i) = \indi(g_i = 1) -p, i \in \cN$. Then,
$$
\begin{array}{l}
\big(\indi( g_{e^+}=1 )-p\big) \big( \indi( g_{e^-}=1 ) -p)  = h(e^{+}) h(e^{-}) \,,\\
\indi(J_e = 1) + \indi(J_e = 2) -p^{2}-q^{2} = 2 h(e^{+}) h(e^{-})+(p-q) \big( h(e^{+}) + h(e^{-}) \big)\, , \\
\indi(g_{e^+} = 1) + \indi(g_{e^-} = 1) -2 p = h(e^{+}) + h(e^{-})\,.
\end{array}
$$
Thus, $W$ can be expressed as
\begin{align*}
W  = & \sum_{e \in G_k} 2 R_e \left(  \frac{ a_{1} }{\sigma_{w}^{\mB}} h(e^{+}) h(e^{-}) + \left( \frac{ a_{2} \sqrt{T} }{\sigma_{\diff}^{\mB}}-\frac{a_{1}(p-q)}{\sigma_{w}^{\mB}(N-2)} \right) \big( h(e^{+}) + h(e^{-}) \big) \right)\\
& + \sum_{i =1 }^N \frac{(a_{3} - a_{2}  \sqrt{T-1}) h(i)}{\sigma^{\mB}} \\
 = & \sum_{e \in G_k}   \frac{ 2 R_e  a_{1} }{\sigma_{w}^{\mB}} h(e^{+}) h(e^{-}) + \left( \frac{ a_{2} \sqrt{T} }{\sigma_{\diff}^{\mB}} - \frac{a_{1}(p-q)}{\sigma_{w}^{\mB}(N-2)} \right) \sum_{i =1}^N 2 R_{i\cdot} h(i)  \\
 & + \sum_{i =1}^N \frac{(a_{3} - a_{2}  \sqrt{T-1}) h(i)}{\sigma^{\mB}} \\
 = & \sum_{e \in G_k}     \frac{2 R_e a_{1} }{\sigma_{w}^{\mB}} h(e^{+}) h(e^{-}) \\
 & + \sum_{i =1}^N \left( \frac{ a_{2} }{\sqrt{p q N(r_1^2 - r_0^2)}}  \left( \frac{ R_{i \cdot} }{N-1} - r_0 \right) - \frac{2 a_{1}(p-q) R_{i \cdot} }{\sigma_{w}^{\mB}(N-2)} +  \frac{ a_{3} }{ \sqrt{p q N } } \right)   h(i) \, ,
\end{align*}
where $R_{i \cdot} = (N-1) \bar R_{i \cdot}$. Let 
$$b_0 = \frac{2 a_1}{\sigma_{w}^{\mB}},  \quad b_i = \frac{ a_{2}  \big(  \bar R_{i \cdot} - r_0 \big) }{\sqrt{p q N(r_1^2 - r_0^2)}}  - \frac{2 a_{1}(p-q) R_{i \cdot} }{\sigma_{w}^{\mB}(N-2)} +  \frac{ a_{3} }{ \sqrt{p q N } } \text{ for } i \in \cN $$
\begin{equation*}
\text{ and }    \xi_{e}=   b_0 R_e h(e^{+}) h(e^{-}) \, , \quad   \xi_{i} = b_{i} h(i)\, .
\end{equation*}
We then have
\begin{equation*}
    W=\sum_{e \in G_k} \xi_{e}+\sum_{i = 1}^N \xi_{i} \, .
\end{equation*}
Plugging in the expressions of $\sigma_{w}^{\mB}$, $\sigma_{\diff}^{\mB}$, $\sigma^{\mB}$, and by
$$R_{i \cdot} ^2 = \sum_{j=1}^N \sum_{l=1}^N R_{ij} R_{il} \leq \frac{1}{2} \sum_{j=1}^N \sum_{l=1}^N \big( R_{ij}^2 + R_{il}^2 \big) = N \sum_{j=1}^N R_{ij}^2  \leq N^2(N-1)r_d^2\, ,$$
we have
$$ \frac{R_{i \cdot} }{\sigma_{w}^{\mB}(N-2)} \precsim \frac{1 }{ \sqrt{ N } }$$
and
$$|b_{0}| \precsim \frac{1}{\sqrt{N^2 r_d^2}}, \quad |b_{i}| \precsim  \frac{  \big|  \bar R_{i \cdot}  - r_0 \big| }{\sqrt{N(r_1^2 - r_0^2)}}   + \frac{1}{\sqrt{N}}\, .$$
Denote $c_0 =  1/{\sqrt{N^2 r_d^2}}$ and $c_i = |\bar R_{i\cdot}  - r_0|/{ \sqrt{N(r_1^2 - r_0^2) }}   + 1/{\sqrt{N}}$, for $i \in \cN$. 
Next, we apply Theorem \ref{theorem:stein} to $\widetilde{W}={W}/{\sqrt{\Vb(W)}}$. 

We now define some notations on the graph $G_k$. Let $G_{k i}$ be the set of edges with one endpoint vertex $i$, $G_{i, 2}$ be the set of edges with at least one endpoint in $G_{ki}$. Besides, we use $\node_{G_{ki}}$ to denote the vertex set connecting by edges in $G_{ki}$ excluding the vertex $i$ and $\node_{G_{i,2}}$ to denote the vertex set connecting by edges in $G_{i,2}$ excluding the vertex $i$. 
For each edge $e=(i, j) \in G_k$, we define $A_{e} = G_{k i} \cup G_{k j}, B_{e} = G_{i,2} \cup G_{j,2}$ and $C_e$ to be the set of edges that share at least one common vertex with an edge in $B_e$. 

Let $\mathcal{J}=G_k \cup \cN$, $K_{e} = A_{e} \cup \{ e^{+}, e^{-} \}$ for each edge $e=(e^{+}, e^{-}) \in G_k$ and $K_{i}=G_{ki} \cup\{i\}$ for each vertex $i \in \cN$. These $K_{e}$'s, $K_{i}$'s obviously satisfy the assumptions in Theorem \ref{theorem:stein} under the bootstrap null distribution. Then, we define $\eta_{e}$'s, $\eta_{i}$'s as follows: 
$$
\eta_{e}=\xi_{e^{+}} + \xi_{e^{-}} + \sum_{e \in A_{e}} \xi_{e}, \text{ for each edge } e \in G_k, \text{ and }
$$
$$
\eta_{i}=\xi_{i}+\sum_{e \in G_{k i}} \xi_{e}, \text{ for each node } i \in \cN. 
$$
By Theorem \ref{theorem:stein}, we have
\begin{equation}
\begin{aligned}
    & \sup _{h \in {\rm Lip}(1)}\big|\Eb h(\widetilde{W})-\Eb h(Z)\big| 
    \\ & \leq  \sqrt{\frac{2}{\pi}} \frac{1}{\Vb(W)} \Eb\Big|\sum_{i  = 1}^N \big\{\xi_{i} \eta_{i}-\Eb(\xi_{i} \eta_{i})\big \}+\sum_{e \in G_k }\big\{\xi_{e} \eta_{e}-\Eb(\xi_{e} \eta_{e})\big\}\Big| \\
    & +\frac{1}{\Vb^{\frac{3}{2}}(W)}\Big(\sum_{i = 1}^N \Eb\big|\xi_{i} \eta_{i}^{2}\big|+\sum_{e \in G_k} \Eb\big|\xi_{e} \eta_{e}^{2}\big|\Big)\, .
\end{aligned}    
\label{eq: Ste}
\end{equation}
Our next goal is to find some conditions under which the right hand side (RHS) of inequality \eqref{eq: Ste} can go to zero. Since the limit of $\Vb(W)$ is bounded above zero when $a_{1}, a_{2}, a_{3}$ are not all zeros, the RHS of inequality \eqref{eq: Ste} goes to zero if the following three terms
\begin{enumerate}
    \item[(A1)] $\Eb\Big|\sum_{i = 1}^N \big(\xi_{i} \eta_{i}-\Eb(\xi_{i} \eta_{i})\big)+\sum_{e \in G_k }\big(\xi_{e} \eta_{e}-\Eb(\xi_{e} \eta_{e})\big)\Big|$ \,,
    \label{A1}
    \item[(A2)]  $\sum_{i = 1}^N \Eb|\xi_{i} \eta_{i}^{2}|$\,,
    \item[(A3)] $\sum_{e \in G_k} \Eb|\xi_{e} \eta_{e}^{2}|$
\end{enumerate}
go to zero. For (A1), we have 
\begin{align*}
& \Eb\Big|\sum_{i = 1}^N \big( \xi_{i} \eta_{i}-\Eb(\xi_{i} \eta_{i})\big) + \sum_{e \in G_k }\big(\xi_{e} \eta_{e}-\Eb(\xi_{e} \eta_{e})\big)\Big| \\
\leq &  \Eb\Big|\sum_{i=1}^N\big\{\xi_{i} \eta_{i}-\Eb(\xi_{i} \eta_{i})\big\}\Big|+\Eb\big|\sum_{e \in G_k} \big( \xi_{e} \eta_{e}-\Eb(\xi_{e} \eta_{e}) \big) \big| \\
\leq &   \sqrt{\sum_{i=1 }^N \Vb\big(\xi_{i} \eta_{i}\big)+\sum_{i, j}^{i \neq j} \Covb\big(\xi_{i} \eta_{i}, \xi_{j} \eta_{j}\big)} \\
& +\sqrt{\sum_{e \in G_k} \Vb\big(\xi_{e} \eta_{e}\big)+\sum_{e, f}^{e \neq f} \Covb\big(\xi_{e} \eta_{e}, \xi_{f} \eta_{f}\big)} \\
= & \sqrt{\sum_{i = 1}^N \Vb\big(\xi_{i} \eta_{i}\big) + \sum_{i=1}^{N} \sum_{j \in \node_{G_{i, 2}} } \Covb\big(\xi_{i} \eta_{i}, \xi_{j} \eta_{j}\big)} \\
&+\sqrt{\sum_{e \in G_k} \Vb(\xi_{e} \eta_{e})+\sum_{e \in G_k} \sum_{f \in C_{e} \backslash\{e\}} \Covb(\xi_{e} \eta_{e}, \xi_{f} \eta_{f}) } \, . 
\end{align*}
The last equality holds as $\xi_{i} \eta_{i}$ and $\big\{\xi_{j} \eta_{j}\big\}_{j \notin \node_{G_{i, 2}}}$ are uncorrelated under the bootstrap null distribution, and $\xi_{e} \eta_{e}$ and $\left\{\xi_{f} \eta_{f}\right\}_{f \notin C_{e}}$ are uncorrelated under the bootstrap null distribution. The covariance part of the edges is a bit complicated to handle directly, so we decompose it into three parts as follows based on the relationship of $e$ and $f$:
\begin{align*}
\sum_{e \in G_k} \sum_{f \in C_{e} \backslash\{e\}} \Covb\big(\xi_{e} \eta_{e}, \xi_{f} \eta_{f}\big)=& \sum_{e \in G_k} \sum_{f \in A_{e} \backslash\{e\}} \Covb\big(\xi_{e} \eta_{e}, \xi_{f} \eta_{f}\big) \\
&+\sum_{e \in G_k} \sum_{f \in B_{e} \backslash A_{e}} \Covb\big(\xi_{e} \eta_{e}, \xi_{f} \eta_{f}\big)
\\ & +\sum_{e \in G_k} \sum_{f \in C_{e} \backslash B_{e}} \Covb\big(\xi_{e} \eta_{e}, \xi_{f} \eta_{f}\big)\, .
\end{align*}
With carefully examining these quantities, we can show the following inequalities \eqref{eq:8}-\eqref{eq:15}. The details of obtaining \eqref{eq:8}-\eqref{eq:15} are provided in Section \ref{sec: ineq}.
\begin{equation}
\sum_{i=1}^{N} \Vb( \xi_{i} \eta_{i} ) \precsim \sum_{i=1}^{N} c_{i}^{4} + c_{0}^{2} \sum_{i=1}^{N} c_{i}^{2} \sum_{j=1}^N R^2_{i j} \,.
\label{eq:8}
\end{equation}

\begin{equation}
\sum_{e \in G_k} \Vb\big(\xi_{e} \eta_{e}\big)  \precsim   c_0^2 \sum_{i=1}^N c_i^2  \sum_{j=1}^N R_{ij}^2  + c_0^3 \sum_{i=1}^N c_i \sum_{j=1}^N R_{ij}^3 + c_0^4  \sum_{i=1}^N \big( \sum_{j=1}^N R_{ij}^2 \big)^2 \, .
\label{eq:9}
\end{equation}

\begin{equation}
\begin{aligned}
 \sum_{i=1}^{N} \sum_{j \in \node_{G_{i, 2}} } \Covb\big(\xi_{i} \eta_{i}, \xi_{j} \eta_{j}\big) & \precsim  \sum_{i=1}^{N} \sum_{j \in \node_{G_{ki}}} \big( c_0 c_i c_j R_{i j} (c_i + c_j) + c_0^2 c_i c_j R_{ij}^2 \big) \\
 & + c_0^2 \Big| \sum_{i=1}^{N}  \sum_{j \in \node_{G_{i, 2}} } b_i b_j \sum_{k=1}^N R_{ik} R_{j k} \Big| \, .
\end{aligned}
\label{eq:10}
\end{equation}

\begin{equation}
\begin{aligned}
& \sum_{e \in G_k} \sum_{f \in A_{e} \backslash\{e\}} \Covb(\xi_{e} \eta_{e}, \xi_{f} \eta_{f}) \\
& \precsim c_0^3  \sum_{i=1}^N \sum_{j,l \in \node_{G_{ki}}}^{j \neq l}  R_{ji} R_{il} \Big(  c_j  \big( R_{jl} + R_{il} \big) +   c_l  \big(R_{ji} +  R_{jl}  \big) +  c_i R_{jl} \Big)\\
& +  c_0^4  \sum_{i=1}^N \sum_{j,l \in \node_{G_{ki}}}^{j \neq l}  R_{ji} R_{il} \Big( R_{j l} \big( R_{ji} + R_{jl} +  R_{il} \big) + \sum_{s=1}^N R_{j s} R_{l s}  \Big) \\
& + c_0^2 \Big| \sum_{i=1}^N \sum_{j,l \in \node_{G_{ki}}}^{j \neq l}  R_{ji} R_{il}  b_j b_l \Big| \, .
\end{aligned}
\label{eq:11}
\end{equation}

\begin{equation}
\begin{aligned}
\sum_{e \in G_k} \sum_{f \in B_{e} \backslash A_{e}} \Covb(\xi_{e} \eta_{e}, \xi_{f} \eta_{f}) \precsim   c_0^4  \sum_{i=1}^N \sum_{j=1}^N  \sum_{l\neq i,j}^N \sum_{s \neq i,j}^N R_{ij} R_{jl}  R_{ls}R_{si} \,.
\end{aligned}
\label{eq:12}
\end{equation}

\begin{equation}
\sum_{e \in G_k} \sum_{f \in C_{e} \backslash B_{e}} \Covb(\xi_{e} \eta_{e}, \xi_{f} \eta_{f}) = 0 \,.
\label{eq:13}
\end{equation}

\begin{equation}
\sum_{i=1}^{N}  \Eb \big( |\xi_{i} \eta_{i}^{2}| \big) \precsim  \sum_{i=1}^{N} c_i^3 + c_0^2 \sum_{i=1}^{N} c_i \sum_{j=1}^N R_{ij}^2 \, .
\label{eq:14}
\end{equation}

\begin{equation}
\begin{aligned}
 \sum_{e \in G_k} \Eb\big( |\xi_e| \eta_e^2 \big) \precsim c_0^3 \sum_{i=1}^N \sum_{j=1}^N   R_{ij}^3 + c_0 \sum_{i=1}^N c_i^2  R_{i \cdot} +  c_0^3  \sum_{i=1}^N R_{i \cdot} \sum_{j=1}^N R_{ij}^2 \, .
\end{aligned}
\label{eq:15}
\end{equation}
Based on facts that $c_{i} \precsim 1$ for all $i$'s, (A1), (A2) and (A3) go to zero as long as the following conditions hold: 
\begin{equation}
    \sum_{i=1}^{N} c_{i}^{3} \rightarrow 0 \,,
\label{eq:18}
\end{equation}
\begin{equation}
    c_{0}^{2} \sum_{i=1}^{N} c_{i} \sum_{j=1}^N R^2_{i j} \rightarrow 0 \,,
\label{eq:19}
\end{equation}
\begin{equation}
 c_0^3 \sum_{i=1}^N  \sum_{j=1}^N R_{ij}^3 \rightarrow 0 \,,
\label{eq:20}
\end{equation}
\begin{equation}
 c_0^4  \sum_{i=1}^N \big( \sum_{j=1}^N R_{ij}^2 \big)^2  \rightarrow 0 \,,
\label{eq:21}
\end{equation}
\begin{equation}
  c_0 \sum_{i=1}^N c_i^2  R_{i \cdot}  \rightarrow 0 \,,
\label{eq:22}
\end{equation}
\begin{equation}
\sum_{i=1}^{N} \sum_{j \in \node_{G_{ki}}} \big( c_0 c_i c_j R_{i j} (c_i + c_j) + c_0^2 c_i c_j R_{ij}^2 \big) \rightarrow 0 \,,
\label{eq:23}
\end{equation}
\begin{equation}
c_0^2  \sum_{i=1}^{N}  \sum_{j \in \node_{G_{i, 2}} } b_i b_j \sum_{l=1}^N R_{il} R_{j l} \rightarrow 0 \,,
\label{eq:24}
\end{equation}
\begin{equation}
c_0^3  \sum_{i=1}^N \sum_{j,l \in \node_{G_{ki}}}^{j \neq l}  R_{ji} R_{il} \big(  c_j  ( R_{jl} + R_{il} ) +   c_l (R_{ji} +  R_{jl}) +  c_i R_{jl}  \big) \rightarrow 0 \,,
\label{eq:25}
\end{equation}
\begin{equation}
c_0^2  \sum_{i=1}^N \sum_{j,l \in \node_{G_{ki}}}^{j \neq l}  R_{ji} R_{il}  b_j b_l  \rightarrow 0 \,,
\label{eq:26}
\end{equation}
\begin{equation}
c_0^4  \sum_{i=1}^N \sum_{j,l \in \node_{G_{ki}}}^{j \neq l}  R_{ji} R_{il} \big(R_{j l} ( R_{ji} + R_{jl} +  R_{il} ) + \sum_{s=1}^N R_{j s} R_{l s}  \big) \rightarrow 0 \,,
\label{eq:27}
\end{equation}
\begin{equation}
   c_0^4  \sum_{i=1}^N \sum_{j=1}^N  \sum_{l\neq i,j}^N \sum_{s \neq i,j}^N R_{ij} R_{jl} R_{ls} R_{si}  \rightarrow 0 \,,
\label{eq:29}
\end{equation}
\begin{equation}
     c_0^3  \sum_{i=1}^N R_{i \cdot} \sum_{j=1}^N R_{ij}^2 \rightarrow 0 \,.
\label{eq:28}
\end{equation}
Next, we show that the conditions in Theorem 3.1 
can ensure \eqref{eq:18}-\eqref{eq:28}. For Condition \eqref{eq:18}, we have
$$\sum_{i=1}^{N} c_{i}^{3} = \sum_{i=1}^{N} \Big( \frac{ \big|\bar R_{i\cdot}  - r_0 \big|  }{ \sqrt{N(r_1^2 - r_0^2) }}  + \frac{1}{\sqrt{N}} \Big)^3 \precsim   \frac{\sum_{i=1}^{N} \big|\bar R_{i\cdot}  - r_0 \big|^3  }{ \big( N(r_1^2 - r_0^2) \big)^{1.5}} + \frac{1}{\sqrt{N}} \, ,$$
so Condition \eqref{eq:18} holds when ${\sum_{i=1}^{N} |\bar R_{i\cdot}  - r_0 |^3  }/{ ( NV_r )^{1.5}} \rightarrow 0$. For Condition \eqref{eq:19}, we have
$$
 c_{0}^{2} \sum_{i=1}^{N} c_{i} \sum_{j=1}^N R^2_{i j} = \frac{1}{N^2 r_d^2}  \sum_{j=1}^N R^2_{i j} \Big( \frac{ \big|\bar R_{i\cdot}  - r_0 \big|  }{ \sqrt{N(r_1^2 - r_0^2) }}  + \frac{1}{\sqrt{N}} \Big) \leq \max_{i \in \cN} \Big( \frac{ \big|\bar R_{i\cdot}  - r_0 \big|  }{ \sqrt{N(r_1^2 - r_0^2) }}  + \frac{1}{\sqrt{N}} \Big)
$$
by $\sum_{i=1}^N \sum_{j=1}^N R^2_{i j} = N(N-1) r_d^2$. Then by Theorem 1 in \cite{hoeffding1951combinatorial} with $r$ taking $3$, we have ${ \max_{i \in \cN} |\bar R_{i\cdot}  - r_0 |  }/{ \sqrt{NV_r }} \rightarrow 0$ when  ${\sum_{i=1}^{N} |\bar R_{i\cdot}  - r_0 |^3  }/{ ( NV_r )^{1.5}} \rightarrow 0$. Condition \eqref{eq:20} holds trivially as 
$$ c_0^3 \sum_{i=1}^N  \sum_{j=1}^N R_{ij}^3 \leq  \frac{N(N-1) r_d^2 K}{ N^3 r_d^3} \leq \frac{K}{ \sqrt{N^2 r_d^2} }  \rightarrow 0 \,.$$
Condition \eqref{eq:21} is equivalent to $ \sum_{i=1}^N \big( \sum_{j=1}^N R_{ij}^2 \big)^2 = o\big( N^4 r_d^4 \big)$. 
For Condition \eqref{eq:22}, we have
\begin{align*}
c_0  \sum_{i=1}^{N} c_i^2 R_{i \cdot} & = \frac{1}{N r_d}  \sum_{i=1}^N R_{i \cdot} \Big( \frac{ \big|\bar R_{i\cdot}  - r_0 \big|  }{ \sqrt{N(r_1^2 - r_0^2) }}  + \frac{1}{\sqrt{N}} \Big)^2 \\
& \precsim \frac{1}{N r_d}  \sum_{i=1}^N R_{i \cdot} \frac{ \big(\bar R_{i\cdot}  - r_0 \big)^2  }{ N(r_1^2 - r_0^2)} 
+ \frac{(N-1) r_0}{N r_d} \\
& = \frac{N-1}{N r_d}  \sum_{i=1}^N \frac{ \big(\bar R_{i\cdot}  - r_0 \big)^3  }{ N(r_1^2 - r_0^2)}  + \frac{2 (N-1) r_0}{N r_d} \, ,
\end{align*}
which goes to zero under the condition ${\sum_{i=1}^N  (\bar R_{i\cdot}  - r_0 )^3  } = o ({ N r_d V_r})$ and $r_0 = o(r_d)$. For Condition \eqref{eq:23}, it is easy to see that $$\sum_{i=1}^{N} \sum_{j \in \node_{G_{ki}}}  c_0 c_i^2 c_j R_{i j} = \sum_{i=1}^{N} \sum_{j \in \node_{G_{ki}}}  c_0 c_i c_j^2 R_{i j} \, .$$
Then by $c_i \precsim 1$, we have
$$\sum_{i=1}^{N} \sum_{j \in \node_{G_{ki}}}  c_0 c_i^2 c_j R_{i j} \precsim \sum_{i=1}^{N} \sum_{j \in \node_{G_{ki}}}  c_0 c_i^2 R_{i j} = c_0  \sum_{i=1}^{N} c_i^2 R_{i \cdot} \,,$$
$$\sum_{i=1}^{N} \sum_{j \in \node_{G_{ki}}}  c_0^2 c_i c_j R_{ij}^2 \precsim \sum_{i=1}^{N} \sum_{j \in \node_{G_{ki}}}  c_0^2 c_i R_{ij}^2 =  c_0^2 \sum_{i=1}^{N}   c_i\sum_{j = 1}^N  R_{ij}^2\,,$$
where both the right hand sides go to zero from \eqref{eq:19} and \eqref{eq:22}. For Condition \eqref{eq:24}, we have
$$
\begin{aligned}
c_0^2 \sum_{i=1}^{N}  \sum_{j \in \node_{G_{i, 2}} } b_i b_j \sum_{l=1}^N R_{il} R_{j l}  =&  \sum_{l=1}^{N} \sum_{i \in \node_{G_{kl}}} \sum_{j \in \node_{G_{kl}}\backslash \{i\} } b_i b_j  R_{i l} R_{j l} \\
= & \sum_{l=1}^{N} \sum_{i,j \in \node_{G_{kl}}}^{i \neq j}  b_i b_j  R_{il} R_{j l} \,,
\end{aligned}
$$
which is the same as the condition \eqref{eq:26}. For Condition \eqref{eq:25}, it is easy to see that $$\sum_{i=1}^N \sum_{j,l \in \node_{G_{ki}}}^{j \neq l}  R_{j i} R_{i l}   c_j ( R_{j l} + R_{il} ) = \sum_{i=1}^N \sum_{j,l \in \node_{G_{ki}}}^{j \neq l}  R_{ji} R_{il}   c_l   (R_{ji} +  R_{jl} ) \, ,$$
which means that we only need to deal with the two parts 
$c_0^3 \sum_{i=1}^N \sum_{j,l \in \node_{G_{ki}}}^{j \neq l}  R_{ji} R_{il}   c_j  ( R_{jl} + R_{il} )$ and $c_0^3  \sum_{i=1}^N \sum_{j,l \in \node_{G_{ki}}}^{j \neq l}  R_{ji} R_{il} c_i R_{jl}$. We have
$$
\begin{aligned}
& c_0^3 \sum_{i=1}^N \sum_{j,l \in \node_{G_{k i}}}^{j \neq l}  R_{ji} R_{il}   c_j  ( R_{jl} + R_{il} ) = c_0^3 \sum_{i=1}^N \sum_{j=1}^N \sum_{l \neq j}^N  c_j R_{ji} R_{il}  ( R_{jl} + R_{il}) \\
& \leq   c_0^3 \sum_{i=1}^N \sum_{j=1}^N \sum_{l = 1}^N  c_j R_{ji} \big( R_{il}^2 + R_{jl}^2\big)  + c_0^3  \sum_{i=1}^N R_{i \cdot} \sum_{j=1}^N R_{ij}^2 
\precsim  c_0^3 \sum_{i=1}^N R_{i \cdot} \sum_{j=1}^N R_{ij}^2 \,,
\end{aligned}
$$
$$
\begin{aligned}
& c_0^3 \sum_{i=1}^N \sum_{j,l \in \node_{G_{ki}}}^{j \neq l}  R_{ji} R_{il} c_i R_{jl}   = c_0^3 \sum_{i=1}^N \sum_{j=1}^N \sum_{l=1}^N  c_i R_{ij} R_{il} R_{jl} \\
& \leq c_0^3 \sum_{i=1}^N \sum_{j=1}^N \sum_{l=1}^N  c_i R_{ij} ( R_{il}^2 + R_{jl}^2) \precsim c_0^3 \sum_{i=1}^N R_{i \cdot} \sum_{j=1}^N R_{ij}^2 \, ,
\end{aligned}
$$
and $c_0^3  \sum_{i=1}^N R_{i \cdot} \sum_{j=1}^N R_{ij}^2$ is bounded by  \eqref{eq:28}. For Condition \eqref{eq:26}, first we have 
\begin{align*}
 b_j b_l  = & \Big( \frac{ a_{2}   \widetilde R_{j \cdot}  }{\sqrt{p q NV_r}}  - \frac{2 a_{1}(p-q) R_{j \cdot} }{\sigma_{w}^{\mB}(N-2)} +  \frac{ a_{3} }{ \sqrt{p q N }} \Big)  \Big(\frac{ a_{2} \widetilde R_{l \cdot} }{\sqrt{p q NV_r}}  - \frac{2 a_{1}(p-q) R_{l \cdot} }{\sigma_{w}^{\mB}(N-2)} +  \frac{ a_{3} }{ \sqrt{p q N }}\Big) \\
 = & \frac{ a_{2}^2  \widetilde R_{j \cdot} \widetilde R_{l \cdot}}{p q NV_r}  +  \frac{ a_{2}   \widetilde R_{j \cdot} }{\sqrt{p q NV_r}} \Big( \frac{ a_{3} }{ \sqrt{p q N }} -  \frac{2 a_{1}(p-q) R_{l \cdot} }{\sigma_{w}^{\mB}(N-2)} \Big)  + \frac{ a_{2}   \widetilde R_{l \cdot} }{\sqrt{p q NV_r}} \Big( \frac{ a_{3} }{ \sqrt{p q N }} -  \frac{2 a_{1}(p-q) R_{j \cdot} }{\sigma_{w}^{\mB}(N-2)} \Big) \\
& + \Big( \frac{ a_{3} }{ \sqrt{p q N }} -  \frac{2 a_{1}(p-q) R_{j \cdot} }{\sigma_{w}^{\mB}(N-2)} \Big)\Big( \frac{ a_{3} }{ \sqrt{p q N }} -  \frac{2 a_{1}(p-q) R_{l \cdot} }{\sigma_{w}^{\mB}(N-2)} \Big) 
\end{align*}
and
\begin{align*}
& \sum_{i=1}^N \sum_{j,l \in \node_{G_{ki}}}^{j \neq l} \frac{  R_{ji} R_{il} |  \widetilde R_{j \cdot}| }{\sqrt{ N^2 V_r}}  \leq   \sum_{i=1}^N \sum_{j = 1}^{N} \sum_{l = 1}^{N} \frac{  R_{ji} R_{il} |  \widetilde R_{j \cdot} | }{\sqrt{ N^2 V_r}} \\
 & =   \sum_{i=1}^N \sum_{j = 1}^{N}  \frac{  R_{ji} R_{i\cdot} |   \widetilde R_{j \cdot}| }{\sqrt{ N^2 V_r}}   \leq \sum_{i=1}^N  \frac{  R_{i\cdot} \sqrt{ \sum_{j  = 1}^{N}  R_{ji}^2 \sum_{j = 1}^{N}  \widetilde R_{j \cdot}^2} }{\sqrt{ N^2 V_r}}  \\
 & =  \frac{ \sum_{i=1}^N  R_{i\cdot}  \sqrt{ \sum_{j = 1}^{N}  R_{ji}^2} }{\sqrt{ N}}\leq  \frac{ \sqrt{ \sum_{i=1}^N  R_{i\cdot}^2  \sum_{i=1}^N \sum_{j = 1}^{N}  R_{ji}^2} }{\sqrt{ N}} \precsim \sqrt{N^4 r_1^2 r_d^2}\,.
\end{align*}
Then 
$$
\begin{aligned}
&| c_0^2  \sum_{i=1}^N \sum_{j,l \in \node_{G_{ki}}}^{j \neq l}  R_{ji} R_{il}  b_j b_l |  \\
& \precsim  \Big| c_0^2  \sum_{i=1}^N \sum_{j,l \in \node_{G_{ki}}}^{j \neq l}  R_{ji} R_{il}   \frac{\widetilde R_{j \cdot} \widetilde  R_{l \cdot}  }{N V_r} \Big|  +   c_0^2  \sum_{i=1}^N \sum_{j,l \in \node_{G_{ki}}}^{j \neq l} \frac{  R_{ji} R_{il}|  \widetilde R_{j \cdot} | }{\sqrt{ N^2 V_r }} +  c_0^2  \sum_{i=1}^N \sum_{j,l \in \node_{G_{ki}}}^{j \neq l} \frac{ R_{ji} R_{il}}{N} \\
& \precsim \frac{\Big| \sum_{i=1}^N \sum_{j,l \in \node_{G_{ki}}}^{j \neq l}  R_{ji} R_{il} \widetilde R_{j \cdot} \widetilde R_{l \cdot} \Big|}{N^3 r_d^2 V_r }+  \frac{  \sqrt{N^4 r_1^2 r_d^2} }{ N^2 r_d^2 }  + \frac{ \sum_{i=1}^N R_{i\cdot}^2 }{N^3 r_d^2} \\
& \precsim  \frac{\Big| \sum_{i=1}^N \sum_{j,l \in \node_{G_{ki}}}^{j \neq l}  R_{ji} R_{il} \widetilde R_{j \cdot} \widetilde R_{l \cdot}\Big|}{N^3 r_d^2 V_r } +  \frac{ r_1 }{r_d} + \frac{ r_1^2 }{r_d^2} \,,
\end{aligned}
$$
which goes to zero when $\Big| \sum_{i=1}^N \sum_{j,l \in \node_{G_{ki}}}^{j \neq l}  R_{ji} R_{il} \widetilde R_{j \cdot} \widetilde R_{l \cdot} \Big| = o( N^3 r_d^2 V_r)$ and $r_1 = o(r_d)$. For Condition \eqref{eq:27}, we have 
\begin{align*}
& c_0^4  \sum_{i=1}^N \sum_{j,l \in \node_{G_{ki}}}^{j \neq l}  R_{ji} R_{il} \big( R_{j l} ( R_{ji} + R_{jl} +  R_{il}) + \sum_{s=1}^N R_{j s} R_{l s}  \big) \\
& \precsim c_0^4  \sum_{i=1}^N \sum_{j=1}^N \sum_{l=1}^N R_{ij}^2 R_{il}R_{jl} + c_0^4  \sum_{i=1}^N \sum_{j=1}^N  \sum_{l\neq i,j}^N  \sum_{s\neq i,j}^N R_{ji} R_{il}  R_{j s} R_{l s}  + c_0^4  \sum_{i=1}^N \sum_{j=1}^N  \sum_{l\neq i,j}^N  R_{ji}^2 R_{il}^2 \\
& \precsim  \frac{\sum_{i=1}^N \sum_{j=1}^N \sum_{l=1}^N R_{ij}^2( R_{il}^2 + R_{jl}^2)}{N^4 r_d^4} \\
& + \frac{\sum_{i=1}^N \sum_{j=1}^N  \sum_{l\neq i,j}^N  \sum_{s\neq i,j}^N R_{ji} R_{il}  R_{j s} R_{l s}}{N^4 r_d^4} +\frac{\sum_{i=1}^N\big(  \sum_{j=1}^N R_{ij}^2 \big)^2}{N^4 r_d^4} \\
& \precsim \frac{ \sum_{i=1}^N \big( \sum_{j=1}^N R_{ij}^2 \big)^2}{N^4 r_d^4}  + \frac{\sum_{i=1}^N \sum_{j=1}^N  \sum_{l\neq i,j}^N  \sum_{s\neq i,j}^N R_{ji} R_{il}  R_{j s} R_{l s}}{N^4 r_d^4} \,,
\end{align*}
where the first term goes to zero when $\sum_{i=1}^N \big( \sum_{j=1}^N R_{ij}^2 \big)^2 =o \big( N^4 r_d^4 \big)$ and the second term is the same as the condition  \eqref{eq:29}. 
The condition \eqref{eq:29} holds when $$\sum_{i=1}^N \sum_{j=1}^N  \sum_{k\neq i,j}^N \sum_{l \neq i,j}^N R_{ij} R_{kl}  \big( R_{ik}R_{jl} + R_{il}R_{jk} \big) = o(N^4 r_d^4) \, .$$ 
For Condition \eqref{eq:28}, we have
$$
\begin{aligned}
c_0^3  \sum_{i=1}^N R_{i \cdot} \sum_{j=1}^N R_{ij}^2 & \leq c_0^3 \sqrt{ \sum_{i=1}^N R_{i \cdot}^2 \sum_{i=1}^N \big(\sum_{j=1}^N R_{ij}^2\big)^2  } \\
& = \frac{\sqrt{N^3 r_1^2 \sum_{i=1}^N \big(\sum_{j=1}^N R_{ij}^2 \big)^2 }}{N^3 r_d^3 } = \frac{r_1}{r_d} \sqrt{ \frac{ \sum_{i=1}^N \big(\sum_{j=1}^N R_{ij}^2 \big)^2 }{N^3 r_d^4 }} \,, 
\end{aligned}
$$
which goes to zero when $r_1 = o(r_d)$ and $ \sum_{i=1}^N \big(\sum_{j=1}^N R_{ij}^2 \big)^2 \precsim N^3 r_d^4$.

\subsection{Proof of Inequalities (\ref{eq:8})-(\ref{eq:15})}

\label{sec: ineq}

\subsubsection{Proof of (\ref{eq:8}) }

For each node $i$, we have
\begin{align*}
\Vb(\xi_{i} \eta_{i}) = & \Vb\Big( \xi_{i} \big(\xi_{i}+\sum_{e \in G_{ki}} \xi_{e}\big) \Big) = \Vb\Big( h(i)^{2}\big( b_{i}^{2}  + b_{0} b_{i} \sum_{j \in \node_{G_{ki}}} R_{i j} h(j) \big) \Big) \\
= &\Eb \big( h(i)^{4} \big) \Eb \Big( \big( b_{i}^{2} + b_{0} b_{i} \sum_{j \in \node_{G_{ki}}}  R_{ij} h(j) \big)^{2} \Big) - \Big( \Eb \big( h(i)^2 b_{i}^{2} \big) \Big)^2\\
 = & ( p q^4 + q p^4) \Eb \Big(b_i^4 + 2 b_i^3 b_0 \sum_{j \in \node_{G_{ki}}} R_{i j} h(j) + b_i^2 b_0^2 \big(\sum_{j \in \node_{G_{ki}}} R_{ij} h(j)\big)^2  \Big) \\
& -  b_{i}^{4} p^{2} q^{2} \\
= &p q (p^3 + q^3 -p q) b_i^4 + p^2 q^2 (p^3 + q^3) b_i^2 b_0^2 \sum_{j \in \node_{G_{ki}}} R^2_{i j}
\end{align*}
Thus,
$$
\sum_{i=1}^{N} \Vb( \xi_{i} \eta_{i} ) \precsim \sum_{i=1}^{N} c_{i}^{4} + c_{0}^{2} \sum_{i=1}^{N} c_{i}^{2} \sum_{j=1}^N R^2_{i j} \,.
$$

\subsubsection{Proof of (\ref{eq:9}) }

For each edge $e=\big(i, j\big) \in G_k$,  we have
$$
\begin{aligned}
\xi_{e} \eta_{e} = & b_{0} R_{ij} h (i) h(j) \big( b_i h(i) + b_j h(j) \big) + b_{0}^{2} R_{ij}^2 h(i)^{2} h(j)^{2} \\
& + b_0^2 R_{ij} h(i)^{2} h(j) \sum_{l \in \node_{G_{ki}} \backslash\{ j \} }R_{i l} h(l) + b_0^2 R_{ij} h(i) h(j)^2 \sum_{l \in \node_{G_{kj}} \backslash\{ i\} }R_{lj} h(l) \,.
\end{aligned}
$$
Then we have $\Eb(\xi_{e} \eta_{e}) = b_{0}^{2} R_{ij}^2 p^2 q^2$ and 
\begin{align*}
\Eb(\xi_{e} \eta_{e})^2  -  b_{0}^{4} R_{ij}^4 p^4 q^4 & \leq b_0^2 R_{ij}^2(b_i^2  + b_j^2 ) + b_0^3( |b_i| + |b_j|) R_{ij}^3 \\
& + b_0^4 R_{ij}^2 \Big( \sum_{l \in \node_{G_{ki}} \backslash\{ j \} }R_{i l}^2 + \sum_{l \in \node_{G_{kj}} \backslash\{ i\} }R_{lj}^2 \Big) \\
& \precsim  c_0^2 R_{ij}^2(c_i^2  + c_j^2 ) + c_0^3( c_i + c_j) R_{ij}^3 \\
& + c_0^4 R_{ij}^2 \Big( \sum_{l \in \node_{G_{ki}}  }R_{i l}^2 + \sum_{l \in \node_{G_{kj}} }R_{lj}^2 \Big) \, .
\end{align*}
Thus, 
\begin{align*}
\sum_{e \in G_k} \Vb\big(\xi_{e} \eta_{e}\big)  \precsim &  \sum_{i=1}^N \sum_{j=1}^N  \Big( c_0^2 R_{ij}^2(c_i^2  + c_j^2 ) + c_0^3( c_i + c_j) R_{ij}^3 \\
& + c_0^4 R_{ij}^2 \big( \sum_{l \in \node_{G_{ki}}  }R_{i l}^2 + \sum_{l \in \node_{G_{kj}} }R_{lj}^2 \big) \Big) \\
\precsim &   \sum_{i=1}^N \sum_{j=1}^N \Big(   c_0^2 R_{ij}^2(c_i^2  + c_j^2 ) + c_0^3( c_i + c_j) R_{ij}^3 + c_0^4 R_{ij}^2 \big( \sum_{l = 1}^NR_{i l}^2 + \sum_{l=1}^N R_{lj}^2 \big) \Big) \\
\precsim  & c_0^2 \sum_{i=1}^N c_i^2 \sum_{j=1}^N R_{ij}^2 + c_0^3 \sum_{i=1}^N c_i  \sum_{j=1}^N R_{ij}^3  + c_0^4  \sum_{i=1}^N \Big( \sum_{j=1}^N R_{ij}^2 \Big)^2 \, .
\end{align*}

\subsubsection{Proof of (\ref{eq:10}) }
We can further decompose \eqref{eq:10} as 
$$
\begin{aligned}
& \sum_{i=1}^{N} \sum_{j \in \node_{G_{i,2}} \backslash\{i\}} \Covb\big( \xi_{i} \eta_{i}, \xi_{j} \eta_{j} \big) \\
=& \sum_{i=1}^{N} \sum_{j \in \node_{G_{ki}}} \Covb\big(\xi_{i} \eta_{i}, \xi_{j} \eta_{j}\big) + \sum_{i=1}^{N} \sum_{j \in \node_{G_{i, 2}} \backslash \node_{G_{ki} }} \Covb\big(\xi_{i} \eta_{i}, \xi_{j} \eta_{j}\big) \, .
\end{aligned}
$$
For $j \in \node_{G_{i}}$ which means node $j$ connects to node $i$ directly, we have
\begin{align*}
& \Eb\big( \xi_{i} \eta_{i} \xi_{j} \eta_{j} \big)\\
= &  \Eb \Big( h(i)^2 h(j)^2\big( b_i^2 + b_0 b_i  \sum_{k_1 \in \node_{G_{ki}} } R_{i k_1} h(k_1) \big) \big( b_j^2 + b_0 b_j  \sum_{k_2 \in \node_{G_{kj}} } R_{j k_2} h(k_2) \big) \Big) \\
 = & \Eb \Big( h(i)^2 h(j)^2\big( b_i^2 + b_0 b_i R_{ij} h(j) \big) \big( b_j^2 + b_0 b_j R_{i j} h(i) \big) \Big) \\
& + \Eb \Big( b_0^2 b_i b_j h(i)^2 h(j)^2  \big( \sum_{k_1 \in \node_{G_{ki}}\backslash \{j\} } R_{i k_1} h(k_1) \big) \big( \sum_{k_2 \in \node_{G_{kj}}\backslash \{i\} } R_{j k_2} h(k_2) \big)  \Big)
\end{align*}
and
$$
\begin{aligned}
\Eb\big( \xi_{i} \eta_{i}\big)  \Eb\big(\xi_{j} \eta_{j} \big) = (b_i^2 p q) (b_j^2 p q) \, .
\end{aligned}
$$
Combining with $\Eb\big( h(i)^3 \big) = pq(q-p)$, we have 
$$
\begin{aligned}
 \Covb\big(\xi_{i} \eta_{i}, \xi_{j} \eta_{j}\big) =&  p^2 q^2 (q-p)  b_0 b_i b_j R_{i j} (b_i + b_j) \\
 & + p^2 q^2 (q-p)^2 b_0^2 b_i b_j R_{ij}^2 +  p^3 q^3 b_0^2 b_i b_j \sum_{l=1}^N R_{il} R_{j l} \, .
\end{aligned}
$$
Thus, we have
$$
\begin{aligned}
& \sum_{i=1}^{N} \sum_{j \in \node_{G_{k i}}} \Covb(\xi_{i} \eta_{i}, \xi_{j} \eta_{j}) -  p^3 q^3 b_0^2 \sum_{i=1}^{N} \sum_{j \in \node_{G_{ki}}} b_i b_j \sum_{l=1}^N R_{il} R_{j l} \\
& \precsim \sum_{i=1}^{N} \sum_{j \in \node_{G_{ki}} } \big( |b_0| |b_i| |b_j| R_{i j} (|b_i| + |b_j|) + b_0^2 |b_i| |b_j| R_{ij}^2 \big) \\
& \precsim \sum_{i=1}^{N} \sum_{j \in \node_{G_{ki}}} \big( c_0 c_i c_j R_{i j} (c_i + c_j) + c_0^2 c_i c_j R_{ij}^2 \big) \,.
\end{aligned}
$$
For $j \in \node_{G_{i, 2}} \backslash \node_{G_{ki}}$ which means node $j$ does not connect to node $i$ directly, we have
$$
\begin{aligned}
& \Eb\big( \xi_{i} \eta_{i} \xi_{j} \eta_{j} \big) \\
=&  \Eb \Big( h(i)^2 h(j)^2\big( b_i^2 + b_0 b_i  \sum_{k_1 \in \node_{G_{ki}} } R_{i k_1} h(k_1) \big) \big( b_j^2 + b_0 b_j  \sum_{k_2 \in \node_{G_{kj}} } R_{j k_2} h(k_2) \big) \Big) \\
= &  \Eb \Big( h(i)^2 h(j)^2 b_i^2  b_j^2  \Big) \\
& + \Eb \Big( b_0^2 b_i b_j h(i)^2 h(j)^2  \big( \sum_{k_1 \in \node_{G_{ki}}} R_{i k_1} h(k_1) \big) \big( \sum_{k_2 \in \node_{G_{kj}} } R_{j k_2} h(k_2) \big)  \Big) \,,
\end{aligned}
$$
which implies that
$$
\Covb\big(\xi_{i} \eta_{i}, \xi_{j} \eta_{j}\big) =  p^3 q^3 b_0^2 b_i b_j \sum_{l=1}^N R_{il} R_{j l} \, .
$$
As a result,
$$
 \sum_{i=1}^{N} \sum_{j \in \node_{G_{i, 2}} \backslash \node_{G_{ki}} } \Covb\big(\xi_{i} \eta_{i}, \xi_{j} \eta_{j}\big) =  p^3 q^3 b_0^2  \sum_{j \in \node_{G_{i, 2}} \backslash \node_{G_{ki}} } b_i b_j \sum_{k=1}^N R_{ik} R_{j k} \, .
$$
Hence,
$$
\begin{aligned}
 \sum_{i=1}^{N} \sum_{j \in \node_{G_{i, 2}} } \Covb\big(\xi_{i} \eta_{i}, \xi_{j} \eta_{j}\big) & \precsim  \sum_{i=1}^{N} \sum_{j \in \node_{G_{ki}}} \big( c_0 c_i c_j R_{i j} (c_i + c_j) + c_0^2 c_i c_j R_{ij}^2 \big) \\
 & + c_0^2 \Big| \sum_{i=1}^{N}  \sum_{j \in \node_{G_{i, 2}} } b_i b_j \sum_{l=1}^N R_{il} R_{j l} \Big| \, .
\end{aligned}
$$

\subsubsection{Proof of (\ref{eq:11}) }

For $f \in A_{e} \backslash \{e\}$ which means $e$ and $f$ have one common node, let's call $e=(1,2)$, $f=(2,3)$. We can firstly write $\xi_{(1,2)} \eta_{(1,2)}$ and $\xi_{(2,3)} \eta_{(2,3)}$ as 
$$
\begin{aligned}
& \xi_{(1,2)} \eta_{(1,2)}\\
 =& b_0 h(1) h(2) \big( b_1 h(1) +  b_2 h(2) \big) R_{12} 
\\
& + b_0^2 h(1)h(2) R_{12} \Big(  h(1) h(2) R_{12} +  h(1) h(3) R_{1 3} +  h(2) h(3) R_{2 3} \Big) \\
& + b_0^2 h(1)^2 h(2) R_{12} \sum_{j \in \node_{G_{k1}} \backslash \{2,3\} } R_{1j} h(j) + b_0^2 h(1) h(2)^2 R_{12} \sum_{j \in \node_{G_{k2}} \backslash \{1,3\} } R_{2j} h(j)  \,,
\end{aligned}
$$
$$
\begin{aligned}
& \xi_{(2,3)} \eta_{(2,3)} \\
= &   b_0 h(2) h(3) \big(  b_2 h(2)  + b_3 h(3) \big) R_{23} 
\\
& + b_0^2 h(2)h(3) R_{2 3} \Big(  h(2) h(3) R_{23} +  h(1) h(3) R_{1 3}  +  h(1) h(2) R_{1 2} \Big) \\
& + b_0^2 h(2)^2 h(3) R_{23} \sum_{j \in \node_{G_{k2}} \backslash \{1,3\} } R_{2j} h(j)+ b_0^2 h(2) h(3)^2 R_{23} \sum_{j \in \node_{G_{k3}} \backslash \{1,2\} } R_{3j} h(j)  \,.
\end{aligned}
$$
Note that
$$\Eb\big( h(i) \big) = 0, \quad \Eb\big( h(i)^2 \big) = pq, \quad \Eb\big( h(i)^3 \big) = pq(q-p), \quad \Eb\big( h(i)^4 \big) = pq (p^3 + q^3) \, ,$$
we have 
\begin{align*}
 & \Eb\big(\xi_{(1,2)} \eta_{(1,2)} \xi_{(2,3)} \eta_{(2,3)} \big) \\
 = &   p^3 q^3 b_0^2 R_{12} R_{23} \Big( b_1 b_3 + (q-p) b_0 b_1  ( R_{13}  + R_{23}) + 2 (q-p) b_0 b_2 R_{13}\\
& + (q-p) b_0 b_3 (R_{12} +  R_{13}  ) + (p^3 + q^3) b_0^2 R_{12} R_{23} \\
& + (q-p)^2 b_0^2 R_{13} ( 2 R_{12} + R_{13} + 2 R_{23}) \\
& + (p^3 + q^3) b_0^2 R_{12}R_{23} + p^4 q^4 b_0^2 \big( \sum_{j=1}^N R_{1 j} R_{3 j} - R_{12}R_{32} \big) \Big)
\end{align*}
and
$$ \Eb\big(\xi_{(1,2)} \eta_{(1,2)}\big)  \Eb\big( \xi_{(2,3)} \eta_{(2,3)} \big) = p^4 q^4 b_0^4 R_{12}^2 R_{23}^2 \,,$$ 
which further implies that
$$
\begin{aligned}
 & \Covb\big(\xi_{(1,2)} \eta_{(1,2)}, \xi_{(2,3)} \eta_{(2,3)} \big) - p^3 q^3 b_0^2 R_{12} R_{23}  b_1 b_3  \\
&  \precsim b_0^2 R_{12} R_{23} \Big( |b_0| |b_1|  ( R_{13} + R_{23} ) +  |b_0| |b_3| (R_{12} +  R_{13} ) +  |b_0| |b_2| R_{13} \\
& + b_0^2 R_{13}( R_{12} + R_{13} +  R_{23} ) +  b_0^2  \sum_{j=1}^N R_{1 j} R_{3 j}  \Big) \\
& \precsim c_0^3 R_{12} R_{23} \Big(  c_1  ( R_{13} + R_{23}) +   c_3  (R_{12} +  R_{13} ) +  c_2 R_{13}\\
& + c_0 R_{13} ( R_{12} + R_{13} +  R_{23} ) +  c_0  \sum_{j=1}^N R_{1 j} R_{3 j}  \Big) \,.
\end{aligned}
$$
As a result,
\begin{align*}
& \sum_{e \in G_k} \sum_{f \in A_{e} \backslash\{e\}} \Covb(\xi_{e} \eta_{e}, \xi_{f} \eta_{f}) =  \sum_{i=1}^N \sum_{j,l \in \node_{G_{ki}}}^{j \neq l}  \Covb\big(\xi_{(j,i)} \eta_{(j,i)}, \xi_{(i,k)} \eta_{(i,k)} \big)\\
\precsim &  c_0^3  \sum_{i=1}^N \sum_{j,l \in \node_{G_{ki}}}^{j \neq l}  R_{ji} R_{ik} \big(  c_j ( R_{jk} + R_{ik} ) +   c_k (R_{ji} +  R_{jk} ) +  c_i R_{jk}  \big)\\
& +  c_0^4  \sum_{i=1}^N \sum_{j,l \in \node_{G_{ki}}}^{j \neq l}  R_{ji} R_{il} \big(R_{j l} ( R_{ji} + R_{jl} +  R_{il} ) + \sum_{s=1}^N R_{j s} R_{l s}  \big) \\
& + c_0^2 \Big| \sum_{i=1}^N \sum_{j,l \in \node_{G_{ki}}}^{j \neq l}  R_{ji} R_{il}  b_j b_l  \Big| 
\end{align*}

\subsubsection{Proof of (\ref{eq:12}) }

For $f \in B_{e} \backslash A_{e}$ which means $f$ and $e$ have no common nodes, let us call $e=(1, 2)$ and $f=(3, 4)$. We can firstly write $\xi_{(1,2)} \eta_{(1,2)}$ and $\xi_{(3,4)} \eta_{(3,4)}$ as 
\begin{align*}
\xi_{(1,2)} \eta_{(1,2)} & = b_0 h(1) h(2) \big( b_1 h(1) +  b_2 h(2) \big) R_{12}  + b_0^2 h(1)^2 h(2)^2 R_{12}^2
\\
& + b_0^2 h(1)h(2) R_{12} \big(   h(1) h(3) R_{1 3}  +  h(1) h(4) R_{1 4} \big) \\
& + b_0^2 h(1)h(2) R_{12} \big(  h(2) h(3) R_{2 3} +   h(2) h(4) R_{2 4}   \big) \\
& + b_0^2 h(1)^2 h(2) R_{12} \sum_{j \in \node_{G_{k1}} \backslash \{2,3,4\} } R_{1j} h(j)\\
& + b_0^2 h(1) h(2)^2 R_{12} \sum_{j \in \node_{G_{k2}} \backslash \{1,3,4\} } R_{2j} h(j)  \,, \\
\xi_{(3,4)} \eta_{(3,4)} & =  b_0 h(3) h(4) \big(  b_3 h(3)  + b_4 h(4) \big) R_{34}  + b_0^2 h(3)^2 h(4)^2 R_{3 4}^2 
\\
& + b_0^2 h(3)h(4) R_{3 4}  \Big(   h(1) h(3) R_{1 3}  +  h(1) h(4) R_{1 4} \Big) \\
& + b_0^2 h(3)h(4) R_{34}  \Big(  h(2) h(3) R_{2 3} +   h(2) h(4) R_{2 4}   \Big) \\
& + b_0^2 h(3)^2 h(4) R_{34} \sum_{j \in \node_{G_{k3}} \backslash \{1,2,4\} } R_{3j} h(j) \\
& + b_0^2 h(3) h(4)^2 R_{34} \sum_{j \in \node_{G_{k4}} \backslash \{1,2,3\} } R_{4j} h(j)  \,.
\end{align*}
As a result, we have
$$
\begin{aligned}
 & \Eb\big(\xi_{(1,2)} \eta_{(1,2)} \xi_{(3,4)} \eta_{(3,4)} \big) = p^4 q^4 b_0^4  R_{12}^2 R_{34}^2 + p^4 q^4 b_0^4  R_{12} R_{34} \big( 2 R_{1 3} R_{24}  + 2  R_{1 4} R_{2 3} \big)
\end{aligned}
$$
and 
$$ 
\begin{aligned}
& \Covb\big(\xi_{(1,2)} \eta_{(1,2)}, \xi_{(3,4)} \eta_{(3,4)} \big)  = 2 p^4 q^4 b_0^4 R_{12} R_{34} \big( R_{1 3} R_{24}   +   R_{1 4} R_{2 3} \big) \,.
\end{aligned}
$$
Then 
$$
\begin{aligned}
\sum_{e \in G} \sum_{f \in B_{e} \backslash A_{e}} \Covb(\xi_{e} \eta_{e}, \xi_{f} \eta_{f}) & \precsim b_0^4  \sum_{e \in G} \sum_{f \in B_{e} \backslash A_{e}} R_e R_f  \big( R_{e^+f^+}R_{e^-f^-} + R_{e^+f^-}R_{e^-f^+} \big) \\
&  \precsim c_0^4  \sum_{i=1}^N \sum_{j=1}^N  \sum_{l \neq i,j}^N \sum_{s \neq i,j}^N R_{ij} R_{jl} R_{ls}R_{si}  \,.
\end{aligned}
$$

\subsubsection{Proof of (\ref{eq:13}) }

When $f \in C_{e} \backslash B_{e}$,  let us call $e=(1, 2)$ and $f=(3, 4)$. We can firstly write $\xi_{(1,2)} \eta_{(1,2)}$ and $\xi_{(3,4)} \eta_{(3,4)}$ as 
\begin{align*}
\xi_{(1,2)} \eta_{(1,2)} = &  b_0 h(1) h(2) \big( b_1 h(1) +  b_2 h(2) \big) R_{12}  + b_0^2 h(1)^2 h(2)^2 R_{12}^2
\\
& + b_0^2 h(1)^2 h(2) R_{12} \sum_{j \in \node_{G_{k1}} \backslash \{2,3,4\} } R_{1j} h(j) \\
& + b_0^2 h(1) h(2)^2 R_{12} \sum_{j \in \node_{G_{k2}} \backslash \{1,3,4\} } R_{2j} h(j)  \,, \\
\xi_{(3,4)} \eta_{(3,4)} = &   b_0 h(3) h(4) \big(  b_3 h(3)  + b_4 h(4) \big) R_{34}  + b_0^2 h(3)^2 h(4)^2 R_{3 4}^2 
\\
& + b_0^2 h(3)^2 h(4) R_{34} \sum_{j \in \node_{G_{k3}} \backslash \{1,2,4\} } R_{3j} h(j) \\
& + b_0^2 h(3) h(4)^2 R_{34} \sum_{j \in \node_{G_{k4}} \backslash \{1,2,3\} } R_{4j} h(j)  \,.
\end{align*}
As a result, we have
$$  \Eb\big(\xi_{(1,2)} \eta_{(1,2)} \xi_{(3,4)} \eta_{(3,4)} \big)    = p^4 q^4 b_0^4  R_{12}^2 R_{34}^2 = \Eb\big(\xi_{(1,2)} \eta_{(1,2)}\big)  \Eb\big(  \xi_{(3,4)} \eta_{(3,4)} \big) \, ,$$
which implies that
$$\sum_{e \in G} \sum_{f \in C_{e} \backslash B_{e}} \Covb(\xi_{e} \eta_{e}, \xi_{f} \eta_{f}) = 0 \,.$$

\subsubsection{Proof of \eqref{eq:14} }

\begin{align*}
 \Eb \big( |\xi_{i} \eta_{i}^{2}| \big) & = \Eb \Big( |b_i h(i)| \big( b_i h(i) + b_0 h(i) \sum_{j \in \node_{G_{ki}}} R_{ij} h(j) \big)^2 \Big) \\
& =  \Eb \big( |b_i h(i)^3| \big)  \Eb \big( b_i  + b_0 \sum_{j \in \node_{G_{ki}}} R_{ij} h(j) \big)^2  \\
& = |b_i| p q (p^2 + q^2) (b_i^2 + pq b_0^2 \sum_{j=1}^N R_{ij}^2)\,,
\end{align*}
which implies that
$$\sum_{i=1}^{N}  \Eb \big( |\xi_{i} \eta_{i}^{2}| \big) =  \sum_{i=1}^{N} |b_i| p q (p^2 + q^2) (b_i^2 + pq b_0^2 \sum_{j=1}^N R_{ij}^2) \precsim  \sum_{i=1}^{N} c_i^3 + c_0^2 \sum_{i=1}^{N} c_i \sum_{j=1}^N R_{ij}^2 \, .$$

\subsubsection{Proof of \eqref{eq:15} }
\begin{align*}
 & \Eb \big( |\xi_{e}| \eta_{e}^{2} \big) \\
= &   \Eb \Big( | b_0   h(e^+)  h(e^-) R_e|  \big( b_{e^+} h(e^+) + b_{e^-} h(e^-) + b_0 h(e^+) h(e^-) R_e \\
 & + b_0h(e^+) \sum_{j \in \node_{G_{k e^+}} \backslash\{e^-\}} R_{e^+ j} h(j) + b_0h(e^-) \sum_{l \in \node_{G_{k e^-}} \backslash\{e^+\}} R_{e^- l} h(l) \big)^2 \Big) \\
= &   \Eb \Big( | b_0   h(e^+)  h(e^-) R_e|  \big( b_{e^+} h(e^+) + b_{e^-} h(e^-) + b_0 h(e^+) h(e^-) R_e \big)^2 \Big)  \\
& + \Eb \Big( | b_0^3   h(e^+)  h(e^-) R_e|  \big( h(e^+) \sum_{j \in \node_{G_{k e^+}} \backslash\{e^-\}} R_{e^+ j} h(j)  + h(e^-) \sum_{l \in \node_{G_{k e^-}} \backslash\{e^+\}} R_{e^- l} h(l) \big)^2 \Big)   \\
= &   \Eb \Big( | b_0   h(e^+)  h(e^-) R_e|  \big( b_{e^+} h(e^+) + b_{e^-} h(e^-) + b_0 h(e^+) h(e^-) R_e \big)^2 \Big)  \\
& + 2 p^3 q^3 (p^2 + q^2) |b_0|^3 R_e \big( \sum_{j=1}^N   R_{e^+ j}^2 + \sum_{j=1}^N   R_{e^- j}^2 - 2 R_e^2 \big) \\
& + 2 p^3 q^3 (q-p)^2 |b_0|^3 R_e \sum_{j=1}^N   R_{e^+ j} R_{e^- j} \\ 
& \precsim |b_0|^3 R_e^3 + |b_0|R_e (b_{e^+}^2 + b_{e^-}^2) + |b_0|^3  R_e \big( \sum_{j=1}^N   R_{e^+ j}^2 + \sum_{j=1}^N   R_{e^- j}^2 \big) \, ,
\end{align*}
which shows that 
\begin{align*}
 \sum_{e \in G_k} \Eb\big( |\xi_e| \eta_e^2 \big) & \precsim  \sum_{e \in G_k} \Big( |b_0|^3 R_e^3 + |b_0|R_e (b_{e^+}^2 + b_{e^-}^2) + |b_0|^3  R_e \big( \sum_{j=1}^N   R_{e^+ j}^2 + \sum_{j=1}^N   R_{e^- j}^2 \big) \Big) \\
 & = \sum_{i=1}^N \sum_{j=1}^N \Big(  |b_0|^3 R_{ij}^3 + |b_0| R_{ij}(b_i^2 + b_j^2) + |b_0|^3 R_{ij} \sum_{l=1}^N \big( R_{il}^2 + R_{jl}^2 \big) \Big) \\
 & \precsim  \sum_{i=1}^N \sum_{j=1}^N \Big( c_0^3  R_{ij}^3 + c_0 R_{ij}(c_i^2 + c_j^2) + c_0^3 R_{ij} \sum_{l=1}^N \big( R_{il}^2 + R_{jl}^2 \big) \Big) \\
 &  \precsim c_0^3 \sum_{i=1}^N \sum_{j=1}^N   R_{ij}^3 + c_0 \sum_{i=1}^N c_i^2  R_{i \cdot} + c_0^3  \sum_{i=1}^N R_{i \cdot} \sum_{j=1}^N R_{ij}^2 \, .
\end{align*}

\end{document}